\documentclass[fleqn,usenatbib]{mnras}
\usepackage[T1]{fontenc}
\DeclareRobustCommand{\VAN}[3]{#2}
\let\VANthebibliography\thebibliography
\def\thebibliography{\DeclareRobustCommand{\VAN}[3]{##3}\VANthebibliography}

\usepackage{graphicx}
\usepackage{amsmath}
\usepackage{amssymb}
\usepackage{xcolor}
\usepackage{newtxtext,newtxmath}
\usepackage{xspace}

\defcitealias{Koposov2019}{K19}
\defcitealias{Erkal2019}{E19}

\newcommand{\Sfive}{${S}^5$\xspace}

\newcommand{\unit}[1]{\ensuremath{\mathrm{\,#1}}\xspace}
\newcommand{\feh}{\unit{[Fe/H]}}
\newcommand{\kms}{\unit{km\,s^{-1}}}
\newcommand{\masyr}{\unit{mas\,yr^{-1}}}
\newcommand{\muone}{\ensuremath{\mu_{\phi,1}}\xspace}
\newcommand{\mutwo}{\ensuremath{\mu_{\phi,2}}\xspace}
\newcommand{\Msun}{\unit{M_\odot}}
\newcommand{\CHANGE}[1]{{{#1}}}

\title[Orphan-Chenab in 6-D]{\Sfive: Probing the Milky Way and Magellanic Clouds potentials with the 6-D map of the Orphan-Chenab stream}

\author[Koposov, Erkal, Li et al.]{
Sergey~E.~Koposov,$^{1,2,3}$\thanks{E-mail: skoposov@ed.ac.uk},
Denis~Erkal$^{4}$\thanks{E-mail: d.erkal@surrey.ac.uk},
Ting~S.~Li$^{5}$,
Gary~S.~Da~Costa$^{6,7}$,
Lara~R.~Cullinane$^{8}$,
\newauthor Alexander~P.~Ji$^{9,10}$,
Kyler~Kuehn$^{11,12}$,
Geraint~F.~Lewis$^{13}$,
Andrew~B.~Pace$^{14}$,
Nora~Shipp$^{15}$,
\newauthor Daniel~B.~Zucker$^{16,17}$,
Joss~Bland-Hawthorn$^{13,7}$,
Sophia~Lilleengen$^{4}$,
and Sarah~L.~Martell$^{18,7}$\newauthor
(\Sfive Collaboration)
\\
$^{1}$ Institute for Astronomy, University of Edinburgh, Royal Observatory, Blackford Hill, Edinburgh EH9 3HJ, UK\\
$^{2}$ Institute of Astronomy, University of Cambridge, Madingley Road, Cambridge CB3 0HA, UK\\
$^{3}$ Kavli Institute for Cosmology, University of Cambridge, Madingley Road, Cambridge CB3 0HA, UK\\
$^{4}$ Department of Physics, University of Surrey, Guildford GU2 7XH, UK\\
$^{5}$ Department of Astronomy and Astrophysics, University of Toronto, 50 St. George Street, Toronto ON, M5S 3H4, Canada\\
$^{6}$ Research School of Astronomy and Astrophysics, Australian National University, Canberra, ACT 2611, Australia\\
$^{7}$ Centre of Excellence for All-Sky Astrophysics in Three Dimensions (ASTRO 3D), Australia\\
$^{8}$ Department of Physics and Astronomy, Johns Hopkins University, 3400 N. Charles St, Baltimore, MD 21218, USA\\
$^{9}$ Department of Astronomy \& Astrophysics, University of Chicago, 5640 S Ellis Avenue, Chicago, IL 60637, USA\\
$^{10}$ Kavli Institute for Cosmological Physics, University of Chicago, Chicago, IL 60637, USA\\
$^{11}$ Lowell Observatory, 1400 W Mars Hill Rd, Flagstaff,  AZ 86001, USA\\
$^{12}$ Australian Astronomical Optics, Faculty of Science and Engineering, Macquarie University, Macquarie Park, NSW 2113, Australia\\
$^{13}$ Sydney Institute for Astronomy, School of Physics, A28, The University of Sydney, NSW 2006, Australia\\
$^{14}$ McWilliams Center for Cosmology, Carnegie Mellon University, 5000 Forbes Ave, Pittsburgh, PA 15213, USA\\
$^{15}$ MIT Kavli Institute for Astrophysics and Space Research, 77 Massachusetts Ave., Cambridge, MA 02139, USA\\
$^{16}$ School of Mathematical and Physical Sciences, Macquarie University, Sydney, NSW 2109, Australia\\
$^{17}$ Macquarie University Research Centre for Astronomy, Astrophysics \& Astrophotonics, Sydney, NSW 2109, Australia\\
$^{18}$ School of Physics, UNSW, Sydney, NSW 2052, Australia\\
}

\date{Accepted XXX. Received YYY; in original form ZZZ}

\pubyear{2022}

\begin{document}
\label{firstpage}
\pagerange{\pageref{firstpage}--\pageref{lastpage}}
\maketitle

\begin{abstract}
We present a 6-D map of the Orphan-Chenab (OC) stream by combining the data from Southern Stellar Stream Spectroscopic Survey (\Sfive) and {\it Gaia}. We reconstruct the proper motion, radial velocity, distance, on-sky track and stellar density along the  stream with spline models.
The stream has a total luminosity of $M_V=-8.2$ and metallicity of $\mathrm{[Fe/H]}=-1.9$, similar to classical Milky Way (MW) satellites like Draco. 
The stream shows drastic changes in its physical width varying from 200 pc to 1 kpc, but a constant line of sight velocity dispersion of 5 \kms. Despite the large apparent variation in the stellar number density along the stream, the flow rate of stars along the stream is remarkably constant. 
We model the 6-D stream track by a Lagrange-point stripping method with a flexible MW potential in the presence of a moving extended Large Magellanic Cloud (LMC). This allows us to constrain the mass profile of the MW within the distance range 15.6\,<\,r\,<\,55.5\, kpc, with the best measured enclosed mass of $(2.85\pm 0.1)\times10^{11}\,\Msun$ within 32.4\,kpc. Our stream measurements are highly sensitive to the LMC mass profile with the most precise measurement of its enclosed mass made at 32.8 kpc, $(7.02\pm 0.9)\times10^{10}\, {\rm M}_\odot$. We also detect that the LMC dark matter halo extends to at least 53\,kpc. The fitting of the OC stream allows us to constrain the past LMC trajectory and the degree of dynamical friction it experienced.
We demonstrate that the stars in the OC stream show large energy and angular momentum spreads caused by LMC perturbation.
\end{abstract}




\section{Introduction}

Stellar streams are one of the most impressive illustrations of the hierarchical nature of galaxy formation predicted by the $\Lambda CDM$ paradigm \citep{white1991}. While Milky Way-like galaxies are expected to have accreted a large number of stellar systems over their assembly history \citep{bullock2005}, only in the last 20 years or so have we started to obtain direct evidence of a large number of disrupting low mass systems in the Milky Way stellar halo \citep[e.g.][]{ibata1994,odenkirchen2003,belokurov2006,grillmair2006}.

Since the discovery of the first streams in the Milky Way halo, it was clear that streams can provide powerful constraints on the Galactic potential \citep{johnston1999,binney2008,Koposov2010} as streams approximately trace the orbit of their progenitor. While initial models of streams were often quite simplistic and relied on the (incorrect) assumption that streams are orbits \citep{sanders2013}, new, more sophisticated ways of modelling stellar streams have since been developed \citep{bovy2014,Gibbons2014,fardal2015}. With the number of known streams growing over time due to increased depth and coverage of Galactic surveys \citep{koposov2014,bernard2014}, this has opened a window for methods that fit multiple stellar streams together with the aim of breaking degeneracies between the parameters of the Galactic potential 
\citep{bovy2016, Bonaca+2018}. However, as yet more streams were discovered with the DES survey \citep{shipp2018} and {\it Gaia} \citep{malhan2018}, the idealistic picture of streams being a clean and easy tracer that can be used to constrain the local Galactic potential has become more muddied. 

As streams were mapped in greater detail using photometric data, it has become clear that streams do not look like clean Gaussian trails of stars. Instead they show stream gaps \citep{carlberg2013,Erkal2017}, broadenings \citep{pricewhelan2016,malhan2018}, bifurcations \citep{belokurov2006}, and "wiggles" \citep{li2021}. The exact attribution of these structures to specific physical effects is still a matter of debate but is likely some combination of perturbations by other MW satellites \citep{deBoer:2020,li2021,Dillamore+2022}, DM subhalos \citep{ibata2002,johnston2002,Yoon2011,carlberg2020}, giant molecular clouds \citep{amorisco2016}, and interaction with the Milky Way bar \citep{hattori2016}. Together  this makes the analysis of streams and their use as a confident tracer of the MW potential more difficult. 

The final nail in the coffin of streams being straightforward tracers of the MW potential was the study of the so-called Orphan stream \citep{grillmair2006,belokurov2007} with {\it Gaia} DR2 data. While initially the stream was traced over approximately 60 degrees on the sky, the RR Lyrae from {\it Gaia} DR2 allowed \cite{Koposov2019} (hereafter \citetalias{Koposov2019}) to trace it for $\sim200$ degrees. It turned out that the stream extension to the South was actually previously detected  and was called "Chenab" by \citet{shipp2018}. The reason for the mis-association of the stream as a new stream, as opposed to the continuation of Orphan, was that their orbital planes were seen as quite distinct. However, the cause for this was found to be that the stream is being actively perturbed by the Large Magellanic Cloud (\citealt{Erkal2019}, hereafter \citetalias{Erkal2019}), causing the stream to twist and the stars in the stream to move sideways. While complicating the analysis of the Milky Way potential, this nevertheless allowed them to put strong constraints on the mass of the Large Magellanic Cloud (LMC), revising its mass to about a fifth or tenth of the Milky Way's. This in turn led to the realisation of a very strong perturbation to the Milky Way potential by the Clouds that manifests itself in multiple forms: direct gravitational pull by the Clouds and second-order effects such as the LMC DM wake \citep{garavito2019,Belokurov:2019,Conroy:2021}, the motion of the MW centre around the LMC+MW centre of mass \citep{Gomez+2015,Erkal:2021,Petersen:2021}, and the deformation of the MW dark matter distribution \citep{lilleengen+2022}. Since then, the effects of the LMC, expressed as stars moving at an angle with respect to the stream itself, have been seen in other streams \citep{shipp2019}.

The effects of the LMC have made stream modelling a much more difficult task, since it requires introducing a time-dependent LMC potential, and accounting for the motion of the MW centre. However, this modelling should now allow us to start probing not only the MW dark matter distribution, but also the LMC dark matter distribution, and how these deform throughout the interaction. Such sophisticated models require extensive, ideally fully 6-D datasets. \cite{Erkal2019} developed the first such model to fit the 5-D data of the OC stream from \cite{Koposov2019}. \cite{Vasiliev+2021} have also demonstrated this with a model of the Sagittarius stream, where model took into account the MW and LMC interaction  to constrain the parameters of both MW and LMC.

In this paper, we focus on the Orphan-Chenab (or OC) stream as it is probably the best stream  for probing the MW potential and the LMC. This stream is very long, probes the Galaxy from $\sim 15$ kpc to $\sim 80$ kpc, and is affected by the LMC. It also does not have the complexity of the Sgr stream with its unexplained bifurcation. Because of this, the OC stream was followed up by the Southern Stellar Stream Spectroscopic Survey (\Sfive) over the last 5 years in order to make a detailed phase space map of the stream \citep{Li2019}. In this paper, we combine the results of the dedicated \Sfive follow-up of the OC stream with other public spectroscopic datasets and recent {\it Gaia} data to model the 6-D phase space of the OC stream. We then use this 6-D track to infer the potential of the Milky Way and LMC through a detailed stream model fitted to the 6-D stream track.

\CHANGE{In the paper we also discuss possible origins of the OC stream and location of the stream progenitor, as despite many claims in the literature of possible association of the stream with known objects \citep{fellhauer2007,grillmair2015,Koposov2019}, no stream progenitor has been identified yet. }

The paper is structured as follows. In Section~\ref{sec:data} we describe the data we rely on in this paper. Section~\ref{sec:6d_track} demonstrates how we model the 6-D phase space track of the stream. \CHANGE{In Section~\ref{sec:analysis} we analyse and interpret observational features extracted in Section~\ref{sec:6d_track}}. Section~\ref{sec:modelling} describes the stream model we employ and in Section~\ref{sec:discussion} we discuss the inference from the model. We conclude in Section ~\ref{sec:conclusions}.

\section{Data}
\label{sec:data}
\subsection{Observations and Target Selection for \Sfive}
\label{sec:s5data}

Most of the data presented in this paper are collected as part of the observations by \Sfive, which couples the Two-degree Field (2dF) fiber positioner \citep{Lewis:2002} with the dual-arm AAOmega spectrograph \citep{Sharp:2006} on the 3.9-m Anglo-Australian Telescope (AAT) to pursue a complete census of known streams in the Southern Hemisphere. For more details about observation, instrument setup and target selection, we refer readers to \citet{Li2019}. The OC stream is one of the 20 streams that have been mapped by \Sfive.

This paper includes \Sfive observations between August 2018 and April 2021. A total of 41 fields dedicated to the OC stream were observed at the AAT, 34 of which were observed in 2018 and 2019, and are listed in Table 2 of \citep{Li2019}; the remaining 7 were added in 2020 and 2021. 
All of the fields have an exposure time between 3000 and 7200 seconds. Since the OC stream covers a large area on the sky and some other streams intersect it, we also include these stream fields in addition to the dedicated OC fields.

As described in \cite{Li2019}, we identify probable OC member stars by applying selections in proper motion space and in colour-magnitude diagram (CMD) space. In particular, we select the targets based on the proper motions and distances of the RR Lyrae stars in OC from \citetalias{Koposov2019}. We compute the difference in proper motions between the targets and the average proper motion from RR Lyrae stars at a given stream longitude, or $|\Delta\mathrm{PM}|$. For the majority of fields we used $|\Delta\mathrm{PM}| < 2$\,\masyr, while close to the Galactic plane (at $|\phi_1|\lesssim15$ deg) we   
used $|\Delta\mathrm{PM}| < 1$\,\masyr .
For most of the fields DES photometry was used for CMD selection; otherwise, {\it Gaia} photometry was used. 

\subsection{Data Reduction for \Sfive}
\label{sec:s5datared}

In this paper, we rely on the \Sfive data from the internal release DR3.1 which differs from the one described in \citet{Li2019} and released in \citet{S5_DR1} in several aspects. We have significantly larger coverage of the sky compared to DR1, as well as improved modelling. For more details of the model fitting updates, we refer readers to \citet{S5_12streams}. In particular, we are now modelling blue and red spectra of all the exposures simultaneously for each object in order to improve stellar parameter estimates. The DR3.1 data also provides the spectro-photometric distance estimates, which are obtained by simultaneous fitting of the spectroscopy, photometry and {\it Gaia} parallaxes by interpolated PHOENIX \citep{Husser2013} spectral templates and MIST isochrones (Koposov et al. 2023 in prep).

\subsection{Other spectroscopic surveys}
\label{sec:specsurveys}

In this paper, we use radial velocity and iron abundance measurements from \Sfive and from several large spectroscopic surveys: APOGEE DR17  \citep{Nidever2015,Ahumada2020,Jonsson2020}, LAMOST DR7 \citep{cui2012,zhao2012}, and SDSS DR14 \citep{Abolfathi2018} measurements by the SSPP spectroscopic pipeline \citep{Lee2008a,Lee2008b,Smolinski2011}. We use the {\tt vhelio\_avg}, {\tt verr} and {\tt fe\_h} columns from allstar table in APOGEE, and {\tt rv}, {\tt rv\_err}, and {\tt feh} columns from LAMOST LRS table\footnote{ We apply a 5 \kms offset to LAMOST radial velocities as this is the typical velocity offset we notice with respect to APOGEE velocities.}. We use the {\tt fehadop}, {\tt elodiervfinal}, and {\tt elodiervfinalerr} columns from SDSS.

\subsection{Gaia}

Throughout the paper, unless specified otherwise, we rely on {\it Gaia } EDR3 data \citep{gaia_edr3}, such as parallaxes, proper motions and BP/RP/G magnitudes.

\section{6-D track extraction} \label{sec:6d_track}
In this section, we describe the modelling of the OC stream stars' positions, velocities, distances and colour-magnitude diagram locations in order to extract the 6-D phase space track of the stream.  

\begin{figure*}
    \centering
    \includegraphics{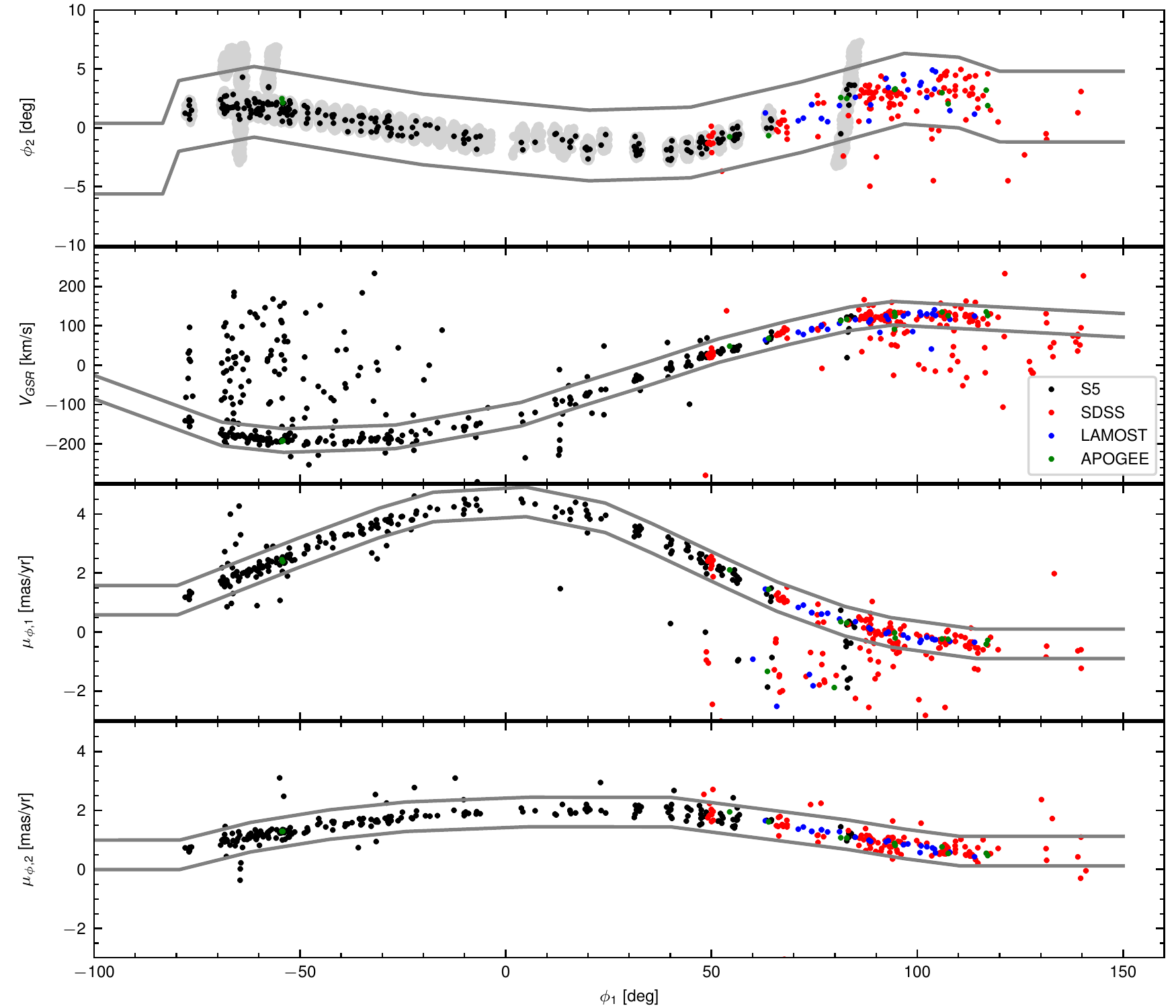}
    \caption{Identification of spectroscopic OC stream members in the \Sfive, SDSS, LAMOST and APOGEE datasets. The 4 panels show the distribution of various properties of stars in our sample (position on the sky $\phi_2$, GSR radial velocity, and proper motion in $\phi_1$ and $\phi_2$) versus the angle along the OC stream ($\phi_1$). The stars plotted in each panel have been selected using all other panels: i.e. the plot of $\phi_2$ vs $\phi_1$ relies on proper motion and radial velocity selection but not on position. The grey lines delineate the selection boundaries for likely OC stream members. Black points show stars from \Sfive, red from SDSS, blue from LAMOST and green from APOGEE. The stream is excellently traced throughout 200 degrees on the sky \citepalias[c.f. Figure 2 from ][]{Koposov2019}. } 
    \label{fig:rv_pm}
\end{figure*}

\subsection{Conventions}
Throughout the paper, we rely on the standard coordinate system aligned with the Orphan stream ($\phi_1$, $\phi_2$) as defined in 
\citetalias{Koposov2019}. As the $(\phi_1,\phi_2)=(0,0)$ point is the location where the stream crosses the MW plane and $\phi_1>0$ part of the stream is in the North Galactic Cap, we will refer the $\phi_1>0$ parts of the stream as Northern, while $\phi_0<0$ as Southern.  We also use 
\muone, \mutwo
which refer to the proper motions in the $\phi_1$, $\phi_2$ heliocentric frame (not corrected for the solar motion). The proper motions in right ascension and $\phi_1$ (unless specified otherwise) always include the cosine term.

In the paper, we occasionally use measurements in the Galactocentric frame. This frame is based on the {\tt astropy} v4.0 frame \citep{Reid2004,Abuter2018,Drimmel2018,Bennett2019}.
\subsection{Kinematic sample}
\label{sec:kinematic_sample}

We proceed to map the OC stream in steps with the goal of fully extracting its track in 6-D phase space. We start with the spectroscopic data as these provide us with the cleanest separation of stream members from foreground stars allowing us to bootstrap our OC member identification.

We first assemble the list of potential OC members from the \Sfive, APOGEE, SDSS, and LAMOST surveys by selecting the stars within 10 degrees of the great circle established by \citetalias{Koposov2019}. On top of the spatial selection, we apply a {\it Gaia} parallax  selection, $\omega < 3 \sigma_\omega + 0.1$ mas, which should remove nearby \CHANGE{(closer than $\sim$ 10 kpc)} stars. \CHANGE{} \CHANGE{The small number of duplicates observed in multiple surveys are removed. When deciding which duplicate to remove we prioritise the surveys such that highest priority is given to APOGEE followed by \Sfive, SDSS and LAMOST.} We require that the spectroscopic metallicity is $\feh<-1.5$.   For the \Sfive stars we also require that the 84-th percentile of spectro-photometric distance is larger than 10\,kpc (this guarantees that the star is  confidently not a nearby star) to further reduce the contamination from nearby stars. Figure~\ref{fig:rv_pm} shows the spatial and kinematic distribution of the sample as a function of the angle along the stream. 
On the plot we also show the linear spline-based regions (grey lines) that we use to make the preliminary selection of possible OC stream members. We call these fiducial splines, and they are broadly based on the stream tracks from \citetalias{Koposov2019}. The splines are provided in the supplementary materials of the paper and the width of the selection region is 3 degrees for the spatial selection, 30\,\kms for the radial velocity selection, and 0.5 \masyr in proper motion. \CHANGE{The total number of possible stream members from different surveys that fall within the spline boundaries in the $-100<\phi_1<150$ range is 213, 110, 25 and 12 for \Sfive, SDSS,  LAMOST and  APOGEE respectively.}

For Figure~\ref{fig:rv_pm}, we use a convention similar to the one adopted in Figure 2 of \citetalias{Koposov2019}, where for each panel of the plot we use the selection from all the other panels. In other words, the plot of radial velocity versus angle only shows stars selected based on positions and proper motions (but not radial velocity), while the plot of the \muone proper motion versus angle includes stars selected based on $RV$ and position on the sky and \mutwo (but not \muone). This type of plot allows us to cleanly see the completeness and contamination of our selections, as well as better assess how prominent the stream signal is.
The coloured points on the plot show measurements from different surveys, with black points representing the \Sfive data. The filled light grey regions in the top panel show areas targeted by the \Sfive survey.

The OC stream track on Figure~\ref{fig:rv_pm} is clearly visible throughout almost 200 degrees on the sky from $\phi_1\sim -80$ deg to $\phi_1 \sim 120$ deg on all panels, matching the behaviour tracked by RR Lyrae in \citetalias{Koposov2019}, but with significantly more stars, higher proper motion precision, and the measurement of radial velocities. We remark that the \Sfive\ data alone maps the stream over almost $\sim 110 $ degrees without any significant gaps, while for the Northern part of the stream ($\phi_1\gtrsim 80$ deg), the 
\Sfive\ data is complemented by SDSS, LAMOST, and somewhat by APOGEE.

\CHANGE{In the next several sections, we will now rely on either the spectroscopic sample of likely stream member stars constructed with the splines boundaries shown on Figure~\ref{fig:rv_pm}, or the same splines selections applied to {\it Gaia}-only data without spectroscopy. The spectroscopic sample of possible OC members based on fiducial splines  shown on Figure~\ref{fig:rv_pm} is provided in Table~\ref{tab:spec_members}.}

\subsection{RR Lyrae Distance model}
\label{sec:distance_model}

In this section, we construct a model of the distance along the stream. We use the catalogue of RR Lyrae from {\it Gaia} DR2 \citep{Holl2018,Clementini2019} complemented by the RR Lyrae from \citet{Stringer2021} and \citet{Sesar2017} cross-matched with {\it Gaia} EDR3. We compute distances from {\it Gaia} magnitudes using the equation from Table 4 of \citet{Muraveva2018} and the extinction coefficient of $A_G/E(B-V)=2.27$ from \citet{Iorio2021}. We use the {\tt phot\_g\_mean\_mag} average magnitudes from {\it Gaia} EDR3, shifted by 0.03 mag due to the typical offset between {\tt phot\_g\_mean\_mag} and the luminosity averaged mean magnitude used by \citet{Muraveva2018}.

\begin{figure}
    \centering
    \includegraphics{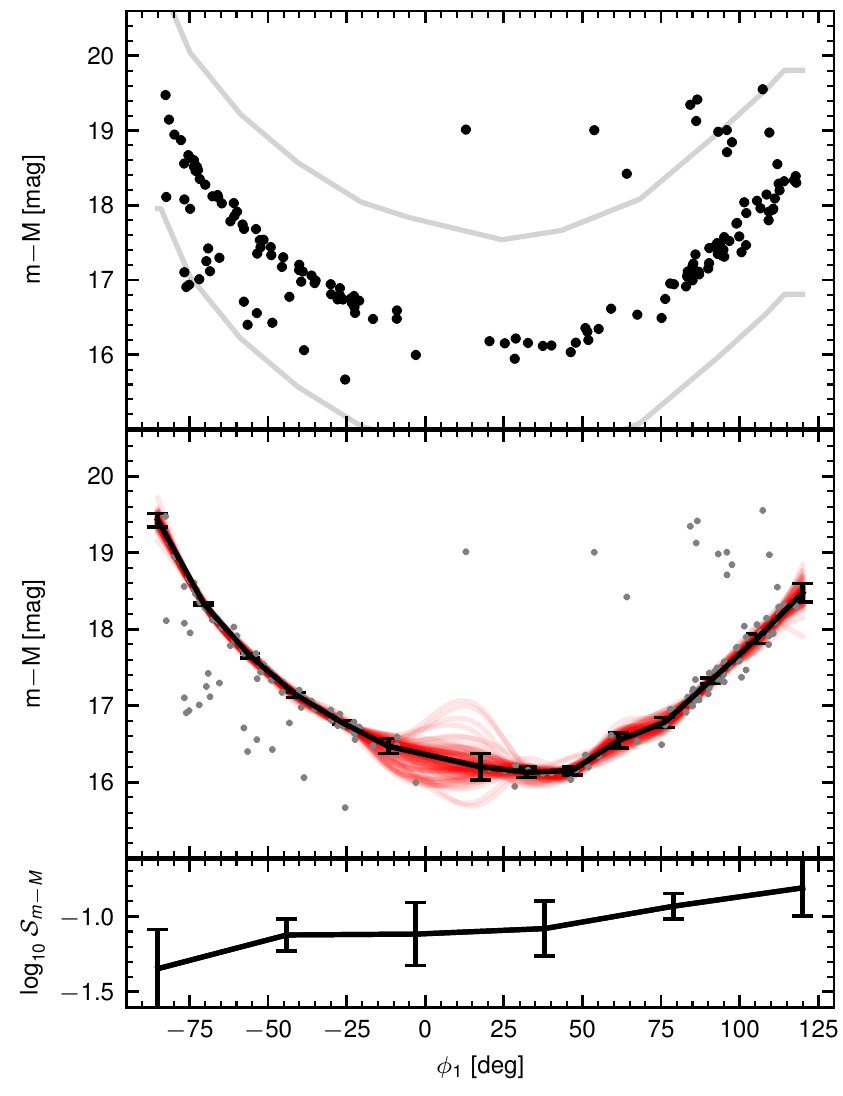}
    \caption{Modelling of the distances to the OC stream stars. {\it Top panel:} The distance moduli to the RR Lyrae selected by proper motion and position on the sky. Grey lines show the selection boundaries for stars considered in the fit. {\it Middle panel:} Grey points show the distance moduli of RR Lyrae selected by proper motion and position on the sky. Black points with error-bars show the distance modulus spline knots. Red lines show sampled models from the posterior.
    {\it Bottom panel:} Base 10 logarithm of intrinsic distance modulus spread at the location of spline knots inferred from the model. We observe an increase of the distance modulus spread to $\sim $ 0.2 mag in the North from about 0.05 mag in the South.} 
    \label{fig:dist_model}
\end{figure}

For the list of assembled RR Lyrae, we compute the positions and proper motions along $\phi_1$ and $\phi_2$, selecting only the ones where $\muone$ and $\mutwo$ are within 0.25 mas$\,$yr$^{-1}$ of the fiducial proper motion tracks shown Figure~\ref{fig:rv_pm} and within 2 degrees of the stream track on the sky.
We then describe the stream distance modulus versus angle along the stream using the spline-based mixture model:

\begin{align}
P(m-M|\phi_1) = (1-f(\phi_1)) U(m-M|{\mathcal D}_{l}(\phi_1),{\mathcal D}_{h}(\phi_1)) +\nonumber\\ f(\phi_1) N^*(m-M|{\mathcal D}(\phi_1),\mathcal{S_{\mathcal D }}(\phi_1)) 
\end{align}
Here, we model the distance modulus distribution as a mixture model of the stream stars and contamination, where the stream stars are modelled as a truncated Gaussian ($N^*$) distributed around the stream track ${\mathcal D}(\phi_1)$ and with the $\phi_1$-dependent width $\mathcal{S_{\mathcal D }}(\phi_1)$, both of which are represented by cubic splines. We model the data with distance moduli between ${\mathcal D}_{l}(\phi_1)$, and ${\mathcal D}_{h}(\phi_1)$ which is a 3 mag wide interval roughly centred on the distance track from \citetalias{Koposov2019}.
The background is modelled as a uniform distribution, and the mixing fraction between the background and the stream $f(\phi_1)$ is modelled by a spline\footnote{Here, and for models defined later, since the cubic spline cannot be made bounded between zero and 1, for the mixing fraction spline we use the transformation from the real line into [0,1] range using either $F(x) = \frac{1}{2} + \frac{1}{\pi}\arctan x $ or \textit{logit} function (depending on the model) and add an appropriate Jacobian to the posterior \citepalias[see e.g.][]{Koposov2019}.}. For this and all spline-based models described later we typically use different number of spline knots for different parameters. \CHANGE{The number of knots is chosen by hand, motivated by constraining power of data and quality of the fit.} 
The model is implemented using the {\tt Stan} probabilistic programming language \citep{Carpenter2017} and is sampled using the {\tt CMDStan} and {\tt CMDstanpy} packages. The splines were implemented using the {\tt stan-splines} package \citep{stan_splines}\footnote{\url{https://github.com/segasai/stan-splines}}. The  {\tt Stan} code of this and other models used in the paper is provided in supplementary materials (see Data availability section). \footnote{Here and for all the {\tt Stan} models described in the next sections, we always run the posterior sampling for at least 10000 iterations using a number of parallel chains (between 6 and 36), and ensure that the convergence statistic $\hat{R}$ is below 1.01.}.

The results of the model are shown in Figure~\ref{fig:dist_model} and are also given in Tables~\ref{tab:dm_meas} and \ref{tab:dmsig_meas} in the Appendix.
We note that our model does not have spline  knots at $\phi_1=0$ deg, which is the point where the stream crosses the disc plane. Our model is not constrained there as we do not have likely RR Lyrae members between $-10\,{\rm deg} <\phi_1<20$ deg. This lack of constraints can be seen in the spread of red curves at $\phi_1=0$ deg in the middle panel of Figure~\ref{fig:dist_model}.
We remark that there is a noticeable increase in the inferred spread of the distance modulus in the North at $\phi_1>60$ deg which corresponds to a $\sim$ 5\% (or $\sim 4\,$kpc) distance spread that was previously noted by \citet{Sesar2013}. The spread can also be visually seen in the top panel of the Figure (especially when compared to a small spread in the South at $\phi_1\sim-50$ deg).

\subsection{Colour magnitude diagram}
\label{sec:cmd_model}

\begin{figure}
    \centering
    \includegraphics{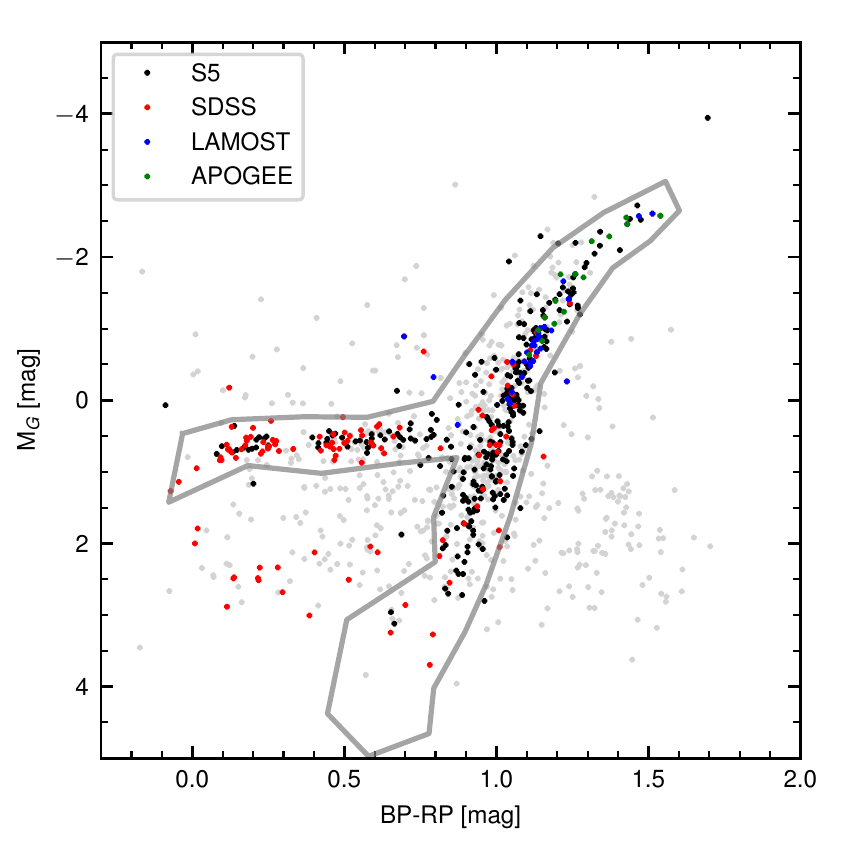}
    \caption{The colour - absolute magnitude diagram of likely OC stream members. The stars have been shifted by the best-fit distance model from Section~\ref{sec:distance_model}. Stars that fall within the proper motion, position and radial velocity selection from Figure~\ref{fig:rv_pm} are shown by points of different colours indicating the origin of the spectroscopy. The greyed points are stars outside the selection region in radial velocity space. The grey contour shows the adopted colour magnitude selection of likely stream stars.} 
    \label{fig:cmd_gaia}
\end{figure}

We can use the sample of stars identified in Figure~\ref{fig:rv_pm} to make a colour-magnitude diagram of likely spectroscopic stream members.
We select all stars falling within the spline boundaries shown in the Figure, and use those to make a {\it Gaia}-based colour-magnitude diagram (CMD). We correct the BP, RP and G magnitudes by the distance modulus model derived in the previous section as well as for extinction. Figure~\ref{fig:cmd_gaia} shows the resulting CMD with points representing stars selected to be within the position, proper motion, radial velocity bounds on Figure~\ref{fig:rv_pm}, and the colour of each point indicating the associated survey. Grey points show the stars that have positions and proper motions within the selection boundaries but are not within the radial velocity boundary of 30\, km\,s$^{-1}$. The spectroscopic members clearly form a well-defined red giant branch and a horizontal branch with an RR Lyrae gap, with only a handful of stars lying far from these branches. The grey points on other hand are very diffusely distributed on the plot, indicating that our sample is very pure. Although we can see stream members on the giant and subgiant branches, the main sequence turn-off and the main sequence are not probed due to the typical distance to OC stream members of more than 20\, kpc and spectroscopic magnitude limits.

To enable an efficient OC candidate member selection, in the next sections, we use Figure~\ref{fig:cmd_gaia} to define a polygon (shown in grey) that encompasses the OC stellar populations in CMD space. We remark that we decided to avoid using the model isochrones to perform CMD selection as we have struggled to find one amongst several isochrone libraries for the \textit{Gaia} photometry that well describes both the RGB and the horizontal branch, and has a metallicity matching our spectroscopic measurements.

\subsection{Radial velocity modelling}
\label{sec:rv_model}

\begin{figure}
    \centering
    \includegraphics{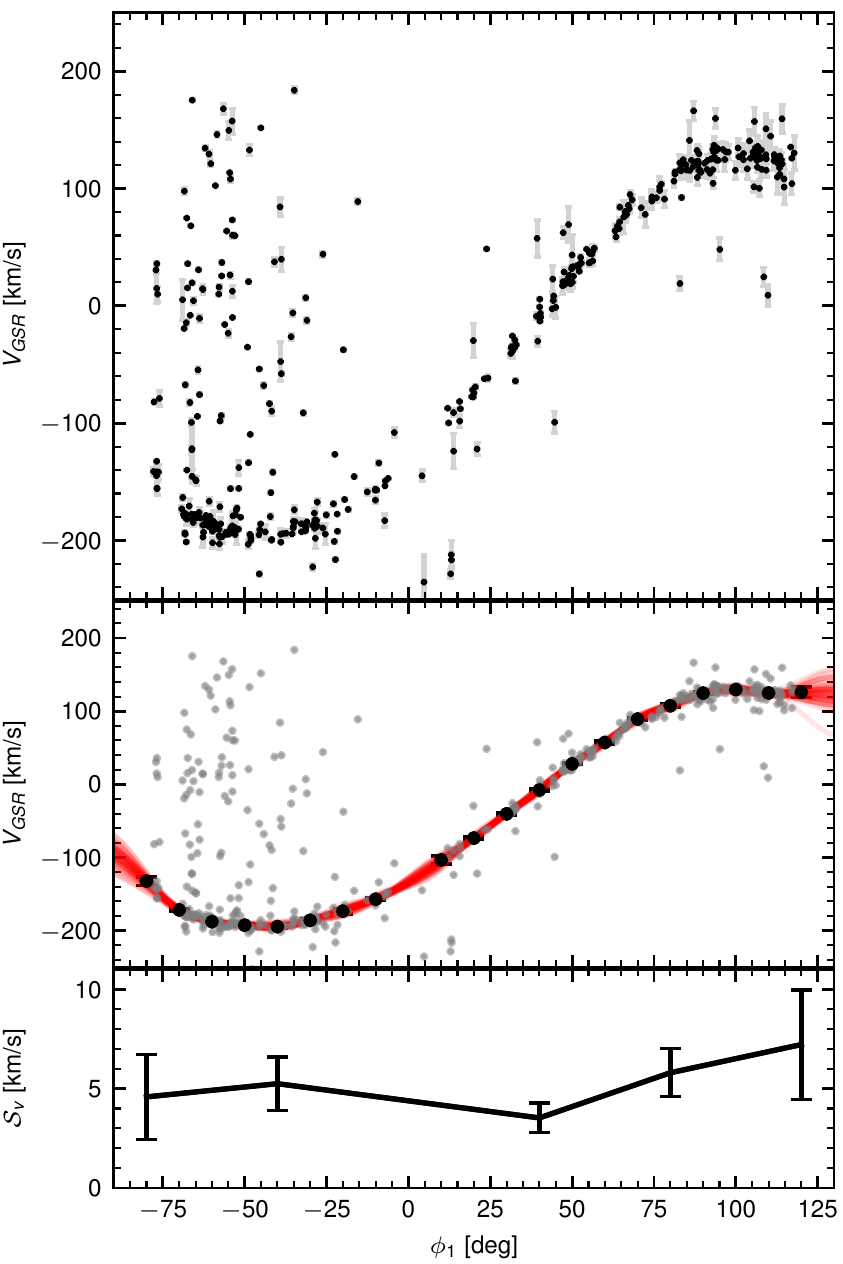}
    \caption{The measurement of radial velocities of the OC stream. {\it Top panel:} The Galactic Standard of rest velocities of the stream as a function of angle along the stream. The stars were selected by proper motions, position on the sky, CMD and spectroscopic metallicity. 
    {\it Middle panel:} The measurement of velocities at spline knots is shown by black points with error-bars. The red curves show the samples from the posterior. The grey points show the fitted sample of stars. {\it Bottom panel:} The velocity dispersion of the OC stream as measured by the spline model. The error-bars show the measurements at the spline knots. }
    \label{fig:rv_model}
\end{figure}

\begin{figure}
    \centering
    \includegraphics{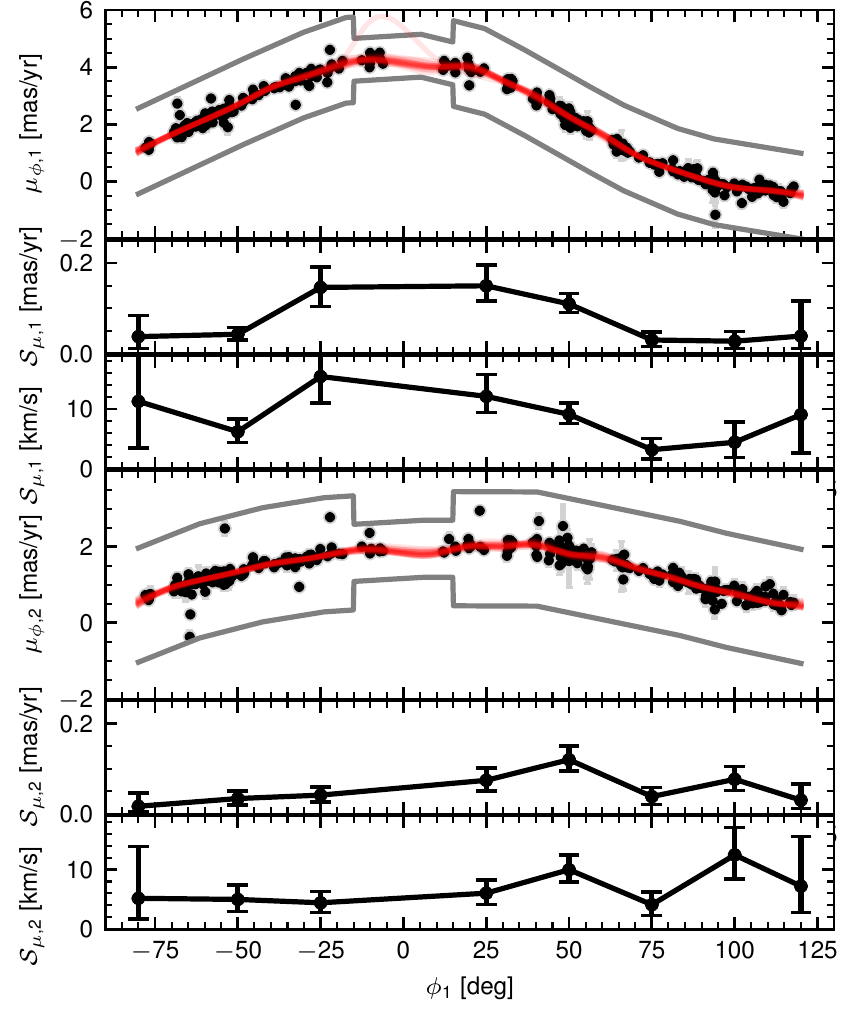}
\caption{OC stream stars proper motions and our models. {\it Top panel:} Black points show the measured (not-reflex corrected) proper motions along $\phi_1$ vs angle along the stream. The stars were selected by CMD, radial velocities, and stream track on the sky. Grey lines show the range of proper motions that were modelled. The red curves show the samples from the posterior. {\it Second panel:} The measurements of intrinsic (i.e. after taking into account measurement errors) proper motion dispersion in \muone from the model.  {\it Third panel:} The intrinsic dispersion of proper motions converted to physical velocity. {\it Bottom three panels:} Same as first three panel but for \mutwo.}
\label{fig:pm_model}
\end{figure}

In this section, we extract the radial velocity track of the OC stream. We use stars with spectroscopic data identified in Section~\ref{sec:kinematic_sample}. We apply the proper motion selection, colour-magnitude diagram selection and stream track on the sky selection. We then construct the model for the radial velocity variation along the stream. 
We model the radial velocities in the Galactic frame as a mixture between the stream, represented by a Gaussian with variable mean and velocity dispersion, and contamination (from the stellar halo) represented by a Gaussian. 
The likelihood function for the radial velocity conditional on $\phi_1$ is given below:

\begin{align}
P(v|\phi_1) = (1-f(\phi_1)) N(v|v_{bg} + d_{bg} \phi_1, {\mathcal S}_{bg}) +\nonumber \\ f(\phi_1) N(v|{\mathcal V}(\phi_1), {\mathcal S}_v(\phi_1))
\end{align}
Here, $f(\phi_1)$ is the mixing fraction, ${\mathcal V}(\phi_1)$ is the velocity of the stream, and ${\mathcal S}_v(\phi_1)$ is the velocity dispersion of the stream, all of which are represented as natural cubic splines with the values at the knots being model parameters. $v_{bg}$ is the centre of the Gaussian for the background component, $d_{bg}$ is the gradient of the mean velocity of the background with respect to $\phi_1$, and ${\mathcal S}_{bg}$ is the velocity dispersion for the background. 
This model (as other models in the paper) is implemented in the {\tt Stan} programming language and the code is provided in the supplementary materials. 
When computing the likelihoods of individual stars, we account for their radial velocity uncertainties by adding them in quadrature to the intrinsic velocity dispersions. The priors on model parameters are mostly non-informative, where the prior for mixing fraction at spline knots is $U(0,1)$,   $N(0,300)$ for the stream velocities, and $N(1.6,1.6)$ for the natural logarithm of velocity dispersion.

The results of the modelling are shown in Figure~\ref{fig:rv_model}. The figure is  structured similarly to Figure~\ref{fig:dist_model}. The top panel shows the radial velocities in the Galactic frame as a function of the angle along the stream for the modelled subset of stars. The middle panel shows the knots of the RV spline with their corresponding error-bars, as well as samples from the posterior for the ${\mathcal V}(\phi_1)$ spline curve. The bottom panel shows the inferred radial velocity dispersion with error-bars. The measurements are provided in Table~\ref{tab:rv_meas} and Table~\ref{tab:rvdisp_meas}. We observe that the radial velocity very gradually changes along 200 degrees of the stream, reaching peak values at $\phi_1\approx -60$ deg and $90$ deg. We trace the stream well down to low Galactic latitudes of $|\phi_1| \sim 10$ deg ($\phi_1=0$ deg corresponds to the point where the stream crosses the Galactic plane).
Our measurements of the velocity dispersion in the stream show a typical dispersion of $\sim$ 5\,\kms without a clear gradient along the stream. This is consistent with the measurement by \citetalias{Koposov2019} based on SDSS data alone (see their Figure 10).

\subsection{Proper motion modelling}
\label{sec:pm_model}

We now proceed to extract the stream proper motions by fitting them with a mixture model similar to the one used for radial velocities. We select stars by combining previous selections. We use stars from the spectroscopic sample defined in Section~\ref{sec:kinematic_sample} that are within 30\, km\,s$^{-1}$ of the radial velocity track from Section~\ref{sec:rv_model}, stars within the CMD mask defined in Section~\ref{sec:cmd_model}, and stars that are within 3 degrees of the fiducial stream track on the sky from Figure~\ref{fig:rv_pm}. We then model the distribution of proper motions in a narrow region around the fiducial proper motion track of the stream. The width of this region is 0.75\masyr at $|\phi_1|<20$ deg and 1.5\,\masyr elsewhere. The reason for modelling a narrow range of proper motions is that the observed \Sfive\ OC stream candidates were selected by proper motion (sometimes in a narrow region around the expected proper motion; see \citealt{Li2019} for details); thus to model the full range of the proper motion distribution we would need to take into account selection effects, which we avoid here.

The likelihood function for proper motion is similar to previous models and is a mixture of the stream and background:

\begin{align}
P(\mu|\phi_1) = (1-f(\phi_1)) U(\mu|\mu_{l}(\phi_1),\mu_{h}(\phi_1)) +\nonumber \\ f(\phi_1) N^*(\mu|{\mathcal M}(\phi_1),\mathcal{S_{\mu}}(\phi_1))
\label{eq:pm_model}
\end{align}
Here, $\mu_{l}$ and $\mu_{t}$ are the $\phi_1$-dependent boundaries of the proper motion fits. $N^*$ is the normal distribution truncated at 
$\mu_{l}$, $\mu_{t}$, and $f(\phi_1)$, ${\mathcal M}(\phi_1)$, ${\mathcal S_\mu}(\phi_1)$ are the splines for the mixing fraction between the stream and background, the mean proper motion, and the spread in proper motions respectively.
This model is fit separately to proper motions in $\phi_1$ and $\phi_2$ and takes into account the individual proper motion uncertainties, obtained by projecting the {\it Gaia} proper motion covariance matrices to the $\phi_1$, $\phi_2$ coordinate system. 

The results of the fits are shown in Figure~\ref{fig:pm_model}. The plot mimics the two bottom panels of Figure~\ref{fig:dist_model} or Figure~\ref{fig:rv_model}. The top panel of Figure~\ref{fig:pm_model} shows the \muone proper motion of individual stars that were modelled with the grey bands showing the range of proper motions fitted. Red curves show the samples from the posterior. The second panel from the top shows the inferred intrinsic spread (standard deviation) of the proper motions in $\phi_1$. The bottom two panels show the same measurements but for $\mutwo$. The central panel shows the proper motion dispersion along the stream in \kms. The measurements of proper motions and proper motion dispersions are also given in Tables \ref{tab:pm_meas} and \ref{tab:pmsig_meas}.
The extracted proper motion tracks are broadly similar to the ones presented in \citetalias{Koposov2019}, albeit significantly more accurate. We also notice that a significant intrinsic (i.e. after taking into account observational errors) proper motion spread in $\muone$\ is measured. \CHANGE{ In the region of $-25\lesssim \phi_1 \lesssim 25$ deg, } where the stream is at the closest distance to the Sun, the proper motion spread in $\phi_1$ is $\sim 0.15$\masyr: which at 15\,kpc corresponds to 10\,\kms. That is higher than the dispersion in the radial velocity of 5\,\kms (see Sec~\ref{sec:rv_model}) and is also higher than the measured velocity dispersion in the $\phi_2$ direction of $\sim 0.05-0.1$\masyr. 
The difference in measured velocity dispersions along different directions is somewhat surprising but is in fact explained by the models; we return to this in  Section~\ref{sec:energy_angm_model}.

\subsection{Stream density model}
\label{sec:density_model}

\begin{figure*}
    \centering
    \includegraphics{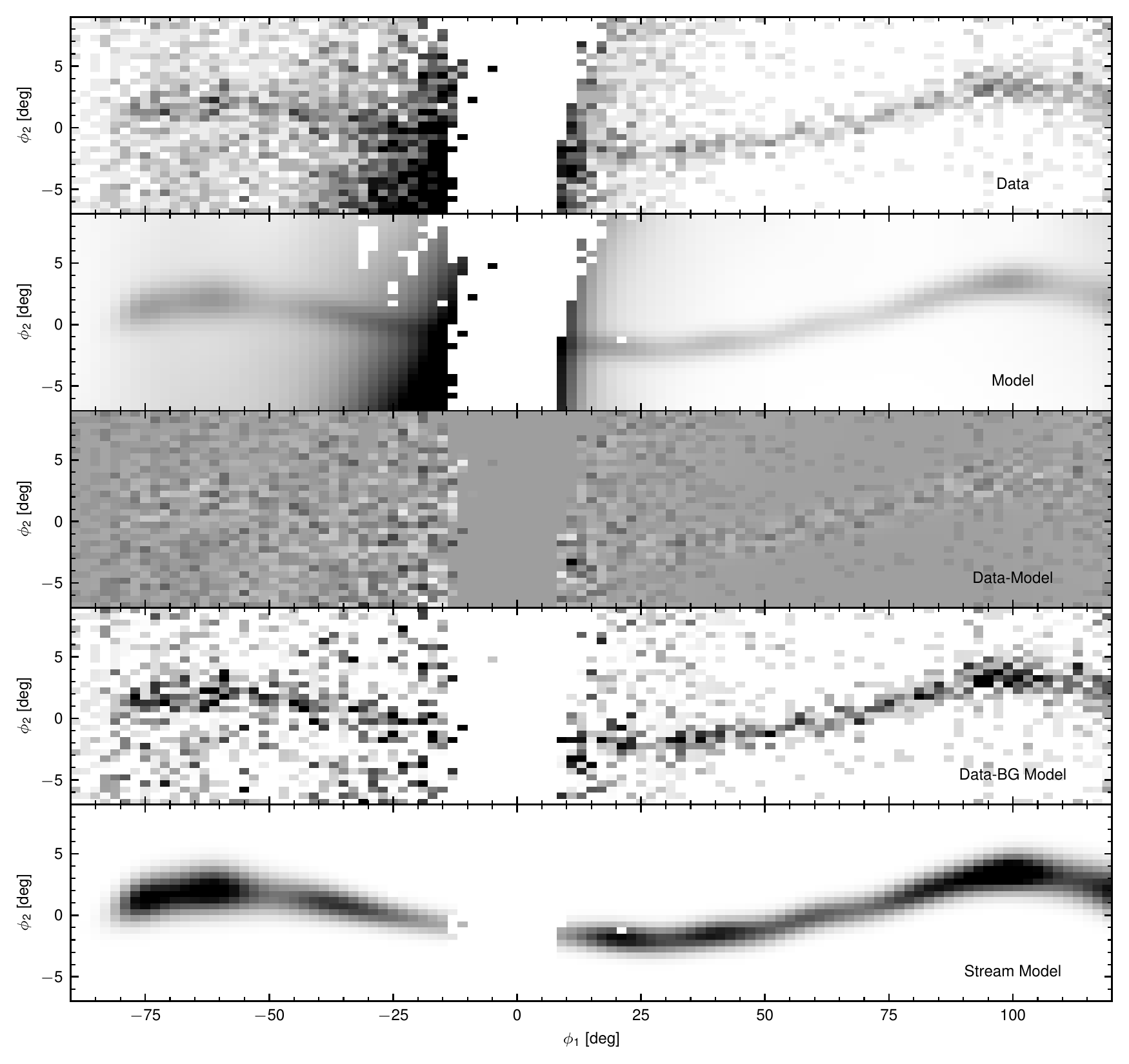}
    \caption{The spatial density model of the OC stream. The plots represent 2-D slices of the 4-D model describing the joint distribution of positions and proper motions. {\it Top panel:} The density of stars selected by colour-magnitude and having proper motions within 0.2 mas\,yr$^{-1}$ of the best-fit proper motion track. {\it Second panel:} Best fit stream density on the sky model corresponding to the same selection of stars as in the top panel. {\it Third panel:} The residuals between the data and the model. {\it Fourth panel:} The density of stars in the data with the best-fit background model subtracted. {\it Bottom panel:} The best-fit model of the stream without the background component of density.
    In all panels we only show the sky pixels that were fitted, so pixels with $|b|<5$ and $E(B-V)>0.25$ are omitted. In all the panels we use linear grey scale, and with the exception of the middle panel white colour corresponds to zero pixel values. 
    }
    \label{fig:pm_dens_map}
\end{figure*}

As a final step in tracing the OC stream, we proceed to model the stream density and track across the whole stream extent. Contrasting with the models described in previous sections, here we work with the entire {\it Gaia} EDR3 sample, rather than focusing on spectroscopic members. We select stars that have $G<19.5$, small parallaxes ($\omega<3 \sigma_{\omega}+0.117$ mas), and that are within the distance corrected CMD mask from Section~\ref{sec:cmd_model} after shifting them by the distance modulus from the distance model of Section~\ref{sec:distance_model}.
We consider stars with $-90\,\,{\rm deg} <\phi_1<120$ deg, $-7\,\,{\rm deg}<\phi_2<9$ deg and within 0.5\,\masyr\ from the proper motion track determined from spectroscopic members.
 
We note that in this section, we re-model the proper motion track of the stream. While we could rely on the proper motion track from Section~\ref{sec:pm_model} and only model the objects selected to lie close to that track, that would prevent us from properly modelling the background contamination. 
Thus we build the following model $\rho(\phi_1, \phi_2, \muone, \mutwo)$ for the stellar density in bins of $\phi_1, \phi_2, \muone, \mutwo$. This is represented by a sum of the background and stream components:
\begin{align}
\rho(\phi_1, \phi_2, \muone, \mutwo) = R_{str}(\phi_1, \phi_2, \muone, \mutwo) + \nonumber \\R_{bg}(\phi_1, \phi_2, \muone, \mutwo)
\label{eqn:bgstream_model}
\end{align}

The stream density is modelled by a Gaussian cross-section in $\phi_2$, and a Gaussian distribution in proper motions (we note that in contrast to previous models, we do not use individual proper motion error-bars in the model: i.e. we model the proper motion distribution as convolved with observational errors). We describe the mean track on the sky, the logairithm of the stream width, the logarithm of surface brightness, the proper motion track, and logarithm of its width by natural cubic splines parametrised by values at knots:

\begin{align}
    R_{str}(\phi_1, \phi_2, \muone, \mutwo) = I(\phi_1) 
    N(\phi_2|\Phi(\phi_1), \Sigma(\phi_1)) \nonumber\\
    N(\muone|{\mathcal M}_1(\phi_1) \Sigma_{\mu,1}(\phi_1))
    N(\mutwo|{\mathcal M}_2(\phi_1) \Sigma_{\mu,2}(\phi_1))
    \label{eqn:stream_dens_model}
\end{align}

Here ${\mathcal M}_1(\phi_1)$, ${\mathcal M}_2(\phi_1)$, $ \Sigma_{\mu,1}(\phi_1)$, $ \Sigma_{\mu,2}(\phi_1)$, $I(\phi_1)$, ${\Phi}(\phi_1)$, $\Sigma(\phi_1)$ are the functions for proper motion tracks ($\muone$ and $\mutwo$), width of proper motion tracks, surface brightness of the stream (per bin), track on the sky, and width of stream on the sky.

The background is modelled similarly to \citet{Erkal2017} and \citetalias{Koposov2019}, except that we add extra terms for the density dependence on proper motions:

\begin{align}
   \log R_{bg}(&\phi_1, \phi_2, \muone, \mutwo) = \nonumber\\ 
   &B_{000}(\phi_1) + 
B_{001}(\phi_1) \phi_2 +
B_{002}(\phi_1) \phi_2^2 + \nonumber\\
&\left (B_{100}(\phi_1) + 
B_{101}(\phi_1) \phi_2 +
B_{102}(\phi_1) \phi_2^2\right)\muone+\nonumber\\
&\left(B_{010}(\phi_1) + 
B_{011}(\phi_1) \phi_2 +
B_{012}(\phi_1) \phi_2^2\right)\mutwo
    \label{eqn:bg_dens_model}
\end{align}

We remark that while for simplicity we write Equations~\ref{eqn:stream_dens_model} and \ref{eqn:bg_dens_model} in terms of $\muone$, and $\mutwo$, in practice we model proper motion residuals with respect to the fiducial proper motion tracks $M_1^0(\phi_1)$, $M_2^0(\phi_1)$ from Section~\ref{sec:kinematic_sample}: $\muone^*=\muone-M_1^0(\phi_1)$, $\mutwo^*=\mutwo-M_2^0(\phi_1)$.

The model given in Equation~\ref{eqn:bgstream_model} is fit to the binned histogram of stars in  $\phi_1, \phi_2, \muone,\mutwo$. We use 2 degree and 0.5 degree bins for $\phi_1$ and $\phi_2$ respectively, and $0.033$\,\masyr\ bins for proper motions. We also excluded pixels that are very close to the Galactic plane, $-5\,\,{\rm deg}<b<5$ deg and pixels with high extinction, $E(B-V)>0.25$ from our modelling.

The priors on the model parameters are mostly non-informative and are available as a part of the {\tt Stan} model in supplementary materials. The only weakly informative priors are the priors on the proper motion offsets with respect to the fiducial proper motion track, ${\mathcal N}(0,1)$, the prior on the logarithm of widths of proper motion track, ${\mathcal N}(\log (0.1), 0.5)$, and the prior on the logarithm of stream width on the sky, ${\mathcal N}(\log (0.9), 0.5)$ (this prior has 16/84 percentiles of 0.25/2.5 deg, thus covering well the expected range of widths of the OC stream).

The resulting model has around 300 parameters and is slow to evaluate; however, as modern versions of {\tt Stan} allow parallelisation across multiple cores with the $\tt reduce\_sum$ functionality, we are still able to sample the model posterior successfully.

Figure~\ref{fig:pm_dens_map} shows the results from the spatial part of the model. Different panels show the spatial distribution of a subset of fitted stars that lie close to the proper motion track (top panel), the best fit model (second from the top panel), the residuals from the model (middle panel), the spatial distribution of stars with the best-fit background model subtracted (second from the bottom panel), and the best fit model without the background (bottom panel). The measurements of the stream track, width, and surface brightness are provided in Tables~\ref{tab:track_meas},
\ref{tab:width_meas}, and \ref{tab:density_meas}. We analyse the details of the linear density along the stream in Section~\ref{sec:lin_dens}.

\begin{figure}
    \centering
    \includegraphics{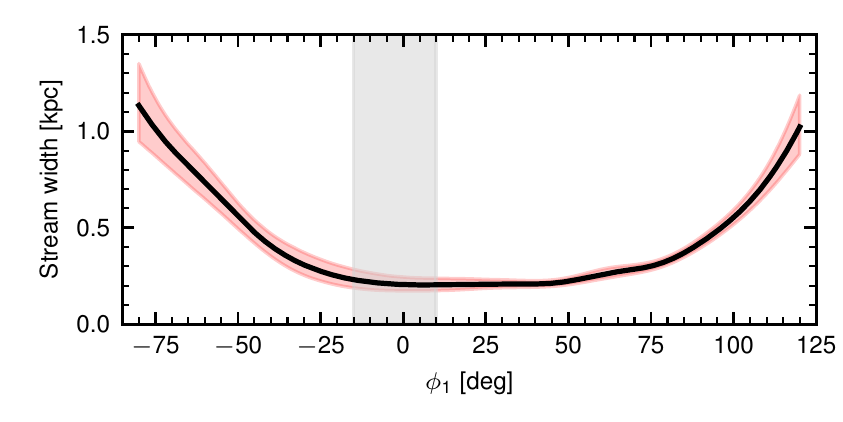}
    \caption{OC stream width in kpc as determined from the best fit model shown in Figure~\ref{fig:pm_dens_map}. The grey band shows the part of the stream hidden behind the MW disc, where the width is not directly constrained.}
    \label{fig:stream width_data}
\end{figure}

We see that the model reproduces the data extremely well. The stream is strongly curved on the sky, both in the North and the South near the stream edges. We also note significant stream width changes, with the stream being  narrowest at the point of the closest approach to the Sun (and the Galactic Centre) at $\phi_1\sim 10-20$ deg with a width of $\sim$ 0.5\, deg. The stream is noticeably wider in the North at $\phi_1\sim 100$ deg, and South at $\phi_1 \sim -60$ deg, with widths of $\gtrsim 1$ deg.   
We note that this implies quite a drastic difference in the width in physical units (200\,pc vs 1\, kpc; see Figure~\ref{fig:stream width_data}) since the edges of the stream are significantly further away at a distance of 60-80 kpc. This on-sky broadening is not expected in a spherical potential but could arise due to a flattened potential \citep[][]{Erkal:2016}. Since the width of the stream is related to the progenitor's mass, it could also suggest that the progenitor of the OC stream was significantly more massive when material near the edges of the stream was stripped than when the material near $\phi_1\sim0$ deg was stripped. 
We also note that the measurements of the stream where it crosses the plane are poorly constrained as the stream there is lost in the contamination. The spectroscopic sample is somewhat better suited to trace the stream in that region due to the availability of radial velocities.

While the model presented in this section gives a good description of stream density, we point out that \citetalias{Koposov2019} modelled the track and density using deeper DECaLS data, covering part of the Northern sky. That model showed a stream gap at $\phi_1=76$ deg. This gap is not clearly visible in our data and model because we rely on significantly shallower {\it Gaia} data and use 2 degree wide bins in $\phi_1$ which precludes us from seeing small-scale density features.

\begin{figure}
    \centering
    \includegraphics{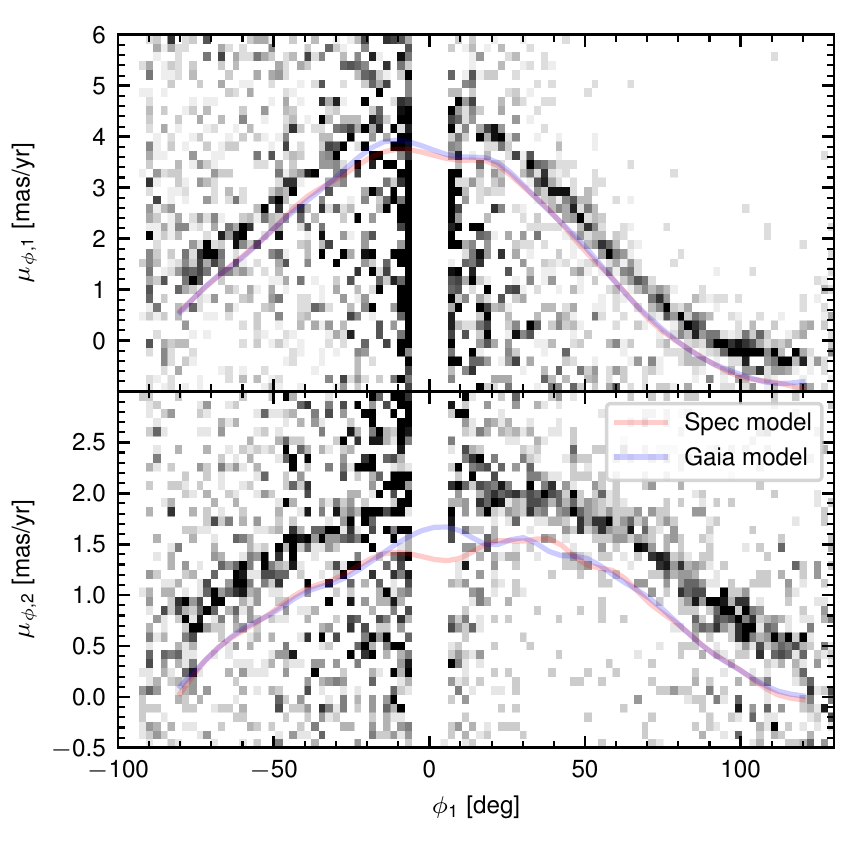}
    \caption{The background subtracted histogram of proper motion distribution of OC stream stars as a function of $\phi_1$ based on the {\it Gaia}-only data. Here, we used stars within 1 degree of the stream track on the sky from Figure~\ref{fig:pm_dens_map} to select stream stars, and stars between 2 and 4 degrees away from the stream track as background stars. The top panel shows $\muone$ vs $\phi_1$ while the bottom shows $\mutwo$. 
    The curves show the best-fit models (shifted down by 0.5 \masyr for clarity). The red curve is the model fitted to the spectroscopic dataset, and the blue curve is the model fitted to the {\it Gaia}-only data. }
    \label{fig:pm_sub_map}
\end{figure}

As a final check of the proper motion/density model, we verify that the proper motion track extracted in this section is accurate and matches the one extracted from spectroscopic data by investigating the proper motion distribution of likely members from the {\it Gaia}-only sample used in this Section. Figure~\ref{fig:pm_sub_map} shows the data-driven proper motion distribution in the stream extracted from the {\it Gaia} sample. To make the figure we take stars within 1 degree of the best-fit stream track as stream stars, and stars that are between 2 and 4 degrees away from the track as background stars. We then subtract  the appropriately scaled $\phi_1$ versus $\mu$ proper motion distribution of background stars from the distribution of likely stream stars. That should give a model-independent proper motion stream track. The stream is clearly visible on both panels as an overdensity. On Figure~\ref{fig:pm_sub_map}, we also overplot our best-fit proper motion tracks (shifted for clarity by 0.5\masyr) from the model based on the spectroscopic sample and the {\it Gaia} fit presented in this Section. We see that the models agree well with each other, giving us confidence in our analysis. The fact that the models also clearly go over the background subtracted {\it Gaia}-only proper-motion tracks also gives us confidence that the \Sfive\ spectroscopic target selection based on proper motions did not bias our measurements.

\section{Analysis of stream measurements} \label{sec:analysis}

The models fitted to the stream data described in the previous section provide us with a wealth of information on the full 6-D trajectory of the stream, stream members, as well as the density of stars along the stream. 
In this section, we look at these measurements in more detail.

\subsection{Stream density} 
\label{sec:lin_dens}

\begin{figure}
    \centering
    \includegraphics{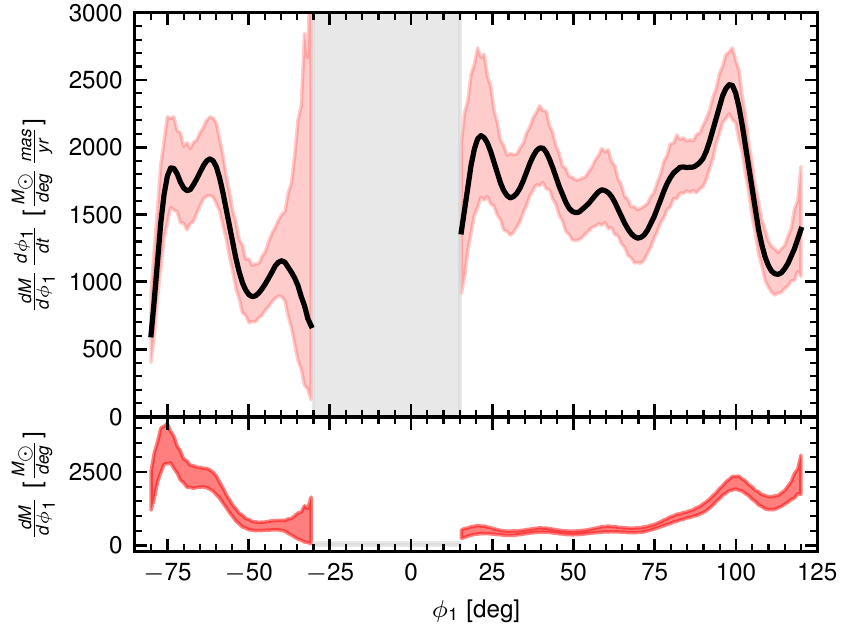}
    \caption{The density of the OC stream as a function of angle along the stream $\phi_1$. The top panel shows the density in solar masses per unit time, i.e. corrected for the distant-dependent sample incompleteness and for the accordion effect caused by the faster/slower motion in the pericentre/apocentre. The bottom panel shows the density only corrected for the sample completeness. The greyed-out region is the region where the OC stream crosses the Galactic plane, and stream densities are not well constrained.}
    \label{fig:lin_dens}
\end{figure}

The first property we look at is the stellar density of the stream. While the map presented in Figure~\ref{fig:pm_dens_map} shows that the stream seems `brighter' at some locations than others, one has to be cautious of several effects that prevent direct interpretation of this. First, the colour-magnitude selection of stars used for the modelling is not the same across the stream: the edges of the stream are significantly further from the Sun and therefore we are only probing the tip of the red giant branch.  \CHANGE{To correct for this distance dependent incompleteness we can assume that the stream consists of a single stellar population occupying an isochrone 
 with a 12.5\,Gyr age, $\feh=-2$ and Chabrier IMF \citep{chabrier2005}. We use MIST 1.2 isochrones \citep{choi2016,dotter2016} and assume that Gaia  data is complete to $G<19.5$ \citep{Boubert2020B}. Since we have measurements of the distance to the stream (see Section~\ref{sec:distance_model}) at each $\phi_1$ we can use it correct the stellar density in  number counts per degree into stellar mass per degree.}  
 
\CHANGE{On top of distance dependent incompleteness}, there are also purely kinematic effects that \CHANGE{affect the observed stream density}, as parts of the stream that are closer to us move significantly faster than the stars at the apocentre. This will lead to an apparent density decrease near the pericentre and a density increase near the apocentre (i.e. concertina/accordion effect).  If we are interested in the physical stream density changes, for example associated with the current or previous location of the progenitor along the stream, we might want to correct for this kinematic stretching/squeezing. 

To do this correction, instead of looking at the density of stars per unit angle $\frac{dN}{d \phi_1}$ along the stream, we need to look at the rate with which stream stars cross a fixed surface perpendicular to the stream, i.e. $\frac{dN}{d \phi_1} \frac{d \phi_1}{d t}$\footnote{Here the derivative of $\phi_1$ needs to be the reflex corrected proper motion along $\phi_1$. 
}. 
This essentially describes the rate of flow of stars at each point along the stream. This is more reflective of the mass loss history of the progenitor than the stellar density since the stream gets stretched (compressed) as it speeds up (slows down) along its orbit.
In Figure \ref{fig:lin_dens}, we show two measures of the density. The top panel shows the linear density of the stream corrected for the distance-dependent sample incompleteness and accordion effect, while the bottom panel shows the density corrected just for the distance-dependent incompleteness.

We notice in the lower panel of Figure \ref{fig:lin_dens} that the densities without kinematic corrections change very significantly along the stream: they are significantly  higher in the South compared to the North. In addition, the density rises by a factor of almost 10 in the North near $\phi_1\sim100$ deg, which could have been interpreted as the location of the progenitor. However, all those features largely disappear in the top panel after correcting for the accordion effect and where the stream density appears much more uniform. There is some increase in stream brightness near $\phi_1=100$ deg, and possibly at $\phi_1=-60$ deg, but overall the density is surprisingly uniform. This uniform density is also somewhat disconcerting as it does not help us constrain where the progenitor of the OC stream is or might have been.

The completeness-corrected linear density map shown in Figure~\ref{fig:lin_dens} also allows us to estimate the total stellar mass by integrating along $\phi_1$. This gives us $\sim 5.6 _{-0.3}^{+0.6}\times10^5 \Msun$ as the mass of the visible part of the stream \CHANGE{(this is  well in agreement with an estimate by  \citet{mendelsohn2022})} . This is equivalent to a luminosity of $M_V\sim -8.2 $ assuming $\feh=-2$ and age = 12.5\,Gyr. This luminosity is somewhat lower (2 $\sigma $) than the luminosity estimate from \citetalias{Koposov2019} based on RR Lyrae count.  The luminosity is however in good agreement with the mean metallicity of $\feh=-1.9$ (see the next section) based on the mass metallicity relationship \citep{kirby2013}, suggesting that we are not observing a small fraction of the actual stream, or missing a massive progenitor. The luminosity of the system together with its metallicity make the progenitor similar in properties to other satellites (classical dwarfs) of the Milky Way like Sextans and Draco \CHANGE{\citep[a similar conclusion was reached based on a stream model by][]{Hendel2018}}.

\subsection{Metallicity gradients}
\label{sec:feh_grad}

Given the lack of evidence about the progenitor location based on the stream density, one may also try to find the location of the stream progenitor through metallicity gradients. 
It is a well-known fact that, for example, the Sagittarius stream shows strong metallicity gradients \citep{Hyde2015,Hayes2020}. In fact, it is expected that streams from dwarf galaxies should show metallicity gradients since most galaxies show negative metallicity gradients with radius (i.e. more metal-poor outskirts relative to their centre: \citealt{Harbeck2001,koleva2011,okamoto2017,mercado2021}). Since the galaxy outskirts should be stripped first, we expect the metallicity in progenitor-less streams to show an inverted V shape as a function of angle along the stream. In other words, the metallicity in the stream at the location of the progenitor should be highest, and then decrease moving away from that point (see e.g. Figure 7 in \citealt{Hayes2020}). 
This motivates us to look for possible metallicity gradients in the OC stream. 
Here we rely on the catalogue selected based on proper motions, radial velocities and positions from the main \Sfive\ catalogue. We use the best-fitting splines from previous sections and also apply additional quality cuts requiring that the best S/N for each star in both blue and red arms of the spectra must be >7 (to ensure uniform quality of metallicities).

\begin{figure}
\includegraphics{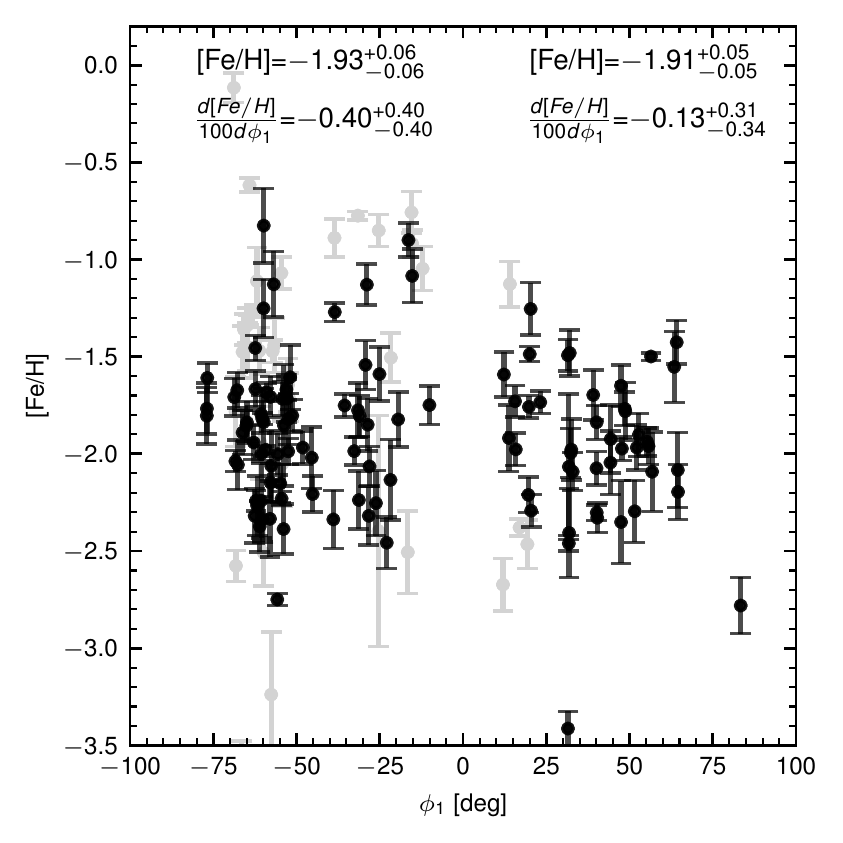}
\caption{The metallicities of likely OC stars measured by \Sfive\ vs angle along the stream. Only stars with high S/N spectra were included. The likely stream stars were selected based on their proper motion, position on the sky and radial velocity (black points). The grey points show stars selected similarly with the exception of a broader proper-motion selection and the requirement that the velocities differ from the $RV(\phi_1)$ model by more than 15 and less than 25 \kms. The metallicity and metallicity gradients shown in the top left/top right  of the figure show measurements from fits to the Southern ($\phi1<0$)/ Northern($\phi_1>0$) parts of the stream respectively. 
}
\label{fig:feh_grad_fig}
\end{figure}

We plot metallicity versus $\phi_1$ for stars in our spectroscopic sample in Figure~\ref{fig:feh_grad_fig}. The black points are very likely OC members, while grey points show the stars selected with the same selection criteria as the members sample, but with velocity deviating from the stream track of OC stream by more than 15\kms and less than 25\kms. This sample is supposed to represent possible background contamination. 
Visually, the metallicities appear roughly constant throughout the stream. To verify that, we model the whole stream, and the Northern ($\phi_1>0$ deg) and Southern parts ($\phi_1<0$ deg) separately, with a Gaussian metallicity distribution where the mean can linearly change with $\phi_1$,
 
$$\feh = \feh_0 + \frac{d \feh}{d \phi_1}(\phi_1 - \phi_{1,\mathrm{ref}})
$$
The results of the modelling for the Northern and Southern parts are shown in the Figure. 
The modelling of the whole dataset gives an $\feh_0=-1.9\pm0.04$, $\feh$ spread of $0.3$ dex, and gradient of $\frac{d\feh}{d 100\phi_1}=0.0\pm0.1$ dex/deg. All the observed gradients are consistent with zero at the $\sim$ 1-sigma level. This is unfortunate, as similarly to the density along the stream, it does not shed light on the progenitor location. There are several possible explanations for this. One is that the progenitor's location is outside the range $-80\,\,{\rm deg}<\phi_1<100$ deg and therefore we are only seeing the leading or trailing tail of the stream. Another explanation is that the stream progenitor may have had an intrinsically small metallicity gradient.

\subsection{Conservation of quantities along the stream} \label{sec:energy_angm}

\begin{figure}
    \centering
    \includegraphics{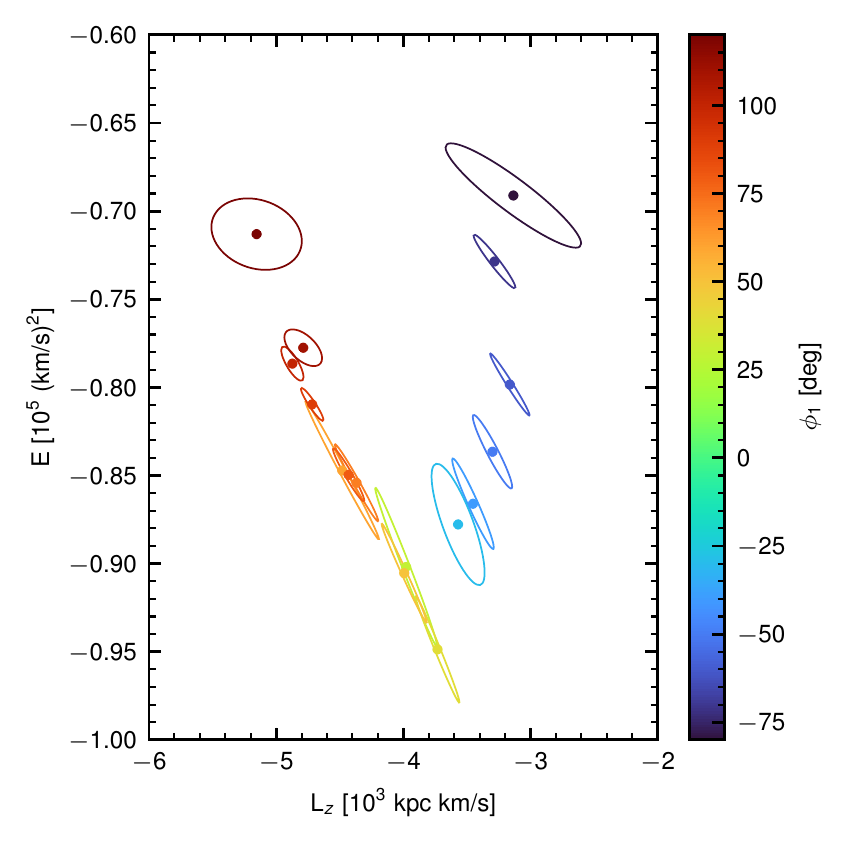}
    \caption{Changes in the energy and the z-component of the angular momentum along the OC stream. Each point shows the inferred energy and angular momentum in 10-degree long segments. The ellipses show the covariances. The colour indicates the $\phi_1$ of the segment; segments with $|\phi_1|\leq20$ deg were excluded. When computing the covariances, the full 6-D phase space uncertainties were taken into account, i.e. on the stream track on the sky, proper motion, distance and radial velocity. } 
    \label{fig:energy_ang}
\end{figure}

Since the first studies of the MW halo \citep{Helmi1999}, but especially since the arrival of {\it Gaia} data, it became useful to look at halo substructures in the space of  orbital invariants -- such as energy, angular momentum and actions \citep{Myeong2018,Koppelman2018,Helmi2020} -- as substructures are expected to be more compact there compared to pure phase space. 
For example, \citet{bonaca2021} and \citet{S5_12streams} looked at a large group of halo-structures/streams in energy and angular momentum space in order to associate streams with different progenitors. 
Also, the lumpiness of the distribution of stars in the space of conserved quantities was proposed as a way of constraining the Milky Way potential \citep{penarrubia2012,Sanderson:2015,reino2021}.

The energy ($E$) and the z-component of angular momentum ($L_z$) are generic orbital invariants as they are expected to be conserved in time-independent and axisymmetric potentials.
In the case of the OC stream, since we do possess the full 6-D phase space measurements spanning continuously more than 120 kpc, we can directly test the conservation of these quantities across a large part of the Galaxy. 
Indeed, when \citet{S5_12streams} divided the OC stream into two components ($\phi_1>0$ deg and $\phi_1<0$ deg) in $E-L_z$ space, they found that these two components did not line up. Here, we take a more detailed look at the change in energy and angular momentum along the stream.

To compute $E$ and $L_z$ as a function of $\phi_1$, we take stream tracks of the stream in 6-D: $D(\phi_1)$, $ \phi_2(\phi_1)$, $\muone(\phi_1)$, $\mutwo(\phi_1)$, $V(\phi_1)$ together with their associated uncertainties measured in Section~\ref{sec:6d_track}. We then sample from the uncertainties and evaluate $E$ and $L_z$ of stream segments on a grid of $\phi_1$. 
Here, we use the Milky Way potential from \citet{McMillan2017}. Figure~\ref{fig:energy_ang} shows the resulting energy and $z$-component of angular momentum measurements at positions along the stream spaced by 10 degrees. The ellipse shapes show uncertainty/covariance in $E$ and $L_z$ (corresponding to 1-$\sigma$ errors). The colour of the ellipse indicates the $\phi_1$ angle of the corresponding stream segment. We note that the uncertainties in $E$ and $L_z$ are often strongly correlated. 

Before examining the figure, we also remark that the stream stars will not necessarily occupy a small region in energy-angular momentum space, as stars in tidal streams naturally produce a distinctly elongated  (bow-tie shaped) E/L distribution \citep{Yoon2011,Gibbons2014}. However, in the case of the OC stream, we expect the intrinsic energy spread in the stream to be of the order of $v_{\rm peri}\sigma_v\approx 0.02\times10^5 ({\rm km}\ {\rm s}^{-1})^2$ (where $v_{\rm peri}$ is the velocity at the pericentre, $\sim 400$\kms, and $\sigma_v$ is the velocity dispersion of the dwarf, $\sim 5$\kms). Figure~\ref{fig:energy_ang} instead shows a very large variation in both $E$ and $L_z$. In particular, $L_z$ changes almost by a factor of two across the stream. The energy variation from the middle of the stream toward the edges is $0.3\times10^5 ({\rm km}\ {\rm s}^{-1})^2$. These large changes are clearly not expected from a normal disruption process, but must be related to the interaction of the stream with the Magellanic Clouds that causes a spreading of stream stars in energy and angular momentum space. We look at the model prediction of that distribution in Section~\ref{sec:energy_angm_model}.

\subsection{Stream direction of motion}

\begin{figure}
    \centering
    \includegraphics{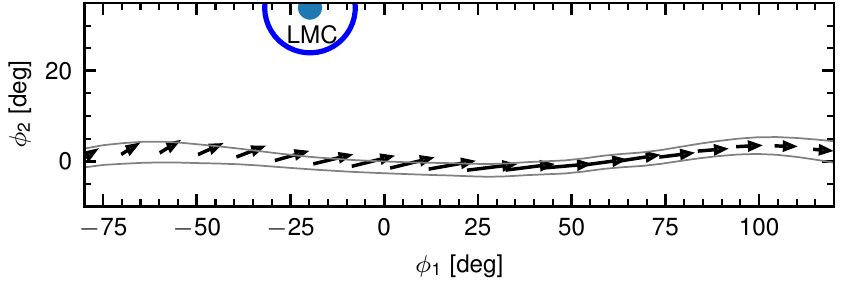}
    \caption{The OC stream proper motions corrected for solar reflex motion compared to the stream track. 
    The grey lines delineate the OC stream as measured in the paper, while the arrows show the evaluation of the best-fit proper-motion splines on a grid of $\phi_1$. We also show the location of the LMC to illustrate that the proper motion deviation is towards it.}
    \label{fig:arrows}
\end{figure}

\begin{figure}
    \centering
    \includegraphics{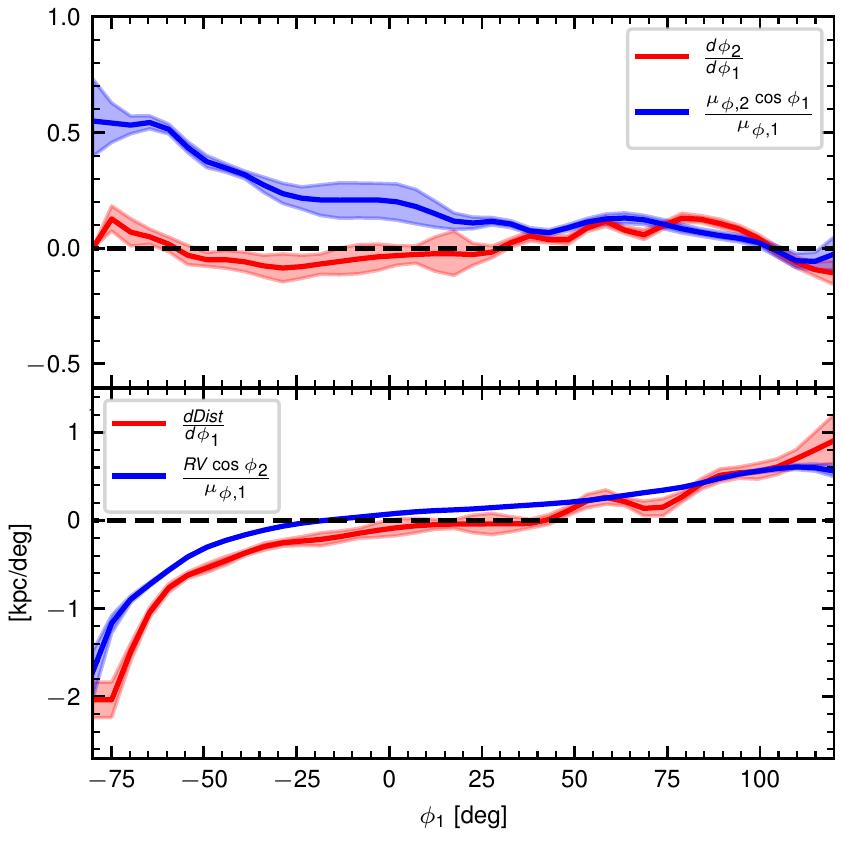}
    \caption{Comparison of stream track orientation and the motion of stars. {\it Top panel:} The stream track direction on the sky $\frac{d \phi_2}{d \phi_1}$ is shown in red, while the direction of motion of stars $\frac{\mutwo \cos \phi_2}{\muone}$ is shown in blue. The shaded areas show the 1-sigma uncertainties from sampling the proper motion, track and distance splines. {\it Bottom panel: } Comparison of the stream track and stellar velocities in the radial direction. The red line shows the distance gradient of the stream track along $\phi_1$, while the blue line shows the ratio of radial velocity to the proper motion in $\phi_1$ direction. Shaded bands show the 1-sigma intervals similarly to the top panel. Both panels show a clear misalignment of the stellar motion and stream direction, meaning that the stars are not moving along the stream, confirming the effects of the perturbation by the LMC.}
    \label{fig:directions}
\end{figure}

In \citetalias{Koposov2019} and \citetalias{Erkal2019}, it was found that the stars in the OC stream are not actually moving along the stream track. This is caused by the gravitational pull of the Large Magellanic Cloud, an effect that has since been detected in several other streams \citep{Shipp+2021}. Here, equipped with the full 6-D track, we can constrain this behaviour much more effectively. To do so, we use the proper motion and distance splines inferred in previous sections. Figure~\ref{fig:arrows} shows the solar reflex corrected proper motion vectors along the OC stream together  with the stream track. 
We notice that, as seen previously, the velocity vectors are aligned with the stream track for the Northern part of the stream ($\phi_1>0$ deg), while in the South the stream stars are moving out of the stream plane, pulled towards the LMC. This motion across the stream is also the reason for the bent-up shape seen in Figure~\ref{fig:pm_dens_map}, as the Southern part of the stream is moving up pulled by the LMC.

While Figure~\ref{fig:arrows} provides an illustration of the misalignment of the stars' motions with that of the stream, we can also directly compare the direction of motion $\frac{\mutwo \cos \phi_2}{\muone}$ of the stars versus stream track direction $\frac{d \phi_2}{d \phi_1}$.  This is shown in the top panel of Figure~\ref{fig:directions} \citepalias[see Fig. 1 of][for comparison]{Erkal2019}. Here, we see that the stream track directions (shown in red) are aligned with the stellar direction of motion (in blue) for the Northern part of the stream at $\phi_1>30$ deg, with deviations starting to show at lower $\phi_1$. On the bottom panel of Figure~\ref{fig:directions}, we show a similar analysis applied to radial velocity measurements, i.e. the comparison between the distance gradient in the stream $\frac{d D}{d \phi_1}$ and the radial velocity to proper motion ratio $\frac{V_r \cos \phi_2}{\mu_1}$. 
If the stream stars were moving along the stream track, then those two quantities would be identical; however, the figure similarly shows that for $\phi_1<30$ deg there is a misalignment between the stars' motions and the stream track, with the stream distances decreasing faster than expected from radial velocities alone.

\section{Dynamical modelling} \label{sec:modelling}

\subsection{Stream generation technique and potential models}
\label{sec:modelling_intro}
In order to fit the exquisite data set presented in Section~\ref{sec:6d_track} with a dynamical model, we use a similar technique to \citetalias{Erkal2019}. In particular, we use the modified Lagrange Cloud stripping technique from \cite{Gibbons2014} which has been modified to include the influence of the LMC in \citetalias{Erkal2019}. Given the significantly improved data set in this work compared to \citetalias{Koposov2019} which \citetalias{Erkal2019} fit, we use a Milky Way and LMC potential with more flexibility. For the \CHANGE{baryonic components of the} Milky Way potential, we assume a Hernquist profile \citep{Hernquist1990} for the bulge with a mass of $5\times10^9 \Msun$ and a scale radius of 500 pc, a Miyamoto-Nagai profile \citep{Miyamoto-Nagai1975} for the disc with a mass of $6.8\times10^{10} \Msun$, a scale radius of $3$ kpc, and a scale height of $0.28$ kpc \CHANGE{based on the \texttt{MWPotential2014} model in \cite{galpy}}. \CHANGE{We model the Milky Way's dark matter halo as} an axisymmetric, generalized NFW \citep{nfw1996} with
\begin{align}
\rho(m') = \frac{\rho_0}{\Big( \frac{m'}{r_s} \Big)^\gamma \Big( 1 + \frac{m'}{r_s}\Big)^{\beta-\gamma}} \exp\Bigg( -\Big(\frac{m'}{r_{\rm cut}}\Big)^2 \Bigg) \,\,,
\label{eqn:gennfw}
\end{align}
where 
\begin{align}
\rho_0 = \frac{M_{\rm NFW}}{4 \pi r_s^3} \frac{1}{\ln(1 + c_{\rm NFW})-\frac{c_{\rm NFW}}{1+c_{\rm NFW}}},
\end{align}
$M_{\rm NFW}$ is the halo mass, $c_{\rm NFW}$ is fixed to 15 to avoid degeneracy with the halo mass, $r_s$ is the scale radius, $m' = \sqrt{x'^2 + y'^2 + z'^2/q^2}$ is the flattened radius, $q$ is the flattening of the halo, $\gamma$ and $\beta$ are the inner and outer slope of the profile respectively, and $r_{\rm cut}$ is the cutoff radius of the profile. The primed coordinates ($\mathbf{x'}$) used here are rotated with respect to the Milky Way Cartesian coordinates ($\mathbf{x}$), 

\begin{align}
\mathbf{x'} = R \mathbf{x},
\label{eqn:rotmat}
\end{align}
so that the axisymmetric halo can be flattened in an arbitrary direction. To parameterise this rotation, we specify the direction in which the halo is flattened/stretched in: $(x_{\rm NFW},y_{\rm NFW},z_{\rm NFW})$. This is done to avoid any period boundaries that would occur if we specified the flattening direction in terms of polar angles. We convert these coordinates into the polar angles with

\begin{align}
\theta &= \tan^{-1}\left( \frac{y_{\rm NFW}}{x_{\rm NFW}}\right), \\
\phi &= \cos^{-1}\left( \frac{z_{\rm NFW}}{\sqrt{x_{\rm NFW}^2+y_{\rm NFW}^2+z_{\rm NFW}^2}} \right).
\end{align}
With these angles, the matrix $R$ from Equation~\ref{eqn:rotmat} is specified as
\begin{align}
R = \begin{pmatrix} \cos\phi\cos\theta & \cos\phi\sin\theta & -\sin\phi\\ -\sin\theta & \cos\theta & 0 \\ \sin\phi\cos\theta & \sin\phi \sin\theta & \cos\phi \end{pmatrix}.
\end{align}

We note that an NFW profile would have $\gamma=1$, $\beta=3$, and an infinite cutoff radius. In this work, we set $r_{\rm cut}=500$ kpc for the MW potential which is significantly beyond the orbital extent of the OC stream. For computational efficiency, we evaluate the forcefield of this flattened halo by rotating into the coordinates where the halo is flattened in the $z$ direction (i.e. the $\mathbf{x}'$ coordinates) and using \textsc{galpot} \citep{DehnenBinney1998}. Altogether, the Milky Way halo potential has 8 free parameters.

For the LMC, we assume a truncated NFW profile:

\[
    M_{\rm LMC}(<r) = \\
\begin{cases}
    M_{\rm LMC} \frac{\log( 1 + \frac{r}{r_{s,{\rm L}}}) - \frac{r}{r+r_{s,{\rm L}}} }{\log( 1 + \frac{r_{\rm max, {\rm L}}}{r_s})- \frac{r_{\rm max,\,L}}{r_{\rm max,L}+r_{s,{\rm L}}}} , & r\leq r_{\rm max, \, L}\\
    M_{\rm LMC} \, ,              & r > r_{\rm max, \, L}
\end{cases}
\]
where $M_{\rm LMC}$ is the total mass of the LMC, $r_{s,{\rm L}}$ is the scale radius of the LMC, and $r_{\rm max, L}$ is the truncation radius of its NFW halo. We include this truncation so that we can explore the extent of the LMC's dark matter halo. The LMC's present-day proper motions, radial velocity, and distance are also free parameters with priors given by observations of these quantities \citep[][respectively]{Kallivayalil+2013,vandermarel+2002,Pietrzynski+2019}. Thus, the LMC model has 7 free parameters. 

For the progenitor, we follow \citetalias{Erkal2019} and place the progenitor's present-day location at $\phi_1 = 6.34$ deg. \citetalias{Erkal2019} found that the inferred Milky Way mass did not depend on the progenitor's location so we do not vary the progenitor's location in our fits. The other parameters describing the present-day phase space coordinates are left as free parameters: $\mu_{\alpha,\, {\rm prog}}^*,\mu_{\delta, \,{\rm prog}},v_{{\rm los}, {\rm prog}},d_{\rm prog},\phi_{2,\,{\rm prog}}$. There are thus 5 free parameters needed to describe the progenitor. All of our parameters, and their respective priors, are given in Table~\ref{tab:priors}.

\begin{table}
\begin{centering}
\begin{tabular}{|c|c|c|}
\hline
Parameter & Prior & Range \\
\hline OC progenitor & & \\
\hline
$\mu_{\alpha,\, \rm prog}^*$ & Uniform & ($-6,0$) \masyr \\
$\mu_{\delta,\, \rm prog}$ & Uniform & $(0,6)$ \masyr \\
$v_{r,\, \rm prog}$ & Uniform & ($-250,250$) \kms \\
$d_{\rm prog}$ & Uniform & (0,) kpc \\
$\phi_{\rm 2,\,prog}$ & Uniform & ($-10$ deg,$10$ deg) \\
\hline Milky Way & & \\
\hline $M_{\rm NFW}$ & Log Uniform & $(1,30)\times10^{11} \,\Msun$ \\
$ r_s$ &  Uniform & $(0,) $ kpc \\
$q_{\rm NFW}$ & Uniform & ($0,1$) or ($1,2$) \\
$x_{\rm NFW}$ & Normal & $0\pm1$ \\
$y_{\rm NFW}$ & Normal & $0\pm1$ \\
$z_{\rm NFW}$ & Normal & $0\pm1$ \\
$\gamma$ & Uniform & ($0,2$) \\
$\beta$ & Uniform & ($2,4$) \\
\hline LMC & & \\
\hline $M_{\rm LMC}$ & Log-Uniform & $(10^8,4\times10^{11})\, \,\Msun$ \\
$r_{s,\,\rm L}$ & Uniform & (1,) kpc\\
$r_{\rm max,\, L}$ & Uniform & (8.7,) kpc \\
$\mu_{\alpha,\, \rm LMC}^*$ & Normal & $1.91\pm0.02$ \masyr \\
$\mu_{\delta,\, \rm LMC}$ & Normal & $0.229\pm0.047$ \masyr \\
$v_{r,\, \rm LMC}$ & Normal & $262.2\pm3.4$ \kms \\
$d_{\rm LMC}$ & Normal & $49.59 \pm 0.547$ kpc \\
\hline Noise nuisance parameters & & \\
\hline $\sigma_{\phi_2}$ & Uniform & ($0^\circ,1^\circ$) \\
$\sigma_{\rm DM}$ & Uniform & ($0,0.25$) \\
$\sigma_{\muone^*}$ & Uniform & ($0,0.25$ \masyr) \\
$\sigma_{\mutwo}$ & Uniform & ($0,0.25$ \masyr) \\
$\sigma_{v_r}$ & Uniform & ($0,30$ \kms) \\
\hline
\end{tabular}
\caption{Parameters and corresponding priors for our OC model. We break the priors down by object. The description of each parameter is provided in Sections \ref{sec:modelling_intro} and \ref{sec:stream_gen}}
\label{tab:priors}
\end{centering}
\end{table}

\subsection{Stream generation} \label{sec:stream_gen}

Given the progenitor, Milky Way, and LMC parameters, we generate a stellar stream using the modified Lagrange Cloud stripping (mLCs) method of \cite{Gibbons2014} which \citetalias{Erkal2019} generalized to include an LMC. We include the dynamical friction from the Milky Way on the LMC using the results of \cite{Jethwa+2016}. We also account for the motion of the Milky Way in response to the LMC \citep[e.g.][]{Gomez+2015} which is known to affect many streams \citep[e.g.][]{Erkal2019,Vasiliev+2021,Ji+2021}. This is done by modelling the Milky Way and LMC as individual particles sourcing their respective potentials. 

We model the stream progenitor as a Plummer sphere \citep{Plummer} with an initial mass of $2.67\times10^{7} \Msun$ and a fixed scale radius of 1 kpc. Since we do not see the  progenitor in the data, the mass of the progenitor is modelled to change linearly from the initial mass to zero at the present day. These parameters were selected to approximately reproduce the width of the stream on the sky. 
Before generating the stream in the simulation, the progenitor is first rewound for 4\,Gyr in the presence of the Milky Way and LMC, and then the system is evolved forwards. Tracer particles are released from the progenitor at the Lagrange points using the mLCs method. 


\subsection{Likelihood and MCMC exploration} \label{sec:mcmc}

In order to compare each stream model with the data (i.e. sky track, distance modulus, proper motions, and radial velocity), we compute the likelihood of the data given the model. As is the case for our data, we use values at spline knots and assume the measurements are uncorrelated. For each observable, we define the likelihood at each knot location as
\[ \log \mathcal L_i = -\frac{1}{2}\log\Big(2\pi \left(\sigma_{i,\rm data} ^2+\sigma_{i,\rm sim}^2\right)\Big) - \frac{1}{2}\frac{\left(m_{i,\rm data} - m_{i,\rm sim}\right)^2}{\sigma_{i,\rm data}^2+\sigma_{i,\rm sim}^2}, \]
where $m_{i,\rm data}$ is the measurement at the $i^{\rm th}$ knot location, $\sigma_{i,\rm data}$ is the uncertainty on this value, and $m_{i,\rm sim}$ is the value in the simulation. To infer the simulated value, we fit a straight line to the simulated stream particles within 5 deg of the knot location to infer the mean ($m_{i,\rm sim}$) and uncertainty on the mean ($\sigma_{i,\rm sim}$). We sum the log-likelihood terms over all knot locations and observables to obtain the full likelihood function of the data given the simulation. 

We note that our model has shot noise since it is represented with a finite number of particles. To ensure that the likelihood function is sufficiently smooth, we require that the shot noise of our stream model is small compared to the observed uncertainties ($\sigma_{i,\rm sim} < \sigma_{i,\rm data}/5$). In order to achieve this, we strip 120,000 particles per pericenter. This number was chosen based on initial testing to meet our uncertainty requirement. We have also checked with the posterior chains and found that it is met $\sim 99\%$ of the time.

Based on initial testing, we found that our best-fit models were not able to fully explain the data, and the typical $\chi^2$ per degree of freedom of our best-fit models is $\sim 2.5$. This can be caused by either inadequacy of our model, or possibly underestimated uncertainties (or ignored correlations in measurements). 
 
To account for this potential problem, we introduced an extra noise nuisance parameter ($\sigma$) for each observable, which is added in quadrature with the errors to give a likelihood of 
\begin{align} 
\log \mathcal L_i = -\frac{1}{2}\log\Big(2\pi \left(\sigma_{i,\rm data} ^2+\sigma_{i,\rm sim}^2 + \sigma^2\right)\Big) \nonumber\\ - \frac{1}{2}\frac{\left(m_{i,\rm data} - m_{i,\rm sim}\right)^2}{\sigma_{i,\rm data}^2+\sigma_{i,\rm sim}^2 + \sigma^2}. 
\end{align}
We add one nuisance parameter for each type of observable, giving us five additional parameters. The maximum of the prior on each parameter is chosen to be slightly bigger than the largest residual between the model and data based on initial testing. 
The final number of parameters is 25 in the case of the oblate and prolate halo, and 21 parameters for a spherical halo.

To sample the posterior obtained by combining the priors shown in Table~\ref{tab:priors} and the likelihood function described above, we use the Markov Chain Monte Carlo (MCMC) sampler \textsc{emcee} \citep{emcee}. We use 250 walkers for 7,500 steps, with a burn-in of 3,750 steps. The post-burn-in posterior chains are available through  Zenodo (see Data Availability section).

\begin{figure*}
    \centering
    \includegraphics[width=\textwidth]{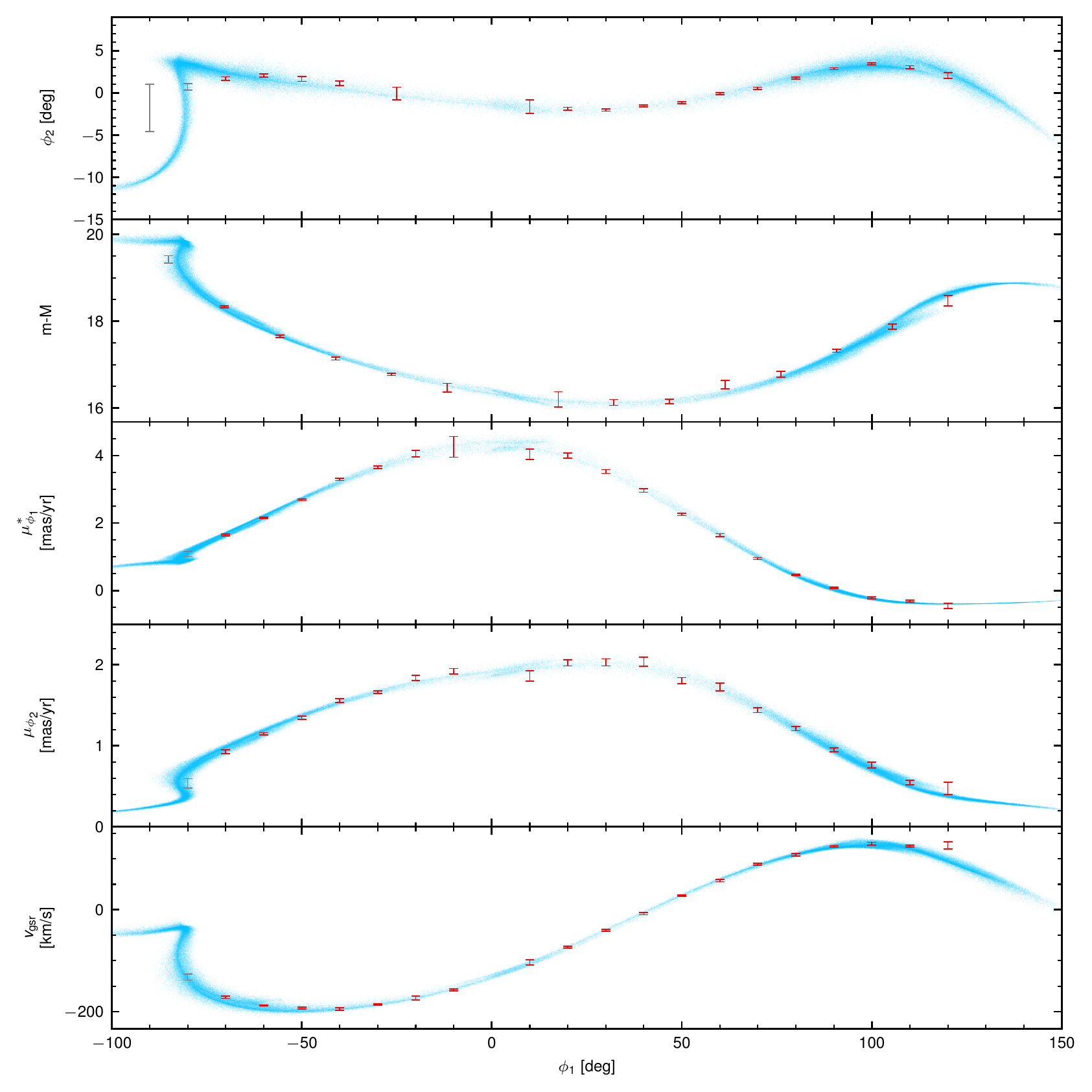}
    \caption{Best-fit OC model compared to the data. The stream measurements are shown with red/grey error-bars and the stream model is shown with blue points. The grey points are not included in the modelling since these occur in a region where the model predicts a kink on the sky, i.e. a large range of $\phi_2$. From top to bottom, the panels show the stream on the sky, in distance modulus, proper motions, and radial velocity.
    The model is a good representation of the data, except for the radial velocity in the right-most bin. A movie showing the evolution of these observables is available \href{https://youtu.be/UIkk1rSuK7M}{here}.}  
    \label{fig:OC_best}
\end{figure*}

The key parameter measurements from the posterior samples for the three MW halo models we consider (spherical, oblate, prolate) are given in Table~\ref{tab:results}. The oblate halo model has the highest log-likelihood and thus represents our best model for the stream, as well as the Milky Way and LMC. In the following discussion, we will take the oblate halo as our fiducial model. For completeness, we also give the nuisance parameters for these three fits in Table~\ref{tab:nuisance}. The best-fit stream model is compared with the data in Figure~\ref{fig:OC_best}. 
\renewcommand{\arraystretch}{1.2}
\begin{table}
\begin{centering}
\begin{tabular}{|c|c|c|c|}
\hline
Parameter & Spherical & Prolate & Oblate \\
\hline Milky Way  & & & \\
\hline $M_{\rm NFW}$ ($10^{11} \Msun$) & $8.41^{+5.11}_{-3.74}$ & $13.78^{+2.88}_{-4.26}$ & $11.22^{+4.77}_{-3.73}$ \\
$r_s$ (kpc) & $13.19^{+6.66}_{-4.96}$ & $10.63^{+3.65}_{-2.16}$ & $16.10^{+5.97}_{-5.17}$ \\
$q_{\rm NFW}$ & 1 & $1.40^{+0.12}_{-0.10}$ & $0.55^{+0.10}_{-0.08}$ \\ 
$l_{\rm NFW}$ & - & $101.49^{+9.48}_{-12.67}$ & $16.19^{+4.83}_{-8.27}$ \\
$b_{\rm NFW}$ & - & $35.99^{+11.26}_{-9.39}$ & $32.98^{+5.19}_{-6.81}$ \\
$\gamma$ & $1.11^{+0.37}_{-0.45}$ & $0.66^{+0.35}_{-0.38}$ & $1.27^{+0.31}_{-0.39}$ \\
$\beta$ & $3.19^{+0.36}_{-0.46}$ & $3.82^{+0.13}_{-0.23}$ & $2.97^{+0.32}_{-0.41}$ \\
$M_{\rm MW}$ (32.4 kpc) & $2.97^{+0.15}_{-0.14}$ & $3.06^{+0.10}_{-0.11}$ & $2.85^{+0.09}_{-0.09}$ \\
$M_{\rm MW}$ (50 kpc) & $3.90^{+0.22}_{-0.23}$ & $3.81^{+0.13}_{-0.13}$ & $3.77^{+0.26}_{-0.19}$ \\
$M_{\rm vir, \, MW}$ ($10^{11} \Msun$) & $7.19^{+3.56}_{-1.12}$ & $5.68^{+0.40}_{-0.26}$ & $7.74^{+4.14}_{-1.48}$\\
$r_{\rm vir, \, MW}$ (kpc) & $236^{+34}_{-13}$ & $219^{+5}_{-3}$ & $242^{+37}_{-17}$\\
\hline LMC & & & \\
\hline $M_{\rm LMC}$ ($10^{10} \Msun$) & $10.28^{+3.51}_{-3.47}$ & $11.70^{+2.79}_{-2.60}$ & $12.85^{+2.76}_{-2.34}$ \\
$r_{s,\,\rm L}$ (kpc) & $2.71^{+2.59}_{-1.18}$ & $3.84^{+1.86}_{-1.56}$ & $2.48^{+1.76}_{-1.06}$ \\
$r_{\rm max,\, L}$ (kpc) & $134.83^{+44.31}_{-82.46}$ & $152.40^{+33.10}_{-44.31}$ & $156.22^{+32.56}_{-41.26}$ \\
$M_{\rm LMC}$(32.8 kpc) ($10^{10} \Msun$) & $5.75^{+0.89}_{-0.71}$ & $5.95^{+0.63}_{-0.64}$ & $7.02^{+0.99}_{-0.86}$ \\
\hline $\Delta \log \mathcal{L}$ & $-23.2$ & $-9.8$ & 0 \\
\hline
\end{tabular}
\caption{Key parameter measurements from the posterior samples. In addition to the model parameters, we also provide some derived quantities such as masses within fixed apertures and the virial mass of the MW. We show parameters for three models of the MW halo: oblate, prolate, and spherical. Note that the values correspond to the median with uncertainties from the 84.1 and 15.9 percentiles.}
\label{tab:results}
\end{centering}
\end{table}
\renewcommand{\arraystretch}{1.}

\subsection{Milky Way potential measurement} \label{sec:mw_pot}

In this section, we look at the inference of the Milky Way potential from the posterior samples presented in Section~\ref{sec:mcmc}. 

\begin{figure}
    \centering
    \includegraphics{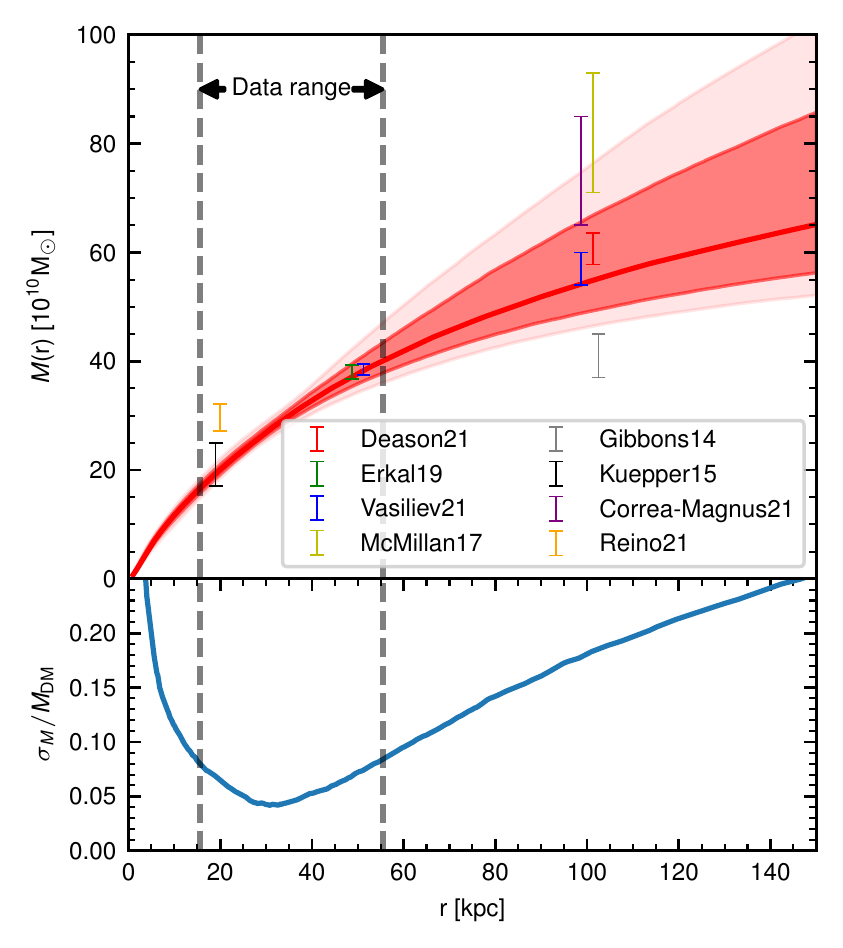}
    \caption{Constraints on the enclosed mass profile of the Milky Way. \textit{Top panel} shows the enclosed mass (including all the Galactic components) as a function of radius. The red curve shows the median enclosed value, while the dark and light red-shaded regions show the 1 and 2$\sigma$ confidence intervals. The vertical dashed lines show the radial extent of our data set. The observations are Deason21 \citep{Deason:21}, Erkal19 \citepalias{Erkal2019}, Vasiliev21 \citep{Vasiliev+2021}, McMillan17 \citep{McMillan2017}, Gibbons14 \citep{Gibbons2014}, Kupper15 \citep{Kuepper:2015}, CorreaMagnus22 \citep{CorreaMagnus21} and Reino21 \citep{reino2021}. \textit{Bottom panel} shows the fractional uncertainty on the enclosed dark matter mass, which demonstrates that the mass is best measured in the region where the OC stream is observed and that the uncertainties are substantially larger outside of this region. The most precise measurement of the Milky Way's enclosed mass is at 32.4 kpc, with $M_{\rm MW}({\rm r}<32.4{\rm kpc})=(2.85\pm0.1)\times 10^{11} \Msun$, which corresponds to a fractional uncertainty of 4.2\% on the enclosed dark matter mass. \CHANGE{The constraints on the mass profile shown on the panel of the figure are provided in digital form in supplementary materials.}}
    \label{fig:mw_mass}
\end{figure}

In order to highlight our mass constraints, we first compute the enclosed mass profile of the Milky Way. This is done by taking individual posterior samples, computing the resulting enclosed mass profile, and then computing the 15.9, 50, and 84.1 percentiles of the enclosed mass profile at each radius. We note that due to the flattening of the mass distribution, we numerically integrate the density profile in order to compute the mass profile. The resulting enclosed mass profile of the Milky Way is shown in the top panel of Figure~\ref{fig:mw_mass}. Note that this enclosed mass includes the bulge, disc, and dark matter halo. For comparison, we show several other recent mass measurements. Our enclosed mass profile is a good match to the other measurements, especially those which fit stream data with dynamical stream models such as \cite{Kuepper:2015,Erkal2019,Vasiliev+2021}. We also highlight the Galactocentric distance range spanned by our stream data. 

Next, we consider the fractional uncertainty of the enclosed dark matter mass in order to determine the region in which we best measure the stream data. This is shown in the bottom panel of Figure~\ref{fig:mw_mass}. Such a fractional uncertainty was originally proposed by \cite{Bonaca+2018}. Using a suite of mock streams, they showed that given a suitably flexible potential, the fractional uncertainty in the acceleration (or equivalently the mass enclosed) should be minimized at the present-day location of the stream. For the first time, we are able to show the same result with real data. In particular, we best measure the Milky Way's dark matter mass at a radius of 32.4\,kpc, with a mass of $M_\mathrm{DM}=(2.18\pm0.1)\times10^{11}\Msun$, corresponding to an uncertainty of 4.2\%, and an enclosed mass of $M_\mathrm{MW}(r<32.4{\rm kpc})=(2.85\pm0.1)\times 10^{11}\,\Msun$. This sits roughly in the middle of the present-day Galactocentric range spanned by the observed parts of the stream.

\begin{figure}
    \centering
    \includegraphics{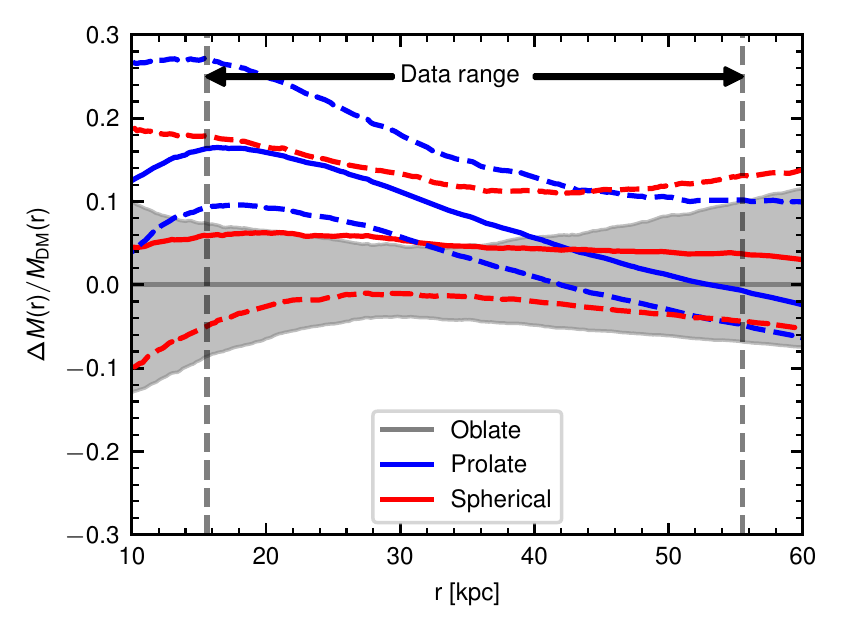}
    \caption{Comparison of the Milky Way's enclosed mass for the models of the MW DM halo. The y-axis shows the fractional difference between the dark matter mass inferred from each fit and that inferred from the oblate fit. The enclosed mass of the Milky Way inferred from these fits is consistent at the $1\sigma$ level over most of the radial range considered, showing that the inferred mass is robust.}
    \label{fig:mw_mass_comparison}
\end{figure}

Next, we compare the enclosed dark matter mass profile of our fiducial (oblate) fits with the prolate and spherical halo fits in Figure~\ref{fig:mw_mass_comparison}. In particular, we show the fractional difference between each enclosed mass profile and the oblate mass profile. We see that over the radial range spanned by the data, our different fits all produce a consistent enclosed mass profile. This shows the robustness of our inferred mass profile. However, we note that beyond this radial range, the different fits give different results.

As a demonstration of this, we compute the virial mass and radius using the standard definition from \cite{Bryan+1998}: i.e. the mass enclosed by a radius within which the average density is $\Delta_c \rho_{\rm crit}$, where $\rho_{\rm crit}=3H_0^2/8\pi G$ is the critical density of the Universe, $H_0=67.37$\,km\,s$^{-1}$\,Mpc$^{-1}$ is the present-day Hubble constant, and the pre-factor $\Delta_c \sim 103$, with cosmological parameters from \cite{Planck+2020}. We present the inferred virial mass of the Milky Way in Table~\ref{tab:results}. While the oblate and spherical haloes have similar virial masses, the mass inferred from the prolate halo is significantly smaller. We stress that the virial masses are extrapolations and that the most robust measurement is within the radial range where the OC stream is observed.

The flexible models used in this work also allow us to measure the density slope of the inner ($\gamma$) and outer ($\beta$) Milky Way dark matter halo, as well as the flattening of the Milky Way halo ($q_{\rm NFW}$). Interestingly, we see that the oblate and spherical halo have inner and outer slopes consistent with an NFW profile, i.e. consistent with 1 and 3 respectively. In contrast, the prolate halo has a much steeper outer profile, $\beta_{\rm prolate}=3.82^{+0.11}_{-0.12}$. This rapid fall-off in density is why the prolate halo has a much smaller virial mass. In addition, we find that both the oblate $\left(q_{\rm NFW} = 0.56^{+0.10}_{-0.08}\right)$ and prolate $\left(q_{\rm NFW} = 1.40^{+0.12}_{-0.10}\right)$ halo are preferred over the spherical halo. We discuss the oblate and prolate halo flattening in more detail in Section~\ref{sec:obl_vs_pro}.

\subsection{LMC mass measurement} \label{sec:lmc_mass}

We now explore the inferred LMC potential. As with the Milky Way potential in Section~\ref{sec:mw_pot}, we consider the enclosed mass profile of the LMC. We note that previous works which used streams to measure the mass of the LMC \citep[e.g.][]{Erkal2019, Shipp+2021, Vasiliev+2021} all considered a family of single parameter models for the LMC. In particular, \cite{Erkal2019, Shipp+2021} assumed a Hernquist profile described by the total mass with a scale radius chosen to satisfy the mass constraint of \cite{vandermarel+2014} at 8.7 kpc. Similarly, \cite{Vasiliev+2021} assumed a smoothly truncated NFW profile described by the total mass with a scale radius chosen to match the mass enclosed in the inner $\sim 9$ kpc of the LMC (see their Fig. 3). 

\begin{figure}
    \centering
    \includegraphics[width=0.45\textwidth]{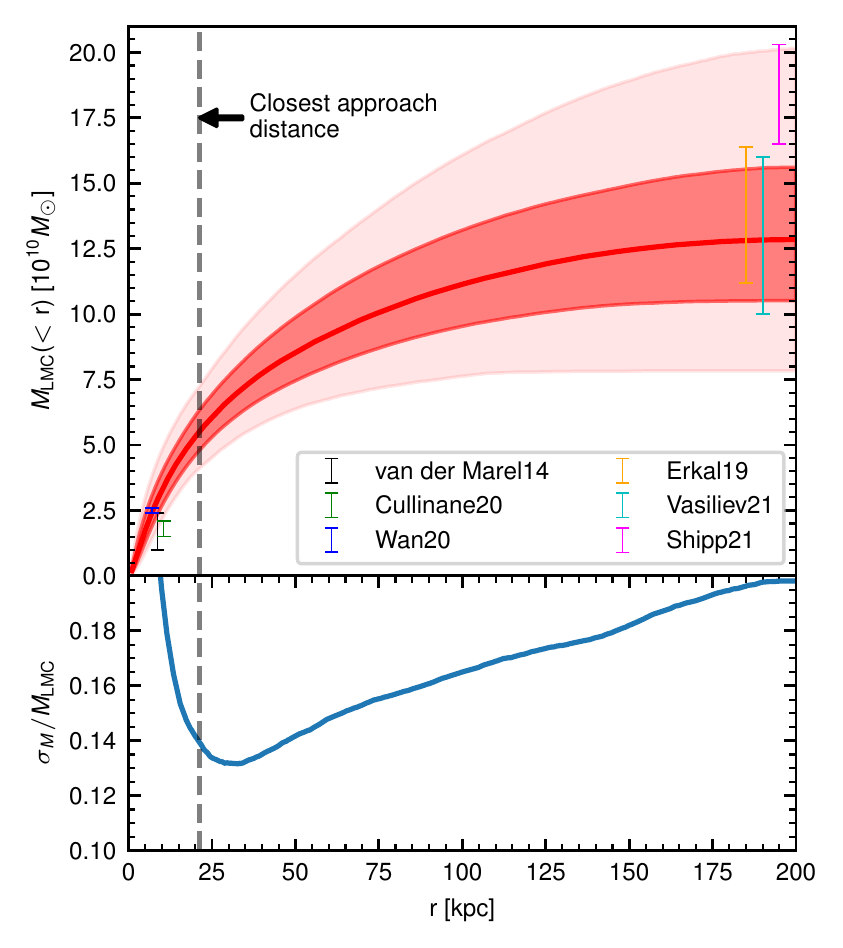}
    \caption{Constraints on the enclosed mass profile of the LMC. \textit{Top panel} shows the enclosed mass profile of the LMC compared to observations. The vertical dashed grey line shows the median closest approach distance of the OC stream to the LMC (21.3 kpc) for our models, restricted to the same $\phi_1$ range as the data, within which we do not expect to be sensitive. The \CHANGE{previous measurements for the inner LMC are} \CHANGE{van der Marel}14 \protect\citep{vandermarel+2014}, \CHANGE{Cullinane}20 \protect\citep{Cullinane+2020}, and Wan20 \protect\citep{Wan+20}. \CHANGE{The previous measurements for the total LMC mass are Erkal19} \protect\citepalias{Erkal2019}, \CHANGE{Vasiliev21} \protect\citep{Vasiliev+2021}, \CHANGE{and Shipp21} \protect\citep{Shipp+2021}. \CHANGE{Note that these outer measurements are staggered radially for readability}. \textit{Bottom panel} shows the fractional uncertainty in the mass. The most precise measurement of the LMC's enclosed mass is made at 32.8 kpc with $M_{\rm LMC}(<32.8 \,{\rm kpc}) =7.02\times10^{10} \Msun$, which corresponds to a fractional uncertainty of 13.2\%. \CHANGE{The constraints on the LMC mass profile shown on the top panel of the figure are provided in digital form in supplementary materials.}} 
    \label{fig:m_lmc}
\end{figure}

In comparison, our flexible model does not put priors on the enclosed mass profile of the LMC. Thus, for the first time, we present a measurement of the enclosed mass profile of the LMC in the top panel of Figure~\ref{fig:m_lmc}. For reference, we show the closest approach distance of the OC stream to the LMC from our best-fit model. Within this radius, we should not be sensitive to the mass distribution. We also show mass measurements of the LMC based on rotation curve measurements for comparison.

Motivated by the results of \cite{Bonaca+2018}, we show the fractional uncertainty of the enclosed LMC mass in the bottom panel of Figure~\ref{fig:m_lmc}. We see that the LMC mass is best measured where we broadly expect, i.e. beyond the closest approach (within which it is only sensitive to the total mass), but still near the closest approach (where the velocity kicks are the largest). 
In particular, there is a large uncertainty on the mass within the closest approach distance, within which we are not sensitive. Similarly, at large distances, the enclosed mass is poorly measured because in this region the acceleration from the LMC is small -- and smaller than the acceleration from the Milky Way. Between these two limits, we achieve the best measurement of the LMC's enclosed mass at 32.8 kpc of $M_{\rm LMC}(<32.8\, {\rm kpc})=7.02^{+0.99}_{-0.86}\times 10^{10} \Msun$.

\begin{figure}
    \centering
    \includegraphics{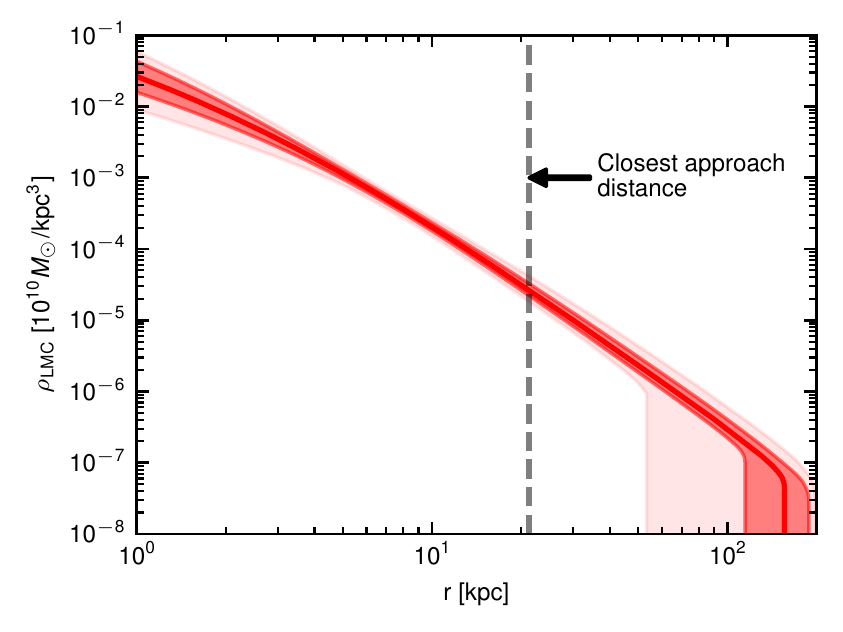}
    \caption{Inferred density of the LMC. The red curve shows the median density and the dark (light) shaded regions show the $1\sigma$ (2$\sigma$) confidence intervals. Interestingly, our fits prefer an extended LMC halo that extends beyond $\sim 53$\,kpc.   
    }
    \label{fig:rho_lmc}
\end{figure}

Interestingly, Figure~\ref{fig:m_lmc} also shows that our fits disfavour compact LMC mass profiles, i.e. where all of the mass sits within the closest approach distance. Instead, our models prefer an extended dark matter halo for the LMC where different parts of the stream experience different LMC masses. In order to better showcase this, we show the inferred density profile of the LMC in Figure~\ref{fig:rho_lmc}. This shows that our models require an LMC whose dark matter halo extends out to at least $\sim 53$ kpc at the 2$\sigma$ level. Such an extended dark matter halo is in agreement with the recently observed extent of the LMC hot corona \citep{krishnarao2022} and is consistent with the LMC being on its first approach to the Milky Way \citep[e.g.][]{Besla+2007}. 

\section{Discussion} \label{sec:discussion}

\subsection{Stream dispersions}

\begin{figure}
    \centering
    \includegraphics{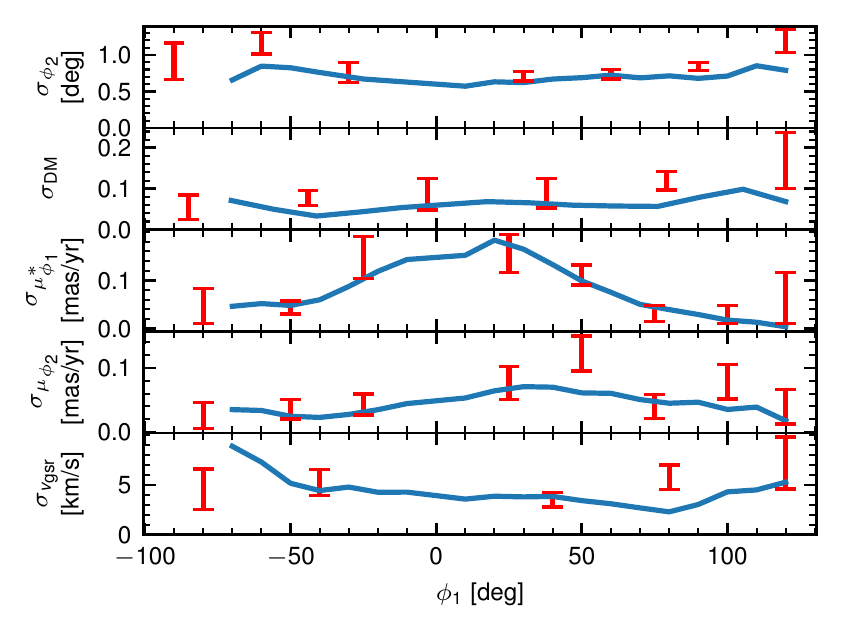}
    \caption{Width of OC stream in each observable. In each panel, the blue line shows the dispersion of the best-fit model from Figure~\protect\ref{fig:OC_best}. From top to bottom, the panels show the dispersion of the stream on the sky, in distance modulus, in proper motions, and in radial velocity. In each panel, the red error bars show the observed dispersions and the blue lines show the dispersion of the best-fit stream model for comparison. While the model was tuned to match the mean width on the sky, it is able to match the other dispersions.}
    \label{fig:stream_width}
\end{figure}

In Sections~\ref{sec:6d_track} \& \ref{sec:analysis}, we showed that with the combination of \textit{Gaia} and the \Sfive survey, we can measure the 6-D track and dispersions of the stream. We also showed in Section~\ref{sec:analysis} that we reliably detect the dispersion of proper motions along the stream which appears to be around 10\,\kms: higher than the dispersion of line-of-sight velocities of 5\,\kms. In order to highlight the reliability of the dispersion measurements, as well as to further validate our stream models, we compare the observed and simulated dispersions in Figure~\ref{fig:stream_width}. As a caveat, we note that the mass of the progenitor was selected to reproduce the average stream width on the sky. That said, the model is able to reproduce the trends seen in the data, such as the rise and fall of the dispersion in the proper motion along the stream, as well as all of the other dispersions. Interestingly, at large $\phi_1$, the width of the distance modulus and the stream on the sky are both larger in the data than in the model. This may be due to a different mass loss history for the real stream (e.g. it was more massive in the past than in our model) or some additional complexity in the potential which could increase these dispersions \citep[e.g.][]{Erkal:2016}.

\subsection{Oblate versus prolate halo} \label{sec:obl_vs_pro}

\begin{figure}
    \centering
    \includegraphics{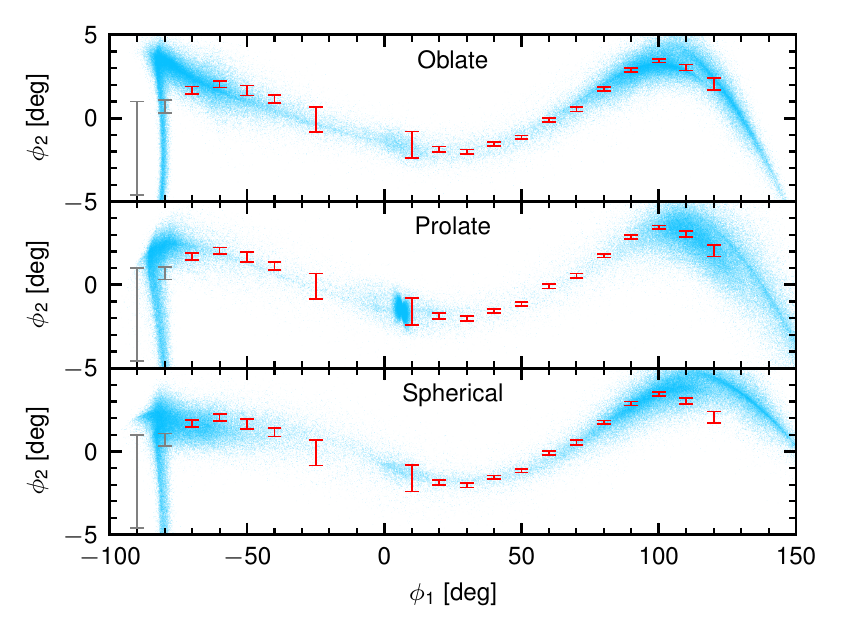}
    \caption{Best-fit stream models for different halo models. From top to bottom, the panels show the best-fit stream in an oblate, prolate, and spherical halo respectively. Compared to the spherical halo, the main improvement of using an oblate or prolate halo is that it allows us to reproduce the observed down-turn in the stream track at $\phi_1 > 100$\,deg. } 
    \label{fig:axi_vs_sph}
\end{figure}

In Section~\ref{sec:mw_pot}, we discussed the Milky Way potential measurement with a focus on the mass enclosed, which is robust across our choice of halo flattening. In this section, we explore why our fits prefer both a prolate and oblate halo over a spherical halo. First, in Figure~\ref{fig:axi_vs_sph}, we focus on what features in the data the oblate and prolate haloes are able to represent. We highlight the on-sky track in this plot since this is the main improvement in the fit; i.e. the improvement in the likelihood between the spherical and flattened halo fits is mainly driven by an improved match to the on-sky track \citepalias[as in][]{Erkal2019}.
In particular, the oblate and prolate halo are able to replicate the downturn in the on-sky track at $\phi_1 > 100$\,deg. In contrast, the spherical halo provides a much shallower downturn. 

\begin{figure*}
    \centering
    \includegraphics{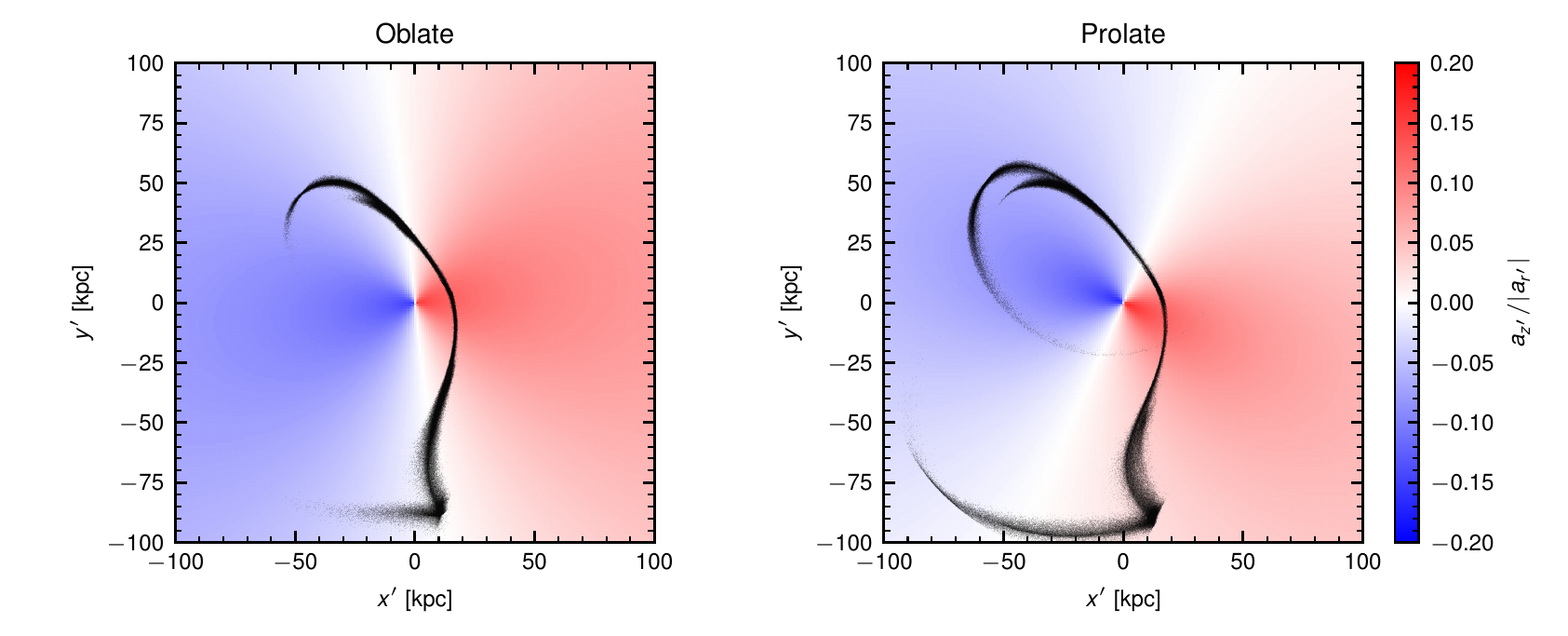}
    \caption{Ratio of out-of-plane to radial acceleration for the oblate and prolate halo fits. The left and right panels show the ratio of the forces out of the stream plane to the radial force in the stream plane. This shows that the oblate and prolate halo fits create a broadly similar force field in the plane of the OC stream. The black points show the best-fit stream model.}
    \label{fig:force_ratio}
\end{figure*}

Based on the downturn on the sky, we next investigate how the acceleration fields of the oblate and prolate halo differ. In order to study this, we consider the acceleration field from the Milky Way halo in the plane of the stream. We define the stream plane using the rotation matrix from \citetalias{Koposov2019}, which defines the stream coordinates and hence the stream plane which goes through the Sun. One advantage of this coordinate system is that a force which points down in these coordinates (i.e. in the negative $z$ direction), will also point down on the sky in the OC coordinate system (i.e. in the negative $\phi_2$ direction). We shift this plane to go through the centre of the Galaxy. Given this stream plane, we evaluate the ratio of the out-of-plane acceleration to the radial acceleration for points on an evenly spaced Cartesian grid in this plane. We average this acceleration ratio across 100 random draws from the posterior chains and show the results in Figure~\ref{fig:force_ratio}. Interestingly, the oblate and prolate halo fits have a similar amplitude and orientation of the acceleration ratio in this plane. Furthermore, in the region where the stream is bending down on the sky (i.e. $\phi_1 > 100$ deg), both haloes have an acceleration field that points down in the stream plane. Thus, despite their seemingly disparate shapes, both the oblate and prolate halo fits seem to be trying to reproduce the same acceleration field which can bend the stream down on the sky. 

\begin{figure}
    \centering
    \includegraphics{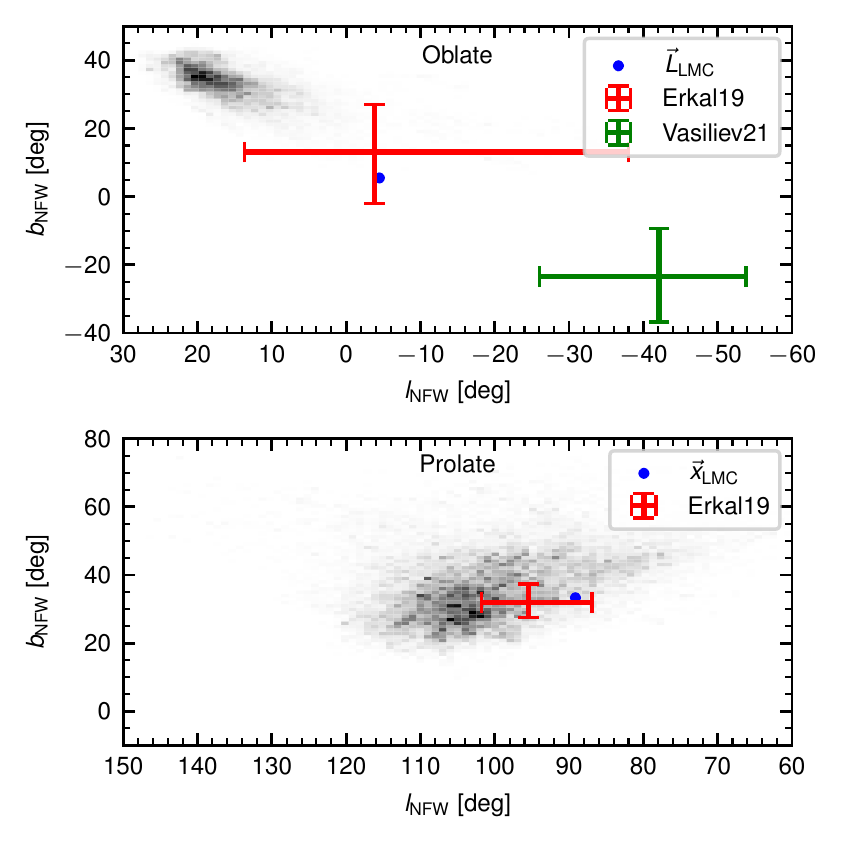}
    \caption{Comparison of the orientation of the oblate and prolate halo inferred in this work.
    The \textit{top} (\textit{bottom}) panel shows the location of the short (long) axis of the oblate (prolate) halo respectively, as viewed from the Milky Way centre. The red error bars show the orientation of the oblate and prolate halo inferred in \protect\citetalias{Erkal2019}. The blue points show the angular momentum vector of the LMC (top panel) and the position of the LMC (bottom panel). Note that the errors in these quantities are smaller than the size of the blue point. In the top panel, we also show the orientation of the short-axis from \protect\cite{Vasiliev+2021} from a fit to the Sagittarius stream.} 
    \label{fig:obl_vs_pro}
\end{figure}

Finally, we compare the orientation of the oblate and prolate haloes in Figure~\ref{fig:obl_vs_pro}. In previous fits to the OC stream, \citetalias{Erkal2019} noted that the oblate halo orientation was consistent with the orientation of the LMC's orbital angular momentum vector, and similarly that the prolate halo orientation was consistent with the present-day vector to the LMC, suggesting a possible connection between the halo shapes and the LMC. We show the results of \citetalias{Erkal2019}, as well as the LMC's angular momentum vector and present-day position vector for comparison in Figure~\ref{fig:obl_vs_pro}. We see that while the prolate halo is still aligned with the LMC's present-day position, the oblate halo is now slightly misaligned with the LMC's angular momentum vector. For comparison, we also show the orientation of the short-axis of the halo inferred with the Sagittarius stream by \cite{Vasiliev+2021}. In that work, they considered a triaxial halo but the intermediate and long axes were nearly identical, consistent with an axisymmetric, oblate halo. With the improved accuracy on the potential in this work, we see that the Milky Way potential shapes inferred with the OC stream and the Sgr stream are now inconsistent. In light of this discrepancy, and also the broadly consistent acceleration field of the oblate and prolate haloes in the OC stream plane, we note that these fits may be trying to describe the deformations of the Milky Way and LMC haloes which are known to affect the OC stream \citep{lilleengen+2022}.

\subsection{Perturbations from classical satellites}

In Section~\ref{sec:modelling}, we focused on the combined influence of the Milky Way and the LMC on the OC stream. Next, we assess whether any of the other classical satellites also could have perturbed the OC stream. In order to do this, we take the same stream generation machinery as in Section~\ref{sec:stream_gen} and add an additional perturber in the form of a Plummer sphere with a mass of $10^9 \Msun$ and a scale radius of 1 kpc. For the present-day location of the dwarfs, we use the distances and radial velocities from \cite{Mcconnachie2012}. For proper motions, we use the results of \cite{GaiaDR2sats} for all the classical dwarf galaxies except Leo~I, Leo~II, and the SMC. For these three, we use proper motion measurements from \cite{LeoIpm}, \cite{LeoIIpm}, and \cite{Kallivayalil+2013} respectively. 

For the parameters of the Milky Way, LMC, and OC stream, we take the best-fit model from the oblate halo fit described in Section~\ref{sec:mcmc}. For each of the classical dwarfs, we sample its present-day phase-space coordinates 10 times and rewind it self-consistently in the combined presence of the Milky Way, LMC, and OC stream progenitor. The system is then evolved forwards while stars are ejected from the OC progenitor to form the OC stream. We then compare the resulting OC stream with the one without a perturbation from the classical satellite (i.e. the fiducial stream). In order to evaluate the changes to the stream observables, we make mock observations of these streams at the locations of the knots and compute the difference in the observables with the fiducial stream. We then compare these differences with the observed uncertainty at each knot to determine how significant they are. We find that the only satellite which results in a $>3\sigma$ change in the observables is the SMC, which perturbs the observed distance by $\sim3\sigma$ in 2 out of the 10 realizations. As a result, we find that the OC stream is likely not strongly affected by the other classical satellites given the present-day data.  

\subsection{Association with dwarfs and globular clusters}

Next, we search for associations between known MW satellites (ultra-faint dwarfs and GCs) with the OC stream. Previous works have searched for such associations by comparing the present-day phase space coordinates of each satellite with observables of the OC stream \citepalias[e.g.][]{Koposov2019}. In contrast, we integrate the orbits of each dwarf and GC backwards in the presence of the stream for 4 Gyr and compute how close each satellite passes to stream particles in phase space. 
In particular, for each satellite, we simultaneously sample from its present-day phase space coordinates and sample the posterior chains for the OC stream and potential parameters. This sampling is performed 100 times for each satellite.

\begin{figure}
    \centering
    \includegraphics{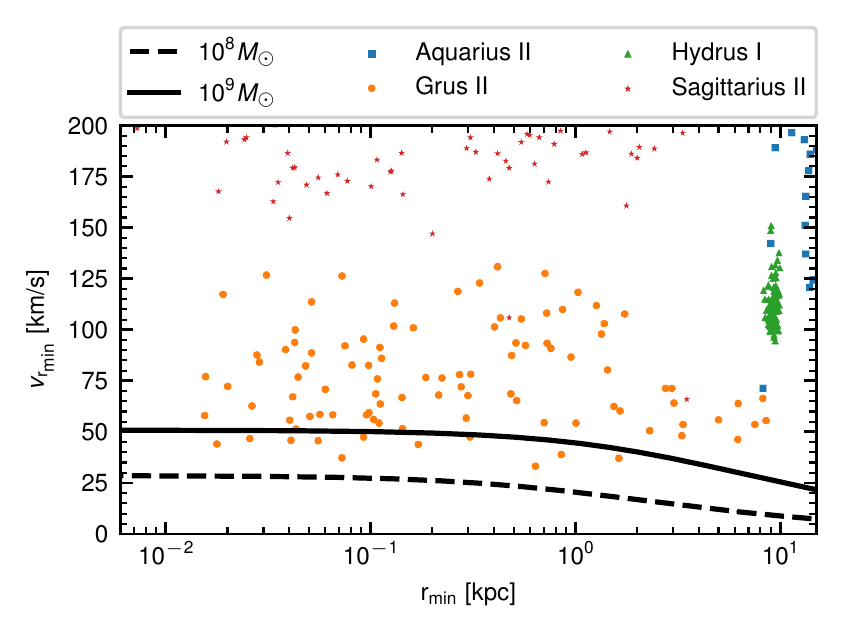}
    \caption{Closest approach distance and relative velocity of each dwarf with the OC stream. The black solid (dashed) curves show the escape velocity of a $10^9 \Msun$ ($10^8 \Msun$) dark matter halo. Grus II stands out as having a high probability of having a very close encounter with the OC stream. In particular, there are many realisations where Grus II could have been bound to a $10^9 \Msun$ host halo of the OC progenitor. }
    \label{fig:oc_ufds}
\end{figure}

For the ultra faint dwarfs, we use the sample of 46 ultra faints in \cite{Pace+2022} that have full 6-D phase-space coordinates. 
In Figure~\ref{fig:oc_ufds}, we show the closest approach distance and relative velocity at this closest approach for the dwarfs that have an approach within 10 kpc and 100 \kms. Interestingly, we see that Grus II stands out as having realizations that have close passages ($\lesssim 0.1$ kpc) with the OC stream particles at low relative velocities ($\lesssim 50$\,\kms). This raises the possibility that Grus II could have been bound to the progenitor of the OC stream. For reference, we show the escape velocity curves of a $10^8\, \Msun$ and $10^9\,\Msun$ subhalo which are modelled as Hernquist profiles, with a halo mass-scale radius relation taken from \cite{Erkal+2016}. Interestingly, we find that 11\% of the realizations would have been bound to a $10^9\,\Msun$ progenitor. 

We note that during each rewinding, Grus II has multiple close approaches to the OC stream. The most recent of these occur $\sim 300-400$ Myr ago, with more ancient passages $\sim 1.75-2.25$ Gyr ago. Interestingly, in some of these earlier passages, Grus II passes very close to the progenitor we have used in this work. Thus, these rewindings are consistent with the picture where Grus II could have been bound to the original host of the OC stream $\sim 2 $ Gyr ago, which subsequently disrupted. Thus, Grus II would be the first ultra-faint dwarf galaxy associated with a classical dwarf galaxy. %

We repeat the same analysis for the sample of 160 Milky Way globular clusters using the 6-D catalogue tabulated by \cite{Vasiliev&Baumgardt2021} and \cite{Baumgardt&Vasiliev2021}. As with the dwarfs, we sample each globular cluster 100 times while simultaneously sampling the stellar stream model. Unlike the dwarfs, we do not find any close and low-velocity passages: i.e. no globular cluster passes within 10\,kpc and 100\,\kms during the rewinding. 

\subsection{Interaction with LMC}

\begin{figure}
    \centering
    \includegraphics{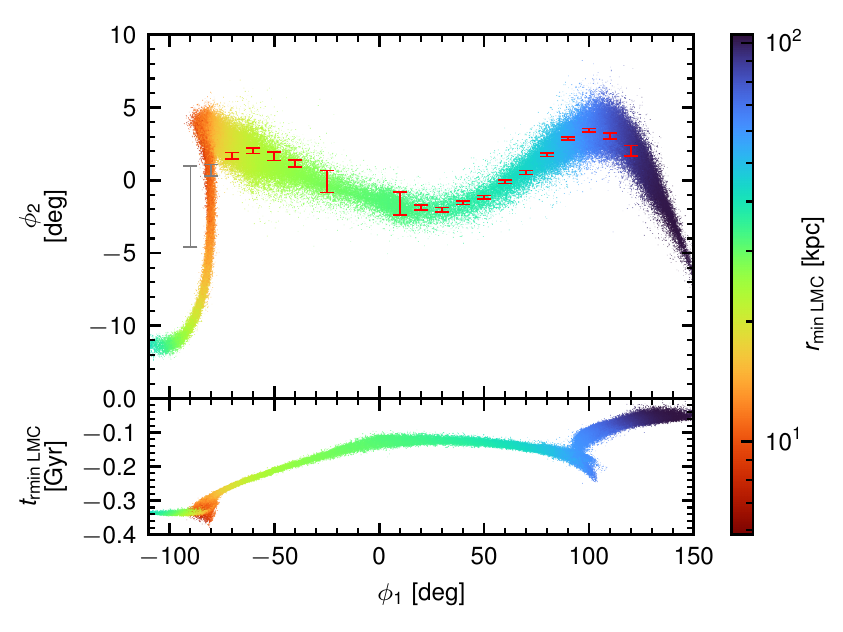}
    \caption{Closest approach to the LMC. \textit{Top panel} shows the stream on the sky, coloured by the closest approach distance of each stream particle to the LMC. \textit{Bottom panel} shows the time of the closest approach. The closest approach occurs at a distance of 5.8 kpc at a time of 366 Myr ago, at $\phi_1 = -82.4$\,deg. For the data, we fit in this work (i.e. $\phi_1 \geq -70$\,deg), the closest approach is in the bin at $\phi_1=-70$\,deg at a distance of 19.1 kpc with a lookback time of 254 Myr.}
    \label{fig:oc_lmc_approach}
\end{figure}

Of the streams modelled to date, the OC stream has one of the closest and strongest encounters with the LMC \citep{Shipp+2021}. In order to better understand this encounter, as well as what we can learn from it, we show the closest approach distance and time of the closest approach in Figure~\ref{fig:oc_lmc_approach} for the best-fit model from Figure~\ref{fig:OC_best}. In particular, we see that the kink in the stream track at $\phi_1 \sim -80$ deg corresponds to the point of closest approach with the LMC. We note that previous fits to the OC stream also predicted such a kink \citep[][]{Erkal2019,Shipp+2021}. For our best-fit model, this approach is very close with some of the particles passing within $\sim6$ kpc of the LMC. Furthermore, this passage occurs $\sim 370$ Myr ago, when the LMC was at a distance of $\sim 107$ kpc from the Milky Way. This suggests that the OC stream should be a sensitive probe of the LMC's past location. 

\begin{figure}
    \centering
    \includegraphics{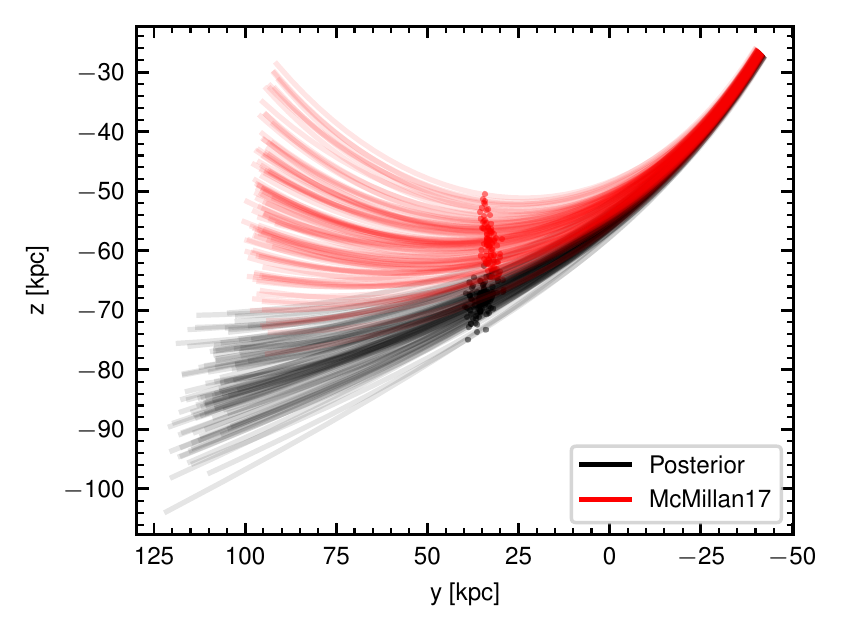}
    \caption{Uncertainty in the LMC's orbit over the past 500 Myr. The black curves show the LMC's past orbit based on the posterior samples in this work in Galactocentric coordinates. The red lines show the LMC's past orbit in the potential from \protect\cite{McMillan2017}. The black and red points show the LMC's location at a lookback time of 254 Myr, i.e. the point of closest approach to the OC stream. This shows that fits to the OC stream can be used to meaningfully constrain the past location of the LMC.}
    \label{fig:lmc_orbit}
\end{figure}

In Figure~\ref{fig:lmc_orbit}, we show how well the fits in this work constrain the past orbit of the LMC. In order to show the uncertainty in the LMC's past orbit without using our OC stream fit, we rewind the LMC in the Milky Way potential from \cite{McMillan2017}. In particular, we sample 100 Milky Way potentials from the posterior chains in that work. For each of these realizations of the Milky Way potential, we also sample the LMC's proper motion, radial velocity, and distance from their observed values \citep[][respectively]{Kallivayalil+2013,vandermarel+2002,Pietrzynski+2019}. We model the LMC as a Hernquist profile \citep{Hernquist1990} and sample its mass from the value measured in \citetalias{Erkal2019}, $M_{\rm LMC}=(1.38\pm0.26)\times10^{11}\Msun$, with a scale radius such that the circular velocity at 8.7 kpc matches the observed value of 91.7\,\kms \citep{vandermarel+2014}. We show the orbit for the past 500 Myr and show the location at a lookback time of 254 Myr, motivated by the median closest approach of the LMC and the OC stream in our models (see Fig.~\ref{fig:oc_lmc_approach}). We also show the uncertainty of the LMC's past orbit from the fits in this work by taking 100 samples from our posterior chains. Figure~\ref{fig:lmc_orbit} shows that fits to the OC stream in this work give a smaller uncertainty in the past trajectory of the LMC. In addition, we see that the LMC's orbit in the potential from \cite{McMillan2017} remains closer to the Milky Way. This is due to the larger mass of the Milky Way in the models of \citet[see Fig~\ref{fig:mw_mass}]{McMillan2017}.

In order to further explore how sensitive the OC stream is to the LMC's past trajectory, we perform two additional tests. First, we take the best-fit OC stream model (i.e. from Fig.~\ref{fig:OC_best}) and turn off the dynamical friction the LMC experiences in the presence of the Milky Way. The resulting stream is a poor fit to the data with a difference in the likelihood of $\Delta \log L \sim -3500$. Without dynamical friction, the LMC's location during the closest approach (254 Myr ago) changes by $\sim 6$ kpc, which is comparable to the LMC's closest approach distance to the observed regions of the OC stream ($\sim 21.3$ kpc). This shows that current observations are already sensitive to the dynamical friction that the LMC has experienced.

As a second test, we re-fit the OC stream with an extra parameter that can scale up or down the dynamical friction the LMC experiences. In particular, we use the same setup as in Section~\ref{sec:mcmc} for the oblate Milky Way halo, with one extra parameter ($\lambda_{\rm DF}$). This parameter multiplies the dynamical friction term the LMC experiences and it has a log uniform prior between $10^{-3}$ to 10. As expected from the first test, this parameter is strongly constrained by the existing data with a posterior of $\log_{10} \lambda_{\rm DF} = 0.31^{+0.20}_{-0.48}$ consistent with $\lambda_{DF}=1$. 

Finally, we note that further work is needed to determine whether this is a robust measurement of dynamical friction. In particular, in the model used for this paper, we have neglected the deformations of both the Milky Way and LMC dark matter haloes which have been predicted to affect the OC stream \citep{lilleengen+2022}. While these effects will need to be accounted for in future work, the tests presented here have demonstrated that the OC stream contains a wealth of information that we have yet to understand. 

\subsection{Evolution in energy and angular momentum}
\label{sec:energy_angm_model}

\begin{figure}
    \centering
    \includegraphics{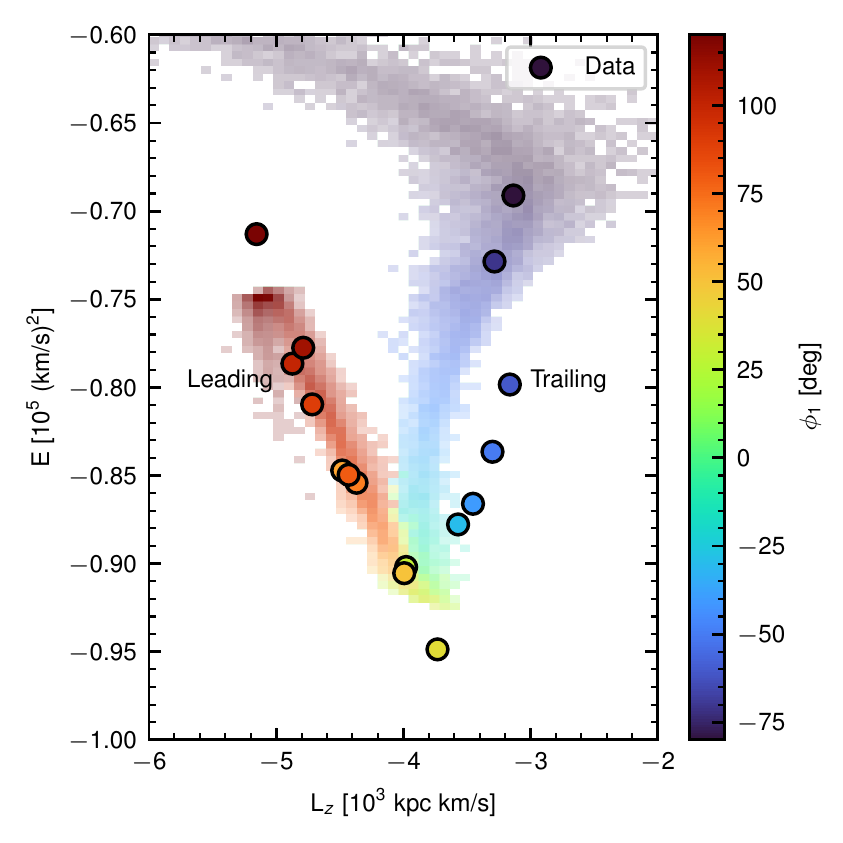}
    \caption{The comparison of the energy vs angular momentum distribution in the simulations vs data. The coloured pixel show the distribution of particles from the best-fit model. The large circles show energy and angular momentum measurements along the stream from this paper as presented in Figure~\ref{fig:energy_ang}. Each point is coloured by $\phi_1$ similarly to Figure~\ref{fig:energy_ang}. 
    }
    \label{fig:energy_ang_mod}
\end{figure}

In order to further understand the LMC's impact on the stream, as well as understand the peculiar ``V'' shape seen in the observed energy versus angular momentum (see Fig.~\ref{fig:energy_ang}), we now investigate the distribution of energy and angular momentum in the model. In Figure~\ref{fig:energy_ang_mod}, we show the observed and simulated streams in the energy versus angular momentum plane. Note that in order to aid this comparison, we use the same potential as we used in Section~\ref{sec:energy_angm} \citep[i.e., the best-fit model from][]{McMillan2017} and not the time-varying potential we have fit in this work. We see that the model reproduces the ``V'' shape. This is not very surprising given the overall good agreement in reproducing the phase-space track of the stream.

In order to understand how peculiar this distribution is, we first consider a stream disrupting in a static potential. For such a stream, we would expect the leading arm ($\phi_1 > 0$ deg) to have lower energies and lower absolute angular momentum, and the trailing arm ($\phi_1 < 0$ deg) to have higher energies and higher absolute angular momentum \citep[see][]{Gibbons2014}. Thus, we would broadly expect the energy and angular momentum distribution to go from the upper left to the bottom right of Figure~\ref{fig:energy_ang_mod}, with the smallest $\phi_1$ in the upper left and the largest $\phi_1$ in the bottom right, and $\phi_1$ smoothly varying in-between. Instead, the leading arm has more absolute angular momentum and the same energy as the trailing arm, and the trailing arm has the wrong correlation along the stream (i.e. $E$ increases as ${\rm L}_z$ increases). Furthermore, while the leading arm seemingly agrees with the static model from \cite{Gibbons2014}, $\phi_1$ should be increasing as ${\rm L}_z$ increases: instead, it does the opposite. 

With our best-fit model, we can understand this remarkable transformation. From looking at different snapshots (see Appendix~\ref{sec:ELz_snaps}), we see that the before the LMC's infall, the stream was arranged as expected from the static disruption model in \cite{Gibbons2014}. During the LMC's close passage, the trailing arm is essentially flipped over in $E-L_z$ space, while the leading arm has a substantial increase in $L_z$ and a decrease in energy. This substantial re-arrangement of the stream in $E-L_z$ is similar to the expected effect of the Sagittarius dwarf on stellar streams \citep{Dillamore+2022}.

\section{Summary and Conclusions} \label{sec:conclusions}

In this paper, for the first time, using the combinations of {\it Gaia}, \Sfive, and other spectroscopic survey data, we mapped the Orphan-Chenab stream over 200 degrees on the sky. Not only were we able to identify over 300 spectroscopic members associated with the OC stream, we also derived the full 6-D phase space track of the stream. We also measured the stream density along the $\sim 120$ kpc of the extent of the structure. Here, we summarise the key findings from this rich dataset:

\begin{itemize}
    \item We observe that while the apparent number density of stars along the stream increases rapidly toward the stream edges, when corrected for the accordion effect (slower speed at apocenter and faster speed at pericenter), the stream density (or stream flow of stars) is mostly constant along the stream.
    \item We do not detect a significant metallicity gradient in either the Northern/Southern arms, or across the whole stream, preventing us from using this to locate the current or previous position of the stream progenitor. 
    \item The mean metallicity of the stream based on \Sfive\ measurements is $-1.9$ with a spread of 0.3\,dex. Together with an estimate of the total stream luminosity of $M_V=-8.2$, this makes the stream similar in properties to classical MW satellites like Draco.
    \item While the stream radial velocity dispersion is constant at around $\sim$ 5\,\kms throughout the stream, we notice a significant change in the intrinsic spread in proper motions, detecting a $\sim $ 10\,\kms dispersion in velocities along the stream near its pericenter.
    \item The stream width changes significantly from 200\,pc for the nearby part of the stream to nearly 1\,kpc for the most distant parts. This is accompanied by a significant distance spread along the line of sight in the North, which has been noticed previously by \citet{Sesar2013}.
    \item  We detect large changes in the energy and the $z$-component of angular momentum across the stream caused by the interaction with the LMC. This spreading of the stream in the space of conserved quantities is expected to significantly hamper our ability to use clustering in this space to identify old accretion events.
    \item The stars in the Southern part of the stream move at a significant angle with respect to the stream track, both on the sky and along the line of sight. Their velocity is clearly deflected in the direction of the LMC. This deflection is the cause of the observed bending of the stream.
    \item The model of the stream disruption in the presence of the LMC can successfully reproduce the full 6-D stream track when the Galactic DM halo model is represented by an oblate NFW-like potential. A prolate shape for the DM halo provides a somewhat worse fit to the data, while a spherical shape is inconsistent with the observations.
    \item The inference of the MW dark matter halo shape has a degeneracy between prolate and oblate shapes that can provide similar force fields in the OC stream plane, and similar mass profiles in the radial distance range.
    \item The modelling of the stream allows us to constrain the mass profile of both the MW and the LMC, with the most precise estimates of enclosed masses for the Milky Way and LMC being $M_\mathrm{MW}(<32.4\,\mathrm{kpc})=(2.85\pm 0.10)\times 10^{11} \Msun$, $M_\mathrm{LMC}(<32.8\,\mathrm{kpc})=(7.02\pm0.9)\times10^{10}\,\Msun$, respectively. The best fit virial mass for the MW is $M_\mathrm{vir}=7.74^{+4.14}_{-1.48}\times 10^{11} \Msun$ and the total mass for the LMC is $M_\mathrm{LMC}=1.29^{+0.28}_{-0.23}\times10^{11}\,\Msun$.
    \item With the closest approach distance of the OC stream to the LMC of 21.3 kpc, we can  \CHANGE{ detect that the DM halo of the LMC extends out to at least 53 kpc. }
    \item The rewinding of MW satellites in the best-fit potential together with the OC stream reveals that Grus II was likely accreted together with the OC progenitor, as it has multiple close and low velocity encounters with the OC stream. 
\end{itemize}

In this paper, we have demonstrated the power of combining {\it Gaia} data with dedicated spectroscopic follow-up by {\Sfive}\ of this very long stellar stream perturbed by the LMC. This has allowed us to put strong constraints on the LMC and MW potential. It also shows that we can directly constrain the past trajectory of the LMC and the amount of dynamical friction it experienced, and potentially probe alternative gravity models that predict a different past orbit of the LMC \citep[e.g. MOND:][]{Wu+2008,Schee+2013}. 

The future of this analysis is manyfold. 
First, even the model presented in this paper does not fully reproduce all of the observable quantities perfectly, requiring a more complex model -- such as a deforming and/or triaxial MW DM halo -- which our data should be able to constrain. Indeed, \cite{lilleengen+2022} have already shown that the effect of the deforming Milky Way on the OC stream is significant. Second, while in this work we used an OC progenitor of fixed mass that was conveniently hidden in the part of the stream located behind the Galactic disc, the properties of the progenitor -- such as its stellar and dark matter mass, and their extent -- will likely have an effect on observables, and can be constrained with our high-quality data. With a detailed stream map and a stellar density measurement, we should also be able to constrain the details of the mass loss history of the progenitor.

While the analysis presented here is potentially applicable to other streams in the MW \citep[e.g.][]{mateu2022}, not many of them have detailed tracks \citep{patrick2022}, especially in 6-D \citep{S5_12streams}. This may however change with the arrival of upcoming spectroscopic surveys like  WEAVE \citep{jin2022}, DESI \citep{cooper2022} and 4MOST \citep{4most}. However, the majority of streams are not as long as the Sgr or OC stream: thus, limiting the amount of information we will be able to extract from each one.

In summary, despite the detailed analysis in this work, the OC stream still contains many mysteries. Where is/was its progenitor? What were the properties of that progenitor? What can it tell us about the Milky Way's dark matter halo and how it deforms in the presence of the LMC? We hope to explore these in future work. 

\section*{Acknowledgements}
SEK, DE, TSL, GFL, DZ \& SM acknowledge funding through ARC DP210100855. 
TSL acknowledges financial support from Natural Sciences and Engineering Research Council
of Canada (NSERC) through grant
RGPIN-2022-04794. This research has also been supported in part by the Australian Research Council Centre of Excellence for All Sky Astrophysics in 3 Dimensions (ASTRO 3D), through project number CE170100013.
The paper includes data obtained with the Anglo-Australian Telescope in Australia. We acknowledge the traditional owners of the land on which the AAT stands, the Gamilaraay people, and pay our respects to elders past, present and emerging.

This work has made use of data from the European Space Agency (ESA) mission
{\it Gaia} (\url{https://www.cosmos.esa.int/gaia}), processed by the {\it Gaia}
Data Processing and Analysis Consortium (DPAC,
\url{https://www.cosmos.esa.int/web/gaia/dpac/consortium}). Funding for the DPAC
has been provided by national institutions, in particular the institutions
participating in the {\it Gaia} Multilateral Agreement.

This project used public archival data from the Dark Energy Survey
(DES). Funding for the DES Projects has been provided by the
U.S. Department of Energy, the U.S. National Science Foundation, the
Ministry of Science and Education of Spain, the Science and Technology
Facilities Council of the United Kingdom, the Higher Education Funding
Council for England, the National Center for Supercomputing
Applications at the University of Illinois at Urbana-Champaign, the
Kavli Institute of Cosmological Physics at the University of Chicago,
the Center for Cosmology and Astro-Particle Physics at the Ohio State
University, the Mitchell Institute for Fundamental Physics and
Astronomy at Texas A\&M University, Financiadora de Estudos e
Projetos, Funda{\c c}{\~a}o Carlos Chagas Filho de Amparo {\`a}
Pesquisa do Estado do Rio de Janeiro, Conselho Nacional de
Desenvolvimento Cient{\'i}fico e Tecnol{\'o}gico and the
Minist{\'e}rio da Ci{\^e}ncia, Tecnologia e Inova{\c c}{\~a}o, the
Deutsche Forschungsgemeinschaft, and the Collaborating Institutions in
the Dark Energy Survey.  The Collaborating Institutions are Argonne
National Laboratory, the University of California at Santa Cruz, the
University of Cambridge, Centro de Investigaciones Energ{\'e}ticas,
Medioambientales y Tecnol{\'o}gicas-Madrid, the University of Chicago,
University College London, the DES-Brazil Consortium, the University
of Edinburgh, the Eidgen{\"o}ssische Technische Hochschule (ETH)
Z{\"u}rich, Fermi National Accelerator Laboratory, the University of
Illinois at Urbana-Champaign, the Institut de Ci{\`e}ncies de l'Espai
(IEEC/CSIC), the Institut de F{\'i}sica d'Altes Energies, Lawrence
Berkeley National Laboratory, the Ludwig-Maximilians Universit{\"a}t
M{\"u}nchen and the associated Excellence Cluster Universe, the
University of Michigan, the National Optical Astronomy Observatory,
the University of Nottingham, The Ohio State University, the OzDES
Membership Consortium, the University of Pennsylvania, the University
of Portsmouth, SLAC National Accelerator Laboratory, Stanford
University, the University of Sussex, and Texas A\&M University.
Based in part on observations at Cerro Tololo Inter-American
Observatory, National Optical Astronomy Observatory, which is operated
by the Association of Universities for Research in Astronomy (AURA)
under a cooperative agreement with the National Science Foundation.

The Legacy Surveys consist of three individual and complementary
projects: the Dark Energy Camera Legacy Survey (DECaLS; NOAO Proposal
ID \# 2014B-0404; PIs: David Schlegel and Arjun Dey), the
Beijing-Arizona Sky Survey (BASS; NOAO Proposal ID \# 2015A-0801; PIs:
Zhou Xu and Xiaohui Fan), and the Mayall z-band Legacy Survey (MzLS;
NOAO Proposal ID \# 2016A-0453; PI: Arjun Dey). DECaLS, BASS and MzLS
together include data obtained, respectively, at the Blanco telescope,
Cerro Tololo Inter-American Observatory, National Optical Astronomy
Observatory (NOAO); the Bok telescope, Steward Observatory, University
of Arizona; and the Mayall telescope, Kitt Peak National Observatory,
NOAO. The Legacy Surveys project is honored to be permitted to conduct
astronomical research on Iolkam Du'ag (Kitt Peak), a mountain with
particular significance to the Tohono O'odham Nation.

NOAO is operated by the Association of Universities for Research in
Astronomy (AURA) under a cooperative agreement with the National
Science Foundation.

The Legacy Surveys imaging of the DESI footprint is supported by the
Director, Office of Science, Office of High Energy Physics of the
U.S. Department of Energy under Contract No. DE-AC02-05CH1123, by the
National Energy Research Scientific Computing Center, a DOE Office of
Science User Facility under the same contract; and by the
U.S. National Science Foundation, Division of Astronomical Sciences
under Contract No. AST-0950945 to NOAO.

Funding for the Sloan Digital Sky 
Survey IV has been provided by the 
Alfred P. Sloan Foundation, the U.S. 
Department of Energy Office of 
Science, and the Participating 
Institutions. SDSS-IV acknowledges support and 
resources from the Center for High 
Performance Computing  at the 
University of Utah. The SDSS 
website is www.sdss.org.

SDSS-IV is managed by the 
Astrophysical Research Consortium 
for the Participating Institutions 
of the SDSS Collaboration including 
the Brazilian Participation Group, 
the Carnegie Institution for Science, 
Carnegie Mellon University, Center for 
Astrophysics | Harvard \& 
Smithsonian, the Chilean Participation 
Group, the French Participation Group, 
Instituto de Astrof\'isica de 
Canarias, The Johns Hopkins 
University, Kavli Institute for the 
Physics and Mathematics of the 
Universe (IPMU) / University of 
Tokyo, the Korean Participation Group, 
Lawrence Berkeley National Laboratory, 
Leibniz Institut f\"ur Astrophysik 
Potsdam (AIP),  Max-Planck-Institut 
f\"ur Astronomie (MPIA Heidelberg), 
Max-Planck-Institut f\"ur 
Astrophysik (MPA Garching), 
Max-Planck-Institut f\"ur 
Extraterrestrische Physik (MPE), 
National Astronomical Observatories of 
China, New Mexico State University, 
New York University, University of 
Notre Dame, Observat\'ario 
Nacional / MCTI, The Ohio State 
University, Pennsylvania State 
University, Shanghai 
Astronomical Observatory, United 
Kingdom Participation Group, 
Universidad Nacional Aut\'onoma 
de M\'exico, University of Arizona, 
University of Colorado Boulder, 
University of Oxford, University of 
Portsmouth, University of Utah, 
University of Virginia, University 
of Washington, University of 
Wisconsin, Vanderbilt University, 
and Yale University.

Guoshoujing Telescope (the Large Sky Area Multi-Object Fiber Spectroscopic Telescope LAMOST) is a National Major Scientific Project built by the Chinese Academy of Sciences. Funding for the project has been provided by the National Development and Reform Commission. LAMOST is operated and managed by the National Astronomical Observatories, Chinese Academy of Sciences.

This paper made use of the Whole Sky Database (wsdb) created by Sergey Koposov and maintained at the Institute of Astronomy, Cambridge by Sergey Koposov  with financial support from the Science \& Technology Facilities Council (STFC) and the European Research Council (ERC).

This paper used {\tt astropy} \citep{astropy2022}, {\tt agama} \citep{agama2019}, {\tt ipython} \citep{ipython2007}, {\tt imf}  \url{https://github.com/keflavich/imf}, {\tt matplotlib} \citep{matplotlib2007}, {\tt minimint} \citep{minimint2021}, {\tt numpy} \citep{numpy2020}, {\tt scipy} \citep{scipy2020}, {\tt sqlutilpy} \citep{sqlutilpy}.

This paper made use of the {\tt Q3C} package \citep{q3c2006}.

For the purpose of open access, the author has applied a Creative Commons Attribution (CC BY) licence to any Author Accepted Manuscript version arising from this submission.

\section*{Data Availability}

This paper relies on publicly available data from {\it Gaia}, SDSS, APOGEE, LAMOST surveys. We also use the data from S5 survey. The latest data is not yet publicly available, but the previous release is available in Zenodo \citep{S5_DR1}. The spectroscopic sample constructed in Section~\ref{sec:kinematic_sample}, {\tt Stan} models, measured parameters, the best fit OC stream model and posterior samples for the stream models are provided either in the appendix or through Zenodo  \url{https://zenodo.org/record/7222654}.


\bibliographystyle{mnras}
\bibliography{main} 

\begin{thebibliography}{}
\makeatletter
\relax
\def\mn@urlcharsother{\let\do\@makeother \do\$\do\&\do\#\do\^\do\_\do\%\do\~}
\def\mn@doi{\begingroup\mn@urlcharsother \@ifnextchar [ {\mn@doi@}
  {\mn@doi@[]}}
\def\mn@doi@[#1]#2{\def\@tempa{#1}\ifx\@tempa\@empty \href
  {http://dx.doi.org/#2} {doi:#2}\else \href {http://dx.doi.org/#2} {#1}\fi
  \endgroup}
\def\mn@eprint#1#2{\mn@eprint@#1:#2::\@nil}
\def\mn@eprint@arXiv#1{\href {http://arxiv.org/abs/#1} {{\tt arXiv:#1}}}
\def\mn@eprint@dblp#1{\href {http://dblp.uni-trier.de/rec/bibtex/#1.xml}
  {dblp:#1}}
\def\mn@eprint@#1:#2:#3:#4\@nil{\def\@tempa {#1}\def\@tempb {#2}\def\@tempc
  {#3}\ifx \@tempc \@empty \let \@tempc \@tempb \let \@tempb \@tempa \fi \ifx
  \@tempb \@empty \def\@tempb {arXiv}\fi \@ifundefined
  {mn@eprint@\@tempb}{\@tempb:\@tempc}{\expandafter \expandafter \csname
  mn@eprint@\@tempb\endcsname \expandafter{\@tempc}}}

\bibitem[\protect\citeauthoryear{{Abolfathi} et~al.,}{{Abolfathi}
  et~al.}{2018}]{Abolfathi2018}
{Abolfathi} B.,  et~al., 2018, \mn@doi [\apjs] {10.3847/1538-4365/aa9e8a},
  \href {https://ui.adsabs.harvard.edu/abs/2018ApJS..235...42A} {235, 42}

\bibitem[\protect\citeauthoryear{{Ahumada} et~al.,}{{Ahumada}
  et~al.}{2020}]{Ahumada2020}
{Ahumada} R.,  et~al., 2020, \mn@doi [\apjs] {10.3847/1538-4365/ab929e}, \href
  {https://ui.adsabs.harvard.edu/abs/2020ApJS..249....3A} {249, 3}

\bibitem[\protect\citeauthoryear{{Amorisco}, {G{\'o}mez}, {Vegetti}  \&
  {White}}{{Amorisco} et~al.}{2016}]{amorisco2016}
{Amorisco} N.~C.,  {G{\'o}mez} F.~A.,  {Vegetti} S.,   {White} S. D.~M.,  2016,
  \mn@doi [\mnras] {10.1093/mnrasl/slw148}, \href
  {https://ui.adsabs.harvard.edu/abs/2016MNRAS.463L..17A} {463, L17}

\bibitem[\protect\citeauthoryear{{Astropy Collaboration} et~al.,}{{Astropy
  Collaboration} et~al.}{2022}]{astropy2022}
{Astropy Collaboration} et~al., 2022, \mn@doi [\apj]
  {10.3847/1538-4357/ac7c74}, \href
  {https://ui.adsabs.harvard.edu/abs/2022ApJ...935..167A} {935, 167}

\bibitem[\protect\citeauthoryear{{Baumgardt} \& {Vasiliev}}{{Baumgardt} \&
  {Vasiliev}}{2021}]{Baumgardt&Vasiliev2021}
{Baumgardt} H.,  {Vasiliev} E.,  2021, \mn@doi [\mnras]
  {10.1093/mnras/stab1474}, \href
  {https://ui.adsabs.harvard.edu/abs/2021MNRAS.505.5957B} {505, 5957}

\bibitem[\protect\citeauthoryear{{Belokurov} et~al.,}{{Belokurov}
  et~al.}{2006}]{belokurov2006}
{Belokurov} V.,  et~al., 2006, \mn@doi [\apjl] {10.1086/504797}, \href
  {https://ui.adsabs.harvard.edu/abs/2006ApJ...642L.137B} {642, L137}

\bibitem[\protect\citeauthoryear{{Belokurov} et~al.,}{{Belokurov}
  et~al.}{2007}]{belokurov2007}
{Belokurov} V.,  et~al., 2007, \mn@doi [\apj] {10.1086/511302}, \href
  {https://ui.adsabs.harvard.edu/abs/2007ApJ...658..337B} {658, 337}

\bibitem[\protect\citeauthoryear{{Belokurov}, {Deason}, {Erkal}, {Koposov},
  {Carballo-Bello}, {Smith}, {Jethwa}  \& {Navarrete}}{{Belokurov}
  et~al.}{2019}]{Belokurov:2019}
{Belokurov} V.,  {Deason} A.~J.,  {Erkal} D.,  {Koposov} S.~E.,
  {Carballo-Bello} J.~A.,  {Smith} M.~C.,  {Jethwa} P.,   {Navarrete} C.,
  2019, \mn@doi [\mnras] {10.1093/mnrasl/slz101}, \href
  {https://ui.adsabs.harvard.edu/abs/2019MNRAS.488L..47B} {488, L47}

\bibitem[\protect\citeauthoryear{{Bennett} \& {Bovy}}{{Bennett} \&
  {Bovy}}{2019}]{Bennett2019}
{Bennett} M.,  {Bovy} J.,  2019, \mn@doi [\mnras] {10.1093/mnras/sty2813},
  \href {https://ui.adsabs.harvard.edu/abs/2019MNRAS.482.1417B} {482, 1417}

\bibitem[\protect\citeauthoryear{{Bernard} et~al.,}{{Bernard}
  et~al.}{2014}]{bernard2014}
{Bernard} E.~J.,  et~al., 2014, \mn@doi [\mnras] {10.1093/mnrasl/slu089}, \href
  {https://ui.adsabs.harvard.edu/abs/2014MNRAS.443L..84B} {443, L84}

\bibitem[\protect\citeauthoryear{{Besla}, {Kallivayalil}, {Hernquist},
  {Robertson}, {Cox}, {van der Marel}  \& {Alcock}}{{Besla}
  et~al.}{2007}]{Besla+2007}
{Besla} G.,  {Kallivayalil} N.,  {Hernquist} L.,  {Robertson} B.,  {Cox} T.~J.,
   {van der Marel} R.~P.,   {Alcock} C.,  2007, \mn@doi [\apj]
  {10.1086/521385}, \href
  {https://ui.adsabs.harvard.edu/abs/2007ApJ...668..949B} {668, 949}

\bibitem[\protect\citeauthoryear{{Binney}}{{Binney}}{2008}]{binney2008}
{Binney} J.,  2008, \mn@doi [\mnras] {10.1111/j.1745-3933.2008.00458.x}, \href
  {https://ui.adsabs.harvard.edu/abs/2008MNRAS.386L..47B} {386, L47}

\bibitem[\protect\citeauthoryear{{Bonaca} \& {Hogg}}{{Bonaca} \&
  {Hogg}}{2018}]{Bonaca+2018}
{Bonaca} A.,  {Hogg} D.~W.,  2018, \mn@doi [\apj] {10.3847/1538-4357/aae4da},
  \href {https://ui.adsabs.harvard.edu/abs/2018ApJ...867..101B} {867, 101}

\bibitem[\protect\citeauthoryear{{Bonaca} et~al.,}{{Bonaca}
  et~al.}{2021}]{bonaca2021}
{Bonaca} A.,  et~al., 2021, \mn@doi [\apjl] {10.3847/2041-8213/abeaa9}, \href
  {https://ui.adsabs.harvard.edu/abs/2021ApJ...909L..26B} {909, L26}

\bibitem[\protect\citeauthoryear{{Boubert} \& {Everall}}{{Boubert} \&
  {Everall}}{2020}]{Boubert2020B}
{Boubert} D.,  {Everall} A.,  2020, \mn@doi [\mnras] {10.1093/mnras/staa2305},
  \href {https://ui.adsabs.harvard.edu/abs/2020MNRAS.497.4246B} {497, 4246}

\bibitem[\protect\citeauthoryear{{Bovy}}{{Bovy}}{2014}]{bovy2014}
{Bovy} J.,  2014, \mn@doi [\apj] {10.1088/0004-637X/795/1/95}, \href
  {https://ui.adsabs.harvard.edu/abs/2014ApJ...795...95B} {795, 95}

\bibitem[\protect\citeauthoryear{{Bovy}}{{Bovy}}{2015}]{galpy}
{Bovy} J.,  2015, \mn@doi [\apjs] {10.1088/0067-0049/216/2/29}, \href
  {https://ui.adsabs.harvard.edu/abs/2015ApJS..216...29B} {216, 29}

\bibitem[\protect\citeauthoryear{{Bovy}, {Bahmanyar}, {Fritz}  \&
  {Kallivayalil}}{{Bovy} et~al.}{2016}]{bovy2016}
{Bovy} J.,  {Bahmanyar} A.,  {Fritz} T.~K.,   {Kallivayalil} N.,  2016, \mn@doi
  [\apj] {10.3847/1538-4357/833/1/31}, \href
  {https://ui.adsabs.harvard.edu/abs/2016ApJ...833...31B} {833, 31}

\bibitem[\protect\citeauthoryear{{Bryan} \& {Norman}}{{Bryan} \&
  {Norman}}{1998}]{Bryan+1998}
{Bryan} G.~L.,  {Norman} M.~L.,  1998, \mn@doi [\apj] {10.1086/305262}, \href
  {https://ui.adsabs.harvard.edu/abs/1998ApJ...495...80B} {495, 80}

\bibitem[\protect\citeauthoryear{{Bullock} \& {Johnston}}{{Bullock} \&
  {Johnston}}{2005}]{bullock2005}
{Bullock} J.~S.,  {Johnston} K.~V.,  2005, \mn@doi [\apj] {10.1086/497422},
  \href {https://ui.adsabs.harvard.edu/abs/2005ApJ...635..931B} {635, 931}

\bibitem[\protect\citeauthoryear{{Carlberg}}{{Carlberg}}{2020}]{carlberg2020}
{Carlberg} R.~G.,  2020, \mn@doi [\apj] {10.3847/1538-4357/ab61f0}, \href
  {https://ui.adsabs.harvard.edu/abs/2020ApJ...889..107C} {889, 107}

\bibitem[\protect\citeauthoryear{{Carlberg} \& {Grillmair}}{{Carlberg} \&
  {Grillmair}}{2013}]{carlberg2013}
{Carlberg} R.~G.,  {Grillmair} C.~J.,  2013, \mn@doi [\apj]
  {10.1088/0004-637X/768/2/171}, \href
  {https://ui.adsabs.harvard.edu/abs/2013ApJ...768..171C} {768, 171}

\bibitem[\protect\citeauthoryear{Carpenter et~al.,}{Carpenter
  et~al.}{2017}]{Carpenter2017}
Carpenter B.,  et~al., 2017, \mn@doi [Journal of Statistical Software]
  {10.18637/jss.v076.i01}, 76, 1–32

\bibitem[\protect\citeauthoryear{{Chabrier}}{{Chabrier}}{2005}]{chabrier2005}
{Chabrier} G.,  2005, in {Corbelli} E.,  {Palla} F.,   {Zinnecker} H.,  eds,
  Astrophysics and Space Science Library Vol. 327, The Initial Mass Function 50
  Years Later. p.~41 (\mn@eprint {arXiv} {astro-ph/0409465}),
  \mn@doi{10.1007/978-1-4020-3407-7\_5}

\bibitem[\protect\citeauthoryear{{Choi}, {Dotter}, {Conroy}, {Cantiello},
  {Paxton}  \& {Johnson}}{{Choi} et~al.}{2016}]{choi2016}
{Choi} J.,  {Dotter} A.,  {Conroy} C.,  {Cantiello} M.,  {Paxton} B.,
  {Johnson} B.~D.,  2016, \mn@doi [\apj] {10.3847/0004-637X/823/2/102}, \href
  {https://ui.adsabs.harvard.edu/abs/2016ApJ...823..102C} {823, 102}

\bibitem[\protect\citeauthoryear{{Clementini} et~al.,}{{Clementini}
  et~al.}{2019}]{Clementini2019}
{Clementini} G.,  et~al., 2019, \mn@doi [\aap] {10.1051/0004-6361/201833374},
  \href {https://ui.adsabs.harvard.edu/abs/2019A&A...622A..60C} {622, A60}

\bibitem[\protect\citeauthoryear{{Conroy}, {Naidu}, {Garavito-Camargo},
  {Besla}, {Zaritsky}, {Bonaca}  \& {Johnson}}{{Conroy}
  et~al.}{2021}]{Conroy:2021}
{Conroy} C.,  {Naidu} R.~P.,  {Garavito-Camargo} N.,  {Besla} G.,  {Zaritsky}
  D.,  {Bonaca} A.,   {Johnson} B.~D.,  2021, \mn@doi [\nat]
  {10.1038/s41586-021-03385-7}, \href
  {https://ui.adsabs.harvard.edu/abs/2021Natur.592..534C} {592, 534}

\bibitem[\protect\citeauthoryear{{Cooper} et~al.,}{{Cooper}
  et~al.}{2022}]{cooper2022}
{Cooper} A.~P.,  et~al., 2022, arXiv e-prints, \href
  {https://ui.adsabs.harvard.edu/abs/2022arXiv220808514C} {p. arXiv:2208.08514}

\bibitem[\protect\citeauthoryear{{Correa Magnus} \& {Vasiliev}}{{Correa Magnus}
  \& {Vasiliev}}{2022}]{CorreaMagnus21}
{Correa Magnus} L.,  {Vasiliev} E.,  2022, \mn@doi [\mnras]
  {10.1093/mnras/stab3726}, \href
  {https://ui.adsabs.harvard.edu/abs/2022MNRAS.511.2610C} {511, 2610}

\bibitem[\protect\citeauthoryear{{Cui} et~al.,}{{Cui} et~al.}{2012}]{cui2012}
{Cui} X.-Q.,  et~al., 2012, \mn@doi [Research in Astronomy and Astrophysics]
  {10.1088/1674-4527/12/9/003}, \href
  {https://ui.adsabs.harvard.edu/abs/2012RAA....12.1197C} {12, 1197}

\bibitem[\protect\citeauthoryear{{Cullinane} et~al.,}{{Cullinane}
  et~al.}{2020}]{Cullinane+2020}
{Cullinane} L.~R.,  et~al., 2020, \mn@doi [\mnras] {10.1093/mnras/staa2048},
  \href {https://ui.adsabs.harvard.edu/abs/2020MNRAS.497.3055C} {497, 3055}

\bibitem[\protect\citeauthoryear{{\VAN{De Boer}{de}{de} }{Boer}, {Erkal}  \&
  {Gieles}}{{\VAN{De Boer}{de}{de} }{Boer} et~al.}{2020}]{deBoer:2020}
{\VAN{De Boer}{de}{de} }{Boer} T.~J.~L.,  {Erkal} D.,   {Gieles} M.,  2020,
  \mn@doi [\mnras] {10.1093/mnras/staa917}, \href
  {https://ui.adsabs.harvard.edu/abs/2020MNRAS.494.5315D} {494, 5315}

\bibitem[\protect\citeauthoryear{{\VAN{De Jong}{de}{de}}~{Jong}
  et~al.,}{{\VAN{De Jong}{de}{de}}~{Jong} et~al.}{2019}]{4most}
{\VAN{De Jong}{de}{de}}~{Jong} R.~S.,  et~al., 2019, \mn@doi [The Messenger]
  {10.18727/0722-6691/5117}, \href
  {https://ui.adsabs.harvard.edu/abs/2019Msngr.175....3D} {175, 3}

\bibitem[\protect\citeauthoryear{{Deason} et~al.,}{{Deason}
  et~al.}{2021}]{Deason:21}
{Deason} A.~J.,  et~al., 2021, \mn@doi [\mnras] {10.1093/mnras/staa3984}, \href
  {https://ui.adsabs.harvard.edu/abs/2021MNRAS.501.5964D} {501, 5964}

\bibitem[\protect\citeauthoryear{{Dehnen} \& {Binney}}{{Dehnen} \&
  {Binney}}{1998}]{DehnenBinney1998}
{Dehnen} W.,  {Binney} J.,  1998, \mn@doi [\mnras]
  {10.1046/j.1365-8711.1998.01282.x}, \href
  {https://ui.adsabs.harvard.edu/abs/1998MNRAS.294..429D} {294, 429}

\bibitem[\protect\citeauthoryear{{Dillamore}, {Belokurov}, {Evans}  \&
  {Price-Whelan}}{{Dillamore} et~al.}{2022}]{Dillamore+2022}
{Dillamore} A.~M.,  {Belokurov} V.,  {Evans} N.~W.,   {Price-Whelan} A.~M.,
  2022, \mn@doi [\mnras] {10.1093/mnras/stac2311}, \href
  {https://ui.adsabs.harvard.edu/abs/2022MNRAS.516.1685D} {516, 1685}

\bibitem[\protect\citeauthoryear{{Dotter}}{{Dotter}}{2016}]{dotter2016}
{Dotter} A.,  2016, \mn@doi [\apjs] {10.3847/0067-0049/222/1/8}, \href
  {https://ui.adsabs.harvard.edu/abs/2016ApJS..222....8D} {222, 8}

\bibitem[\protect\citeauthoryear{{Drimmel} \& {Poggio}}{{Drimmel} \&
  {Poggio}}{2018}]{Drimmel2018}
{Drimmel} R.,  {Poggio} E.,  2018, \mn@doi [Research Notes of the American
  Astronomical Society] {10.3847/2515-5172/aaef8b}, \href
  {https://ui.adsabs.harvard.edu/abs/2018RNAAS...2..210D} {2, 210}

\bibitem[\protect\citeauthoryear{{Erkal}, {Sanders}  \& {Belokurov}}{{Erkal}
  et~al.}{2016a}]{Erkal:2016}
{Erkal} D.,  {Sanders} J.~L.,   {Belokurov} V.,  2016a, \mn@doi [\mnras]
  {10.1093/mnras/stw1400}, \href
  {https://ui.adsabs.harvard.edu/abs/2016MNRAS.461.1590E} {461, 1590}

\bibitem[\protect\citeauthoryear{{Erkal}, {Belokurov}, {Bovy}  \&
  {Sanders}}{{Erkal} et~al.}{2016b}]{Erkal+2016}
{Erkal} D.,  {Belokurov} V.,  {Bovy} J.,   {Sanders} J.~L.,  2016b, \mn@doi
  [\mnras] {10.1093/mnras/stw1957}, \href
  {https://ui.adsabs.harvard.edu/abs/2016MNRAS.463..102E} {463, 102}

\bibitem[\protect\citeauthoryear{{Erkal}, {Koposov}  \& {Belokurov}}{{Erkal}
  et~al.}{2017}]{Erkal2017}
{Erkal} D.,  {Koposov} S.~E.,   {Belokurov} V.,  2017, \mn@doi [\mnras]
  {10.1093/mnras/stx1208}, \href
  {https://ui.adsabs.harvard.edu/abs/2017MNRAS.470...60E} {470, 60}

\bibitem[\protect\citeauthoryear{{Erkal} et~al.,}{{Erkal}
  et~al.}{2019}]{Erkal2019}
{Erkal} D.,  et~al., 2019, \mn@doi [\mnras] {10.1093/mnras/stz1371}, \href
  {https://ui.adsabs.harvard.edu/abs/2019MNRAS.487.2685E} {487, 2685}

\bibitem[\protect\citeauthoryear{{Erkal} et~al.,}{{Erkal}
  et~al.}{2021}]{Erkal:2021}
{Erkal} D.,  et~al., 2021, \mn@doi [\mnras] {10.1093/mnras/stab1828}, \href
  {https://ui.adsabs.harvard.edu/abs/2021MNRAS.506.2677E} {506, 2677}

\bibitem[\protect\citeauthoryear{{Fardal}, {Huang}  \& {Weinberg}}{{Fardal}
  et~al.}{2015}]{fardal2015}
{Fardal} M.~A.,  {Huang} S.,   {Weinberg} M.~D.,  2015, \mn@doi [\mnras]
  {10.1093/mnras/stv1198}, \href
  {https://ui.adsabs.harvard.edu/abs/2015MNRAS.452..301F} {452, 301}

\bibitem[\protect\citeauthoryear{{Fellhauer} et~al.,}{{Fellhauer}
  et~al.}{2007}]{fellhauer2007}
{Fellhauer} M.,  et~al., 2007, \mn@doi [\mnras]
  {10.1111/j.1365-2966.2006.11404.x}, \href
  {https://ui.adsabs.harvard.edu/abs/2007MNRAS.375.1171F} {375, 1171}

\bibitem[\protect\citeauthoryear{{Foreman-Mackey}, {Hogg}, {Lang}  \&
  {Goodman}}{{Foreman-Mackey} et~al.}{2013}]{emcee}
{Foreman-Mackey} D.,  {Hogg} D.~W.,  {Lang} D.,   {Goodman} J.,  2013, \mn@doi
  [\pasp] {10.1086/670067}, \href
  {https://ui.adsabs.harvard.edu/abs/2013PASP..125..306F} {125, 306}

\bibitem[\protect\citeauthoryear{{Gaia Collaboration} et~al.,}{{Gaia
  Collaboration} et~al.}{2018}]{GaiaDR2sats}
{Gaia Collaboration} et~al., 2018, \mn@doi [\aap]
  {10.1051/0004-6361/201832698}, \href
  {https://ui.adsabs.harvard.edu/abs/2018A&A...616A..12G} {616, A12}

\bibitem[\protect\citeauthoryear{{Gaia Collaboration} et~al.,}{{Gaia
  Collaboration} et~al.}{2021}]{gaia_edr3}
{Gaia Collaboration} et~al., 2021, \mn@doi [\aap]
  {10.1051/0004-6361/202039657}, \href
  {https://ui.adsabs.harvard.edu/abs/2021A&A...649A...1G} {649, A1}

\bibitem[\protect\citeauthoryear{{Garavito-Camargo}, {Besla}, {Laporte},
  {Johnston}, {G{\'o}mez}  \& {Watkins}}{{Garavito-Camargo}
  et~al.}{2019}]{garavito2019}
{Garavito-Camargo} N.,  {Besla} G.,  {Laporte} C. F.~P.,  {Johnston} K.~V.,
  {G{\'o}mez} F.~A.,   {Watkins} L.~L.,  2019, \mn@doi [\apj]
  {10.3847/1538-4357/ab32eb}, \href
  {https://ui.adsabs.harvard.edu/abs/2019ApJ...884...51G} {884, 51}

\bibitem[\protect\citeauthoryear{{Gibbons}, {Belokurov}  \& {Evans}}{{Gibbons}
  et~al.}{2014}]{Gibbons2014}
{Gibbons} S.~L.~J.,  {Belokurov} V.,   {Evans} N.~W.,  2014, \mn@doi [\mnras]
  {10.1093/mnras/stu1986}, \href
  {https://ui.adsabs.harvard.edu/abs/2014MNRAS.445.3788G} {445, 3788}

\bibitem[\protect\citeauthoryear{{G{\'o}mez}, {Besla}, {Carpintero},
  {Villalobos}, {O'Shea}  \& {Bell}}{{G{\'o}mez} et~al.}{2015}]{Gomez+2015}
{G{\'o}mez} F.~A.,  {Besla} G.,  {Carpintero} D.~D.,  {Villalobos} {\'A}.,
  {O'Shea} B.~W.,   {Bell} E.~F.,  2015, \mn@doi [\apj]
  {10.1088/0004-637X/802/2/128}, \href
  {https://ui.adsabs.harvard.edu/abs/2015ApJ...802..128G} {802, 128}

\bibitem[\protect\citeauthoryear{{Gravity Collaboration} et~al.,}{{Gravity
  Collaboration} et~al.}{2018}]{Abuter2018}
{Gravity Collaboration} et~al., 2018, \mn@doi [\aap]
  {10.1051/0004-6361/201833718}, \href
  {https://ui.adsabs.harvard.edu/abs/2018A&A...615L..15G} {615, L15}

\bibitem[\protect\citeauthoryear{{Grillmair} \& {Dionatos}}{{Grillmair} \&
  {Dionatos}}{2006}]{grillmair2006}
{Grillmair} C.~J.,  {Dionatos} O.,  2006, \mn@doi [\apjl] {10.1086/505111},
  \href {https://ui.adsabs.harvard.edu/abs/2006ApJ...643L..17G} {643, L17}

\bibitem[\protect\citeauthoryear{{Grillmair}, {Hetherington}, {Carlberg}  \&
  {Willman}}{{Grillmair} et~al.}{2015}]{grillmair2015}
{Grillmair} C.~J.,  {Hetherington} L.,  {Carlberg} R.~G.,   {Willman} B.,
  2015, \mn@doi [\apjl] {10.1088/2041-8205/812/2/L26}, \href
  {https://ui.adsabs.harvard.edu/abs/2015ApJ...812L..26G} {812, L26}

\bibitem[\protect\citeauthoryear{{Harbeck} et~al.,}{{Harbeck}
  et~al.}{2001}]{Harbeck2001}
{Harbeck} D.,  et~al., 2001, \mn@doi [\aj] {10.1086/324232}, \href
  {https://ui.adsabs.harvard.edu/abs/2001AJ....122.3092H} {122, 3092}

\bibitem[\protect\citeauthoryear{Harris et~al.,}{Harris
  et~al.}{2020}]{numpy2020}
Harris C.~R.,  et~al., 2020, \mn@doi [Nature] {10.1038/s41586-020-2649-2}, 585,
  357

\bibitem[\protect\citeauthoryear{{Hattori}, {Erkal}  \& {Sanders}}{{Hattori}
  et~al.}{2016}]{hattori2016}
{Hattori} K.,  {Erkal} D.,   {Sanders} J.~L.,  2016, \mn@doi [\mnras]
  {10.1093/mnras/stw1006}, \href
  {https://ui.adsabs.harvard.edu/abs/2016MNRAS.460..497H} {460, 497}

\bibitem[\protect\citeauthoryear{{Hayes} et~al.,}{{Hayes}
  et~al.}{2020}]{Hayes2020}
{Hayes} C.~R.,  et~al., 2020, \mn@doi [\apj] {10.3847/1538-4357/ab62ad}, \href
  {https://ui.adsabs.harvard.edu/abs/2020ApJ...889...63H} {889, 63}

\bibitem[\protect\citeauthoryear{{Helmi}}{{Helmi}}{2020}]{Helmi2020}
{Helmi} A.,  2020, \mn@doi [\araa] {10.1146/annurev-astro-032620-021917}, \href
  {https://ui.adsabs.harvard.edu/abs/2020ARA&A..58..205H} {58, 205}

\bibitem[\protect\citeauthoryear{{Helmi}, {White}, {de Zeeuw}  \&
  {Zhao}}{{Helmi} et~al.}{1999}]{Helmi1999}
{Helmi} A.,  {White} S. D.~M.,  {de Zeeuw} P.~T.,   {Zhao} H.,  1999, \mn@doi
  [\nat] {10.1038/46980}, \href
  {https://ui.adsabs.harvard.edu/abs/1999Natur.402...53H} {402, 53}

\bibitem[\protect\citeauthoryear{{Hendel} et~al.,}{{Hendel}
  et~al.}{2018}]{Hendel2018}
{Hendel} D.,  et~al., 2018, \mn@doi [\mnras] {10.1093/mnras/sty1455}, \href
  {https://ui.adsabs.harvard.edu/abs/2018MNRAS.479..570H} {479, 570}

\bibitem[\protect\citeauthoryear{{Hernquist}}{{Hernquist}}{1990}]{Hernquist1990}
{Hernquist} L.,  1990, \mn@doi [\apj] {10.1086/168845}, \href
  {https://ui.adsabs.harvard.edu/abs/1990ApJ...356..359H} {356, 359}

\bibitem[\protect\citeauthoryear{{Holl} et~al.,}{{Holl}
  et~al.}{2018}]{Holl2018}
{Holl} B.,  et~al., 2018, \mn@doi [\aap] {10.1051/0004-6361/201832892}, \href
  {https://ui.adsabs.harvard.edu/abs/2018A&A...618A..30H} {618, A30}

\bibitem[\protect\citeauthoryear{Hunter}{Hunter}{2007}]{matplotlib2007}
Hunter J.~D.,  2007, \mn@doi [Computing in Science \& Engineering]
  {10.1109/MCSE.2007.55}, 9, 90

\bibitem[\protect\citeauthoryear{{Husser}, {Wende-von Berg}, {Dreizler},
  {Homeier}, {Reiners}, {Barman}  \& {Hauschildt}}{{Husser}
  et~al.}{2013}]{Husser2013}
{Husser} T.~O.,  {Wende-von Berg} S.,  {Dreizler} S.,  {Homeier} D.,  {Reiners}
  A.,  {Barman} T.,   {Hauschildt} P.~H.,  2013, \mn@doi [\aap]
  {10.1051/0004-6361/201219058}, \href
  {https://ui.adsabs.harvard.edu/abs/2013A&A...553A...6H} {553, A6}

\bibitem[\protect\citeauthoryear{{Hyde} et~al.,}{{Hyde}
  et~al.}{2015}]{Hyde2015}
{Hyde} E.~A.,  et~al., 2015, \mn@doi [\apj] {10.1088/0004-637X/805/2/189},
  \href {https://ui.adsabs.harvard.edu/abs/2015ApJ...805..189H} {805, 189}

\bibitem[\protect\citeauthoryear{{Ibata}, {Gilmore}  \& {Irwin}}{{Ibata}
  et~al.}{1994}]{ibata1994}
{Ibata} R.~A.,  {Gilmore} G.,   {Irwin} M.~J.,  1994, \mn@doi [\nat]
  {10.1038/370194a0}, \href
  {https://ui.adsabs.harvard.edu/abs/1994Natur.370..194I} {370, 194}

\bibitem[\protect\citeauthoryear{{Ibata}, {Lewis}, {Irwin}  \& {Quinn}}{{Ibata}
  et~al.}{2002}]{ibata2002}
{Ibata} R.~A.,  {Lewis} G.~F.,  {Irwin} M.~J.,   {Quinn} T.,  2002, \mn@doi
  [\mnras] {10.1046/j.1365-8711.2002.05358.x}, \href
  {https://ui.adsabs.harvard.edu/abs/2002MNRAS.332..915I} {332, 915}

\bibitem[\protect\citeauthoryear{{Iorio} \& {Belokurov}}{{Iorio} \&
  {Belokurov}}{2021}]{Iorio2021}
{Iorio} G.,  {Belokurov} V.,  2021, \mn@doi [\mnras] {10.1093/mnras/stab005},
  \href {https://ui.adsabs.harvard.edu/abs/2021MNRAS.502.5686I} {502, 5686}

\bibitem[\protect\citeauthoryear{{Jethwa}, {Erkal}  \& {Belokurov}}{{Jethwa}
  et~al.}{2016}]{Jethwa+2016}
{Jethwa} P.,  {Erkal} D.,   {Belokurov} V.,  2016, \mn@doi [\mnras]
  {10.1093/mnras/stw1343}, \href
  {https://ui.adsabs.harvard.edu/abs/2016MNRAS.461.2212J} {461, 2212}

\bibitem[\protect\citeauthoryear{{Ji} et~al.,}{{Ji} et~al.}{2021}]{Ji+2021}
{Ji} A.~P.,  et~al., 2021, \mn@doi [\apj] {10.3847/1538-4357/ac1869}, \href
  {https://ui.adsabs.harvard.edu/abs/2021ApJ...921...32J} {921, 32}

\bibitem[\protect\citeauthoryear{{Jin} et~al.,}{{Jin} et~al.}{2022}]{jin2022}
{Jin} S.,  et~al., 2022, \mn@doi [arXiv e-prints] {10.48550/arXiv.2212.03981},
  \href {https://ui.adsabs.harvard.edu/abs/2022arXiv221203981J} {p.
  arXiv:2212.03981}

\bibitem[\protect\citeauthoryear{{Johnston}, {Zhao}, {Spergel}  \&
  {Hernquist}}{{Johnston} et~al.}{1999}]{johnston1999}
{Johnston} K.~V.,  {Zhao} H.,  {Spergel} D.~N.,   {Hernquist} L.,  1999,
  \mn@doi [\apjl] {10.1086/311876}, \href
  {https://ui.adsabs.harvard.edu/abs/1999ApJ...512L.109J} {512, L109}

\bibitem[\protect\citeauthoryear{{Johnston}, {Spergel}  \& {Haydn}}{{Johnston}
  et~al.}{2002}]{johnston2002}
{Johnston} K.~V.,  {Spergel} D.~N.,   {Haydn} C.,  2002, \mn@doi [\apj]
  {10.1086/339791}, \href
  {https://ui.adsabs.harvard.edu/abs/2002ApJ...570..656J} {570, 656}

\bibitem[\protect\citeauthoryear{{J{\"o}nsson} et~al.,}{{J{\"o}nsson}
  et~al.}{2020}]{Jonsson2020}
{J{\"o}nsson} H.,  et~al., 2020, \mn@doi [\aj] {10.3847/1538-3881/aba592},
  \href {https://ui.adsabs.harvard.edu/abs/2020AJ....160..120J} {160, 120}

\bibitem[\protect\citeauthoryear{{Kallivayalil}, {van der Marel}, {Besla},
  {Anderson}  \& {Alcock}}{{Kallivayalil} et~al.}{2013}]{Kallivayalil+2013}
{Kallivayalil} N.,  {van der Marel} R.~P.,  {Besla} G.,  {Anderson} J.,
  {Alcock} C.,  2013, \mn@doi [\apj] {10.1088/0004-637X/764/2/161}, \href
  {https://ui.adsabs.harvard.edu/abs/2013ApJ...764..161K} {764, 161}

\bibitem[\protect\citeauthoryear{{Kirby}, {Cohen}, {Guhathakurta}, {Cheng},
  {Bullock}  \& {Gallazzi}}{{Kirby} et~al.}{2013}]{kirby2013}
{Kirby} E.~N.,  {Cohen} J.~G.,  {Guhathakurta} P.,  {Cheng} L.,  {Bullock}
  J.~S.,   {Gallazzi} A.,  2013, \mn@doi [\apj] {10.1088/0004-637X/779/2/102},
  \href {https://ui.adsabs.harvard.edu/abs/2013ApJ...779..102K} {779, 102}

\bibitem[\protect\citeauthoryear{{Koleva}, {Prugniel}, {De Rijcke}  \&
  {Zeilinger}}{{Koleva} et~al.}{2011}]{koleva2011}
{Koleva} M.,  {Prugniel} P.,  {De Rijcke} S.,   {Zeilinger} W.~W.,  2011,
  \mn@doi [\mnras] {10.1111/j.1365-2966.2011.19057.x}, \href
  {https://ui.adsabs.harvard.edu/abs/2011MNRAS.417.1643K} {417, 1643}

\bibitem[\protect\citeauthoryear{Koposov}{Koposov}{2021}]{minimint2021}
Koposov S.,  2021, segasai/minimint: Minimint 0.3.0,
  \mn@doi{10.5281/zenodo.5610692}, \url
  {https://doi.org/10.5281/zenodo.5610692}

\bibitem[\protect\citeauthoryear{Koposov}{Koposov}{2022b}]{sqlutilpy}
Koposov S.,  2022b, segasai/sqlutilpy: sqlutilpy v0.19.0,
  \mn@doi{10.5281/zenodo.6867957}, \url
  {https://doi.org/10.5281/zenodo.6867957}

\bibitem[\protect\citeauthoryear{Koposov}{Koposov}{2022a}]{stan_splines}
Koposov S.,  2022a, segasai/stan-splines: v1.0.0,
  \mn@doi{10.5281/zenodo.7193910}, \url
  {https://doi.org/10.5281/zenodo.7193910}

\bibitem[\protect\citeauthoryear{{Koposov} \& {Bartunov}}{{Koposov} \&
  {Bartunov}}{2006}]{q3c2006}
{Koposov} S.,  {Bartunov} O.,  2006, in {Gabriel} C.,  {Arviset} C.,  {Ponz}
  D.,   {Enrique} S.,  eds,  Astronomical Society of the Pacific Conference
  Series Vol. 351, Astronomical Data Analysis Software and Systems XV. p.~735

\bibitem[\protect\citeauthoryear{{Koposov}, {Rix}  \& {Hogg}}{{Koposov}
  et~al.}{2010}]{Koposov2010}
{Koposov} S.~E.,  {Rix} H.-W.,   {Hogg} D.~W.,  2010, \mn@doi [\apj]
  {10.1088/0004-637X/712/1/260}, \href
  {https://ui.adsabs.harvard.edu/abs/2010ApJ...712..260K} {712, 260}

\bibitem[\protect\citeauthoryear{{Koposov}, {Irwin}, {Belokurov},
  {Gonzalez-Solares}, {Yoldas}, {Lewis}, {Metcalfe}  \& {Shanks}}{{Koposov}
  et~al.}{2014}]{koposov2014}
{Koposov} S.~E.,  {Irwin} M.,  {Belokurov} V.,  {Gonzalez-Solares} E.,
  {Yoldas} A.~K.,  {Lewis} J.,  {Metcalfe} N.,   {Shanks} T.,  2014, \mn@doi
  [\mnras] {10.1093/mnrasl/slu060}, \href
  {https://ui.adsabs.harvard.edu/abs/2014MNRAS.442L..85K} {442, L85}

\bibitem[\protect\citeauthoryear{{Koposov} et~al.,}{{Koposov}
  et~al.}{2019}]{Koposov2019}
{Koposov} S.~E.,  et~al., 2019, \mn@doi [\mnras] {10.1093/mnras/stz457}, \href
  {https://ui.adsabs.harvard.edu/abs/2019MNRAS.485.4726K} {485, 4726}

\bibitem[\protect\citeauthoryear{{Koppelman}, {Helmi}  \&
  {Veljanoski}}{{Koppelman} et~al.}{2018}]{Koppelman2018}
{Koppelman} H.,  {Helmi} A.,   {Veljanoski} J.,  2018, \mn@doi [\apjl]
  {10.3847/2041-8213/aac882}, \href
  {https://ui.adsabs.harvard.edu/abs/2018ApJ...860L..11K} {860, L11}

\bibitem[\protect\citeauthoryear{{Krishnarao} et~al.,}{{Krishnarao}
  et~al.}{2022}]{krishnarao2022}
{Krishnarao} D.,  et~al., 2022, \mn@doi [\nat] {10.1038/s41586-022-05090-5},
  \href {https://ui.adsabs.harvard.edu/abs/2022Natur.609..915K} {609, 915}

\bibitem[\protect\citeauthoryear{{K{\"u}pper}, {Balbinot}, {Bonaca},
  {Johnston}, {Hogg}, {Kroupa}  \& {Santiago}}{{K{\"u}pper}
  et~al.}{2015}]{Kuepper:2015}
{K{\"u}pper} A. H.~W.,  {Balbinot} E.,  {Bonaca} A.,  {Johnston} K.~V.,  {Hogg}
  D.~W.,  {Kroupa} P.,   {Santiago} B.~X.,  2015, \mn@doi [\apj]
  {10.1088/0004-637X/803/2/80}, \href
  {https://ui.adsabs.harvard.edu/abs/2015ApJ...803...80K} {803, 80}

\bibitem[\protect\citeauthoryear{{Lee} et~al.,}{{Lee} et~al.}{2008a}]{Lee2008a}
{Lee} Y.~S.,  et~al., 2008a, \mn@doi [\aj] {10.1088/0004-6256/136/5/2022},
  \href {https://ui.adsabs.harvard.edu/abs/2008AJ....136.2022L} {136, 2022}

\bibitem[\protect\citeauthoryear{{Lee} et~al.,}{{Lee} et~al.}{2008b}]{Lee2008b}
{Lee} Y.~S.,  et~al., 2008b, \mn@doi [\aj] {10.1088/0004-6256/136/5/2050},
  \href {https://ui.adsabs.harvard.edu/abs/2008AJ....136.2050L} {136, 2050}

\bibitem[\protect\citeauthoryear{{Lewis} et~al.,}{{Lewis}
  et~al.}{2002}]{Lewis:2002}
{Lewis} I.~J.,  et~al., 2002, \mn@doi [\mnras]
  {10.1046/j.1365-8711.2002.05333.x}, \href
  {https://ui.adsabs.harvard.edu/#abs/2002MNRAS.333..279L} {333, 279}

\bibitem[\protect\citeauthoryear{{Li} \& {S5 Collaboration}}{{Li} \& {S5
  Collaboration}}{2021}]{S5_DR1}
{Li} T.,  {S5 Collaboration} 2021, {Southern Stellar Stream Spectroscopic
  Survey: The First Public Data Release}, \mn@doi{10.5281/zenodo.4695135}, \url
  {https://doi.org/10.5281/zenodo.4695135}

\bibitem[\protect\citeauthoryear{{Li} et~al.,}{{Li} et~al.}{2019}]{Li2019}
{Li} T.~S.,  et~al., 2019, \mn@doi [\mnras] {10.1093/mnras/stz2731}, \href
  {https://ui.adsabs.harvard.edu/abs/2019MNRAS.490.3508L} {490, 3508}

\bibitem[\protect\citeauthoryear{{Li} et~al.,}{{Li} et~al.}{2021}]{li2021}
{Li} T.~S.,  et~al., 2021, \mn@doi [\apj] {10.3847/1538-4357/abeb18}, \href
  {https://ui.adsabs.harvard.edu/abs/2021ApJ...911..149L} {911, 149}

\bibitem[\protect\citeauthoryear{{Li} et~al.,}{{Li}
  et~al.}{2022}]{S5_12streams}
{Li} T.~S.,  et~al., 2022, \mn@doi [\apj] {10.3847/1538-4357/ac46d3}, \href
  {https://ui.adsabs.harvard.edu/abs/2022ApJ...928...30L} {928, 30}

\bibitem[\protect\citeauthoryear{{Lilleengen} et~al.,}{{Lilleengen}
  et~al.}{2023}]{lilleengen+2022}
{Lilleengen} S.,  et~al., 2023, \mn@doi [\mnras] {10.1093/mnras/stac3108},
  \href {https://ui.adsabs.harvard.edu/abs/2023MNRAS.518..774L} {518, 774}

\bibitem[\protect\citeauthoryear{{Malhan}, {Ibata}  \& {Martin}}{{Malhan}
  et~al.}{2018}]{malhan2018}
{Malhan} K.,  {Ibata} R.~A.,   {Martin} N.~F.,  2018, \mn@doi [\mnras]
  {10.1093/mnras/sty2474}, \href
  {https://ui.adsabs.harvard.edu/abs/2018MNRAS.481.3442M} {481, 3442}

\bibitem[\protect\citeauthoryear{{Mateu}}{{Mateu}}{2022}]{mateu2022}
{Mateu} C.,  2022, arXiv e-prints, \href
  {https://ui.adsabs.harvard.edu/abs/2022arXiv220410326M} {p. arXiv:2204.10326}

\bibitem[\protect\citeauthoryear{{McConnachie}}{{McConnachie}}{2012}]{Mcconnachie2012}
{McConnachie} A.~W.,  2012, \mn@doi [\aj] {10.1088/0004-6256/144/1/4}, \href
  {https://ui.adsabs.harvard.edu/abs/2012AJ....144....4M} {144, 4}

\bibitem[\protect\citeauthoryear{{McMillan}}{{McMillan}}{2017}]{McMillan2017}
{McMillan} P.~J.,  2017, \mn@doi [\mnras] {10.1093/mnras/stw2759}, \href
  {https://ui.adsabs.harvard.edu/abs/2017MNRAS.465...76M} {465, 76}

\bibitem[\protect\citeauthoryear{{Mendelsohn}, {Newberg}, {Shelton}, {Widrow},
  {Thompson}  \& {Grillmair}}{{Mendelsohn} et~al.}{2022}]{mendelsohn2022}
{Mendelsohn} E.~J.,  {Newberg} H.~J.,  {Shelton} S.,  {Widrow} L.~M.,
  {Thompson} J.~M.,   {Grillmair} C.~J.,  2022, \mn@doi [\apj]
  {10.3847/1538-4357/ac498a}, \href
  {https://ui.adsabs.harvard.edu/abs/2022ApJ...926..106M} {926, 106}

\bibitem[\protect\citeauthoryear{{Mercado} et~al.,}{{Mercado}
  et~al.}{2021}]{mercado2021}
{Mercado} F.~J.,  et~al., 2021, \mn@doi [\mnras] {10.1093/mnras/staa3958},
  \href {https://ui.adsabs.harvard.edu/abs/2021MNRAS.501.5121M} {501, 5121}

\bibitem[\protect\citeauthoryear{{Miyamoto} \& {Nagai}}{{Miyamoto} \&
  {Nagai}}{1975}]{Miyamoto-Nagai1975}
{Miyamoto} M.,  {Nagai} R.,  1975, \pasj, \href
  {https://ui.adsabs.harvard.edu/abs/1975PASJ...27..533M} {27, 533}

\bibitem[\protect\citeauthoryear{{Muraveva}, {Delgado}, {Clementini}, {Sarro}
  \& {Garofalo}}{{Muraveva} et~al.}{2018}]{Muraveva2018}
{Muraveva} T.,  {Delgado} H.~E.,  {Clementini} G.,  {Sarro} L.~M.,   {Garofalo}
  A.,  2018, \mn@doi [\mnras] {10.1093/mnras/sty2241}, \href
  {https://ui.adsabs.harvard.edu/abs/2018MNRAS.481.1195M} {481, 1195}

\bibitem[\protect\citeauthoryear{{Myeong}, {Evans}, {Belokurov}, {Sanders}  \&
  {Koposov}}{{Myeong} et~al.}{2018}]{Myeong2018}
{Myeong} G.~C.,  {Evans} N.~W.,  {Belokurov} V.,  {Sanders} J.~L.,   {Koposov}
  S.~E.,  2018, \mn@doi [\apjl] {10.3847/2041-8213/aab613}, \href
  {https://ui.adsabs.harvard.edu/abs/2018ApJ...856L..26M} {856, L26}

\bibitem[\protect\citeauthoryear{{Navarro}, {Frenk}  \& {White}}{{Navarro}
  et~al.}{1996}]{nfw1996}
{Navarro} J.~F.,  {Frenk} C.~S.,   {White} S. D.~M.,  1996, \mn@doi [\apj]
  {10.1086/177173}, \href
  {https://ui.adsabs.harvard.edu/abs/1996ApJ...462..563N} {462, 563}

\bibitem[\protect\citeauthoryear{{Nidever} et~al.,}{{Nidever}
  et~al.}{2015}]{Nidever2015}
{Nidever} D.~L.,  et~al., 2015, \mn@doi [\aj] {10.1088/0004-6256/150/6/173},
  \href {https://ui.adsabs.harvard.edu/abs/2015AJ....150..173N} {150, 173}

\bibitem[\protect\citeauthoryear{{Odenkirchen} et~al.,}{{Odenkirchen}
  et~al.}{2003}]{odenkirchen2003}
{Odenkirchen} M.,  et~al., 2003, \mn@doi [\aj] {10.1086/378601}, \href
  {https://ui.adsabs.harvard.edu/abs/2003AJ....126.2385O} {126, 2385}

\bibitem[\protect\citeauthoryear{{Okamoto}, {Arimoto}, {Tolstoy}, {Jablonka},
  {Irwin}, {Komiyama}, {Yamada}  \& {Onodera}}{{Okamoto}
  et~al.}{2017}]{okamoto2017}
{Okamoto} S.,  {Arimoto} N.,  {Tolstoy} E.,  {Jablonka} P.,  {Irwin} M.~J.,
  {Komiyama} Y.,  {Yamada} Y.,   {Onodera} M.,  2017, \mn@doi [\mnras]
  {10.1093/mnras/stx086}, \href
  {https://ui.adsabs.harvard.edu/abs/2017MNRAS.467..208O} {467, 208}

\bibitem[\protect\citeauthoryear{{Pace}, {Erkal}  \& {Li}}{{Pace}
  et~al.}{2022}]{Pace+2022}
{Pace} A.~B.,  {Erkal} D.,   {Li} T.~S.,  2022, \mn@doi [\apj]
  {10.3847/1538-4357/ac997b}, \href
  {https://ui.adsabs.harvard.edu/abs/2022ApJ...940..136P} {940, 136}

\bibitem[\protect\citeauthoryear{{Patrick}, {Koposov}  \& {Walker}}{{Patrick}
  et~al.}{2022}]{patrick2022}
{Patrick} J.~M.,  {Koposov} S.~E.,   {Walker} M.~G.,  2022, \mn@doi [\mnras]
  {10.1093/mnras/stac1478}, \href
  {https://ui.adsabs.harvard.edu/abs/2022MNRAS.514.1757P} {514, 1757}

\bibitem[\protect\citeauthoryear{{Pe{\~n}arrubia}, {Koposov}  \&
  {Walker}}{{Pe{\~n}arrubia} et~al.}{2012}]{penarrubia2012}
{Pe{\~n}arrubia} J.,  {Koposov} S.~E.,   {Walker} M.~G.,  2012, \mn@doi [\apj]
  {10.1088/0004-637X/760/1/2}, \href
  {https://ui.adsabs.harvard.edu/abs/2012ApJ...760....2P} {760, 2}

\bibitem[\protect\citeauthoryear{P\'erez \& Granger}{P\'erez \&
  Granger}{2007}]{ipython2007}
P\'erez F.,  Granger B.~E.,  2007, \mn@doi [Computing in Science and
  Engineering] {10.1109/MCSE.2007.53}, 9, 21

\bibitem[\protect\citeauthoryear{{Petersen} \& {Pe{\~n}arrubia}}{{Petersen} \&
  {Pe{\~n}arrubia}}{2021}]{Petersen:2021}
{Petersen} M.~S.,  {Pe{\~n}arrubia} J.,  2021, \mn@doi [Nature Astronomy]
  {10.1038/s41550-020-01254-3}, \href
  {https://ui.adsabs.harvard.edu/abs/2021NatAs...5..251P} {5, 251}

\bibitem[\protect\citeauthoryear{{Piatek}, {Pryor}  \& {Olszewski}}{{Piatek}
  et~al.}{2016}]{LeoIIpm}
{Piatek} S.,  {Pryor} C.,   {Olszewski} E.~W.,  2016, \mn@doi [\aj]
  {10.3847/0004-6256/152/6/166}, \href
  {https://ui.adsabs.harvard.edu/abs/2016AJ....152..166P} {152, 166}

\bibitem[\protect\citeauthoryear{{Pietrzy{\'n}ski} et~al.,}{{Pietrzy{\'n}ski}
  et~al.}{2019}]{Pietrzynski+2019}
{Pietrzy{\'n}ski} G.,  et~al., 2019, \mn@doi [\nat]
  {10.1038/s41586-019-0999-4}, \href
  {https://ui.adsabs.harvard.edu/abs/2019Natur.567..200P} {567, 200}

\bibitem[\protect\citeauthoryear{{Planck Collaboration} et~al.,}{{Planck
  Collaboration} et~al.}{2020}]{Planck+2020}
{Planck Collaboration} et~al., 2020, \mn@doi [\aap]
  {10.1051/0004-6361/201833910}, \href
  {https://ui.adsabs.harvard.edu/abs/2020A&A...641A...6P} {641, A6}

\bibitem[\protect\citeauthoryear{{Plummer}}{{Plummer}}{1911}]{Plummer}
{Plummer} H.~C.,  1911, \mn@doi [\mnras] {10.1093/mnras/71.5.460}, \href
  {https://ui.adsabs.harvard.edu/abs/1911MNRAS..71..460P} {71, 460}

\bibitem[\protect\citeauthoryear{{Price-Whelan}, {Sesar}, {Johnston}  \&
  {Rix}}{{Price-Whelan} et~al.}{2016}]{pricewhelan2016}
{Price-Whelan} A.~M.,  {Sesar} B.,  {Johnston} K.~V.,   {Rix} H.-W.,  2016,
  \mn@doi [\apj] {10.3847/0004-637X/824/2/104}, \href
  {https://ui.adsabs.harvard.edu/abs/2016ApJ...824..104P} {824, 104}

\bibitem[\protect\citeauthoryear{{Reid} \& {Brunthaler}}{{Reid} \&
  {Brunthaler}}{2004}]{Reid2004}
{Reid} M.~J.,  {Brunthaler} A.,  2004, \mn@doi [\apj] {10.1086/424960}, \href
  {https://ui.adsabs.harvard.edu/abs/2004ApJ...616..872R} {616, 872}

\bibitem[\protect\citeauthoryear{{Reino}, {Rossi}, {Sanderson}, {Sellentin},
  {Helmi}, {Koppelman}  \& {Sharma}}{{Reino} et~al.}{2021}]{reino2021}
{Reino} S.,  {Rossi} E.~M.,  {Sanderson} R.~E.,  {Sellentin} E.,  {Helmi} A.,
  {Koppelman} H.~H.,   {Sharma} S.,  2021, \mn@doi [\mnras]
  {10.1093/mnras/stab304}, \href
  {https://ui.adsabs.harvard.edu/abs/2021MNRAS.502.4170R} {502, 4170}

\bibitem[\protect\citeauthoryear{{Sanders} \& {Binney}}{{Sanders} \&
  {Binney}}{2013}]{sanders2013}
{Sanders} J.~L.,  {Binney} J.,  2013, \mn@doi [\mnras] {10.1093/mnras/stt806},
  \href {https://ui.adsabs.harvard.edu/abs/2013MNRAS.433.1813S} {433, 1813}

\bibitem[\protect\citeauthoryear{{Sanderson}, {Helmi}  \& {Hogg}}{{Sanderson}
  et~al.}{2015}]{Sanderson:2015}
{Sanderson} R.~E.,  {Helmi} A.,   {Hogg} D.~W.,  2015, \mn@doi [\apj]
  {10.1088/0004-637X/801/2/98}, \href
  {https://ui.adsabs.harvard.edu/abs/2015ApJ...801...98S} {801, 98}

\bibitem[\protect\citeauthoryear{{Schee}, {Stuchl{\'\i}k}  \&
  {Petr{\'a}sek}}{{Schee} et~al.}{2013}]{Schee+2013}
{Schee} J.,  {Stuchl{\'\i}k} Z.,   {Petr{\'a}sek} M.,  2013, \mn@doi [\jcap]
  {10.1088/1475-7516/2013/12/026}, \href
  {https://ui.adsabs.harvard.edu/abs/2013JCAP...12..026S} {2013, 026}

\bibitem[\protect\citeauthoryear{{Sesar} et~al.,}{{Sesar}
  et~al.}{2013}]{Sesar2013}
{Sesar} B.,  et~al., 2013, \mn@doi [\apj] {10.1088/0004-637X/776/1/26}, \href
  {https://ui.adsabs.harvard.edu/abs/2013ApJ...776...26S} {776, 26}

\bibitem[\protect\citeauthoryear{{Sesar} et~al.,}{{Sesar}
  et~al.}{2017}]{Sesar2017}
{Sesar} B.,  et~al., 2017, \mn@doi [\aj] {10.3847/1538-3881/aa661b}, \href
  {https://ui.adsabs.harvard.edu/abs/2017AJ....153..204S} {153, 204}

\bibitem[\protect\citeauthoryear{{Sharp} et~al.,}{{Sharp}
  et~al.}{2006}]{Sharp:2006}
{Sharp} R.,  et~al., 2006, in Society of Photo-Optical Instrumentation
  Engineers (SPIE) Conference Series. p. 62690G (\mn@eprint {}
  {astro-ph/0606137}), \mn@doi{10.1117/12.671022}

\bibitem[\protect\citeauthoryear{{Shipp} et~al.,}{{Shipp}
  et~al.}{2018}]{shipp2018}
{Shipp} N.,  et~al., 2018, \mn@doi [\apj] {10.3847/1538-4357/aacdab}, \href
  {https://ui.adsabs.harvard.edu/abs/2018ApJ...862..114S} {862, 114}

\bibitem[\protect\citeauthoryear{{Shipp} et~al.,}{{Shipp}
  et~al.}{2019}]{shipp2019}
{Shipp} N.,  et~al., 2019, \mn@doi [\apj] {10.3847/1538-4357/ab44bf}, \href
  {https://ui.adsabs.harvard.edu/abs/2019ApJ...885....3S} {885, 3}

\bibitem[\protect\citeauthoryear{{Shipp} et~al.,}{{Shipp}
  et~al.}{2021}]{Shipp+2021}
{Shipp} N.,  et~al., 2021, \mn@doi [\apj] {10.3847/1538-4357/ac2e93}, \href
  {https://ui.adsabs.harvard.edu/abs/2021ApJ...923..149S} {923, 149}

\bibitem[\protect\citeauthoryear{{Smolinski} et~al.,}{{Smolinski}
  et~al.}{2011}]{Smolinski2011}
{Smolinski} J.~P.,  et~al., 2011, \mn@doi [\aj] {10.1088/0004-6256/141/3/89},
  \href {https://ui.adsabs.harvard.edu/abs/2011AJ....141...89S} {141, 89}

\bibitem[\protect\citeauthoryear{{Sohn}, {Besla}, {van der Marel},
  {Boylan-Kolchin}, {Majewski}  \& {Bullock}}{{Sohn} et~al.}{2013}]{LeoIpm}
{Sohn} S.~T.,  {Besla} G.,  {van der Marel} R.~P.,  {Boylan-Kolchin} M.,
  {Majewski} S.~R.,   {Bullock} J.~S.,  2013, \mn@doi [\apj]
  {10.1088/0004-637X/768/2/139}, \href
  {https://ui.adsabs.harvard.edu/abs/2013ApJ...768..139S} {768, 139}

\bibitem[\protect\citeauthoryear{{Stringer} et~al.,}{{Stringer}
  et~al.}{2021}]{Stringer2021}
{Stringer} K.~M.,  et~al., 2021, \mn@doi [\apj] {10.3847/1538-4357/abe873},
  \href {https://ui.adsabs.harvard.edu/abs/2021ApJ...911..109S} {911, 109}

\bibitem[\protect\citeauthoryear{\VAN{Marel}{van der}{van der}~{Marel} \&
  {Kallivayalil}}{\VAN{Marel}{van der}{van der}~{Marel} \&
  {Kallivayalil}}{2014}]{vandermarel+2014}
\VAN{Marel}{van der}{van der}~{Marel} R.~P.,  {Kallivayalil} N.,  2014, \mn@doi
  [\apj] {10.1088/0004-637X/781/2/121}, \href
  {https://ui.adsabs.harvard.edu/abs/2014ApJ...781..121V} {781, 121}

\bibitem[\protect\citeauthoryear{\VAN{Marel}{van der}{van der}~{Marel},
  {Alves}, {Hardy}  \& {Suntzeff}}{\VAN{Marel}{van der}{van der}~{Marel}
  et~al.}{2002}]{vandermarel+2002}
\VAN{Marel}{van der}{van der}~{Marel} R.~P.,  {Alves} D.~R.,  {Hardy} E.,
  {Suntzeff} N.~B.,  2002, \mn@doi [\aj] {10.1086/343775}, \href
  {https://ui.adsabs.harvard.edu/abs/2002AJ....124.2639V} {124, 2639}

\bibitem[\protect\citeauthoryear{{Vasiliev}}{{Vasiliev}}{2019}]{agama2019}
{Vasiliev} E.,  2019, \mn@doi [\mnras] {10.1093/mnras/sty2672}, \href
  {https://ui.adsabs.harvard.edu/abs/2019MNRAS.482.1525V} {482, 1525}

\bibitem[\protect\citeauthoryear{{Vasiliev} \& {Baumgardt}}{{Vasiliev} \&
  {Baumgardt}}{2021}]{Vasiliev&Baumgardt2021}
{Vasiliev} E.,  {Baumgardt} H.,  2021, \mn@doi [\mnras]
  {10.1093/mnras/stab1475}, \href
  {https://ui.adsabs.harvard.edu/abs/2021MNRAS.505.5978V} {505, 5978}

\bibitem[\protect\citeauthoryear{{Vasiliev}, {Belokurov}  \&
  {Erkal}}{{Vasiliev} et~al.}{2021}]{Vasiliev+2021}
{Vasiliev} E.,  {Belokurov} V.,   {Erkal} D.,  2021, \mn@doi [\mnras]
  {10.1093/mnras/staa3673}, \href
  {https://ui.adsabs.harvard.edu/abs/2021MNRAS.501.2279V} {501, 2279}

\bibitem[\protect\citeauthoryear{Virtanen et~al.,}{Virtanen
  et~al.}{2020}]{scipy2020}
Virtanen P.,  et~al., 2020, \mn@doi [Nature Methods]
  {10.1038/s41592-019-0686-2}, \href {https://rdcu.be/b08Wh} {17, 261}

\bibitem[\protect\citeauthoryear{{Wan}, {Guglielmo}, {Lewis}, {Mackey}  \&
  {Ibata}}{{Wan} et~al.}{2020}]{Wan+20}
{Wan} Z.,  {Guglielmo} M.,  {Lewis} G.~F.,  {Mackey} D.,   {Ibata} R.~A.,
  2020, \mn@doi [\mnras] {10.1093/mnras/stz3493}, \href
  {https://ui.adsabs.harvard.edu/abs/2020MNRAS.492..782W} {492, 782}

\bibitem[\protect\citeauthoryear{{White} \& {Frenk}}{{White} \&
  {Frenk}}{1991}]{white1991}
{White} S. D.~M.,  {Frenk} C.~S.,  1991, \mn@doi [\apj] {10.1086/170483}, \href
  {https://ui.adsabs.harvard.edu/abs/1991ApJ...379...52W} {379, 52}

\bibitem[\protect\citeauthoryear{{Wu}, {Famaey}, {Gentile}, {Perets}  \&
  {Zhao}}{{Wu} et~al.}{2008}]{Wu+2008}
{Wu} X.,  {Famaey} B.,  {Gentile} G.,  {Perets} H.,   {Zhao} H.,  2008, \mn@doi
  [\mnras] {10.1111/j.1365-2966.2008.13198.x}, \href
  {https://ui.adsabs.harvard.edu/abs/2008MNRAS.386.2199W} {386, 2199}

\bibitem[\protect\citeauthoryear{{Yoon}, {Johnston}  \& {Hogg}}{{Yoon}
  et~al.}{2011}]{Yoon2011}
{Yoon} J.~H.,  {Johnston} K.~V.,   {Hogg} D.~W.,  2011, \mn@doi [\apj]
  {10.1088/0004-637X/731/1/58}, \href
  {https://ui.adsabs.harvard.edu/abs/2011ApJ...731...58Y} {731, 58}

\bibitem[\protect\citeauthoryear{{Zhao}, {Zhao}, {Chu}, {Jing}  \&
  {Deng}}{{Zhao} et~al.}{2012}]{zhao2012}
{Zhao} G.,  {Zhao} Y.-H.,  {Chu} Y.-Q.,  {Jing} Y.-P.,   {Deng} L.-C.,  2012,
  \mn@doi [Research in Astronomy and Astrophysics]
  {10.1088/1674-4527/12/7/002}, \href
  {https://ui.adsabs.harvard.edu/abs/2012RAA....12..723Z} {12, 723}

\makeatother
\end{thebibliography}

\section*{APPENDIX A}
Table~\ref{tab:spec_members} contains the list of possible spectroscopic members from \Sfive, APOGEE, SDSS and LAMOST selected based on position, proper motion, metallicity and \CHANGE{radial velocity as shown on Figure~\ref{fig:rv_pm}. The stars in that table were not filtered by colour-magnitude, but we provide a column indicating whether the star is in the CMD mask shown on Figure~\ref{fig:cmd_gaia}.}

\begin{table*}
    \centering
    \begin{tabular}{ccc ccc cccc c}
\hline
 GDR3 source\_id & $\alpha$ & $\delta$ & $\phi_1$ & $\phi_2$  & RV  &
 $V_{GSR}$ & $\sigma_V$ & $\mu_{\phi, 1}$ & $\mu_{\phi,2}$ & CMD \\
  & deg & deg & deg & deg & origin & km\,s$^{-1}$&km\,s$^{-1}$ & mas\,yr$^{-1}$  &mas\,yr$^{-1}$ & selected \\
\hline 
3888500490279502464 & 155.87168 & 15.86127 & 83.3819 & 1.9210 & S5 & 121.2 & 1.3 & 0.36 & 1.23 & 1 \\
6451721201411909760 & 318.52474 & -62.09448 & -48.9244 & 1.8850 & S5 & -203.3 & 3.8 & 2.65 & 1.23 & 1 \\
6558441247408890240 & 327.06713 & -52.19549 & -59.8363 & 2.1710 & S5 & -191.8 & 0.2 & 2.18 & 1.18 & 1 \\
\hline 
\end{tabular}

    \caption{Possible spectroscopic OC member stars from S5, APOGEE, LAMOST and SDSS datasets used in the analysis. The full table is provided in supplementary materials.\CHANGE{ The {\tt source\_id} is  Gaia DR3 identifier. The star coordinates are are given in equatorial and $\phi_1,\phi_2$ coordinate systems. We also provide the radial velocity in the GSR frame with the uncertainty as well as the survey that is the source of the radial velocity. The proper motions are provided in the $\phi_1,\phi_2$ frame. The last column indicates if the star was selected through with the CMD mask shown on Figure~\ref{fig:cmd_gaia}.}}
    \label{tab:spec_members}
\end{table*}

\section*{APPENDIX B:}
Here, we provide the table of measurements from the spline models used to extract the stream 6-D track. The full versions of these tables are also provided as {\tt FITS} files in supplementary materials, and can be used as inputs to natural cubic splines (i.e. {\tt CubicSpline} with  {\tt bc\_type='natural'} option from {\tt scipy.interpolate} in {\tt Python} programming language).
The stream distance measurements are in Table~\ref{tab:dm_meas}, while distance spread measurements are in Table~\ref{tab:dmsig_meas}. The radial velocity measurements are provided in Table~\ref{tab:rv_meas}. The stream velocity dispersion is provided in Table~\ref{tab:rvdisp_meas}. The proper motion stream track measurements are in Table~\ref{tab:pm_meas}. The intrinsic dispersions in proper motions are provided in Table~\ref{tab:pmsig_meas}. The stream track on the sky is provided in Table~\ref{tab:track_meas}. The stream width measurements are in Table~\ref{tab:width_meas}. The stream surface brightness measurements are in Table~\ref{tab:density_meas}. The values of nuisance noise parameters used for the stream modelling are given in Table~\ref{tab:nuisance}.

\begin{table}
    \centering
    \begin{tabular}{ccc}
\hline 
$\phi_1$ & $m-M$ & $\sigma_{m-M}$ \\
deg & mag & mag \\
\hline 
-85.0 & 19.42 & 0.09 \\
-70.4 & 18.33 & 0.02 \\
-55.7 & 17.65 & 0.03 \\
\hline 
\end{tabular}

    \caption{\CHANGE{Measurements of the stream distance modulus through spline models. We provide the distance modulus and it uncertainty at the set of $\phi_1$ points. The full table is included in the supplementary materials.}}
    \label{tab:dm_meas}
\end{table}

\begin{table}
    \centering
    \begin{tabular}{ccc}
\hline 
$\phi_1$ & $ \ln {\mathcal S}_{m-M}$ & $\sigma_{ \ln {\mathcal S},{m-M}}$ \\
deg & mag & mag \\
\hline 
-85.0 & -3.10 & 0.60 \\
-44.0 & -2.59 & 0.25 \\
-3.0 & -2.57 & 0.48 \\
\hline 
\end{tabular}

    \caption{\CHANGE{Measurements of the intrinsic spread of stream distance moduli through spline models.  We provide the logarithm of $m-M$ spread and its uncertainty. The  full table is included in the supplementary materials.}}
    \label{tab:dmsig_meas}
\end{table}
\begin{table}
    \centering
    \begin{tabular}{ccc}
\hline 
$\phi_1$ & $V_{GSR}$ & $\sigma_V$ \\
deg & km\,s$^{-1}$ & km\,s$^{-1}$ \\
\hline 
-80.0 & -132.4 & 5.7 \\
-70.0 & -171.8 & 2.1 \\
-60.0 & -187.8 & 1.1 \\
\hline 
\end{tabular}

    \caption{\CHANGE{Measurements of stream radial velocity track through spline models. We provide the radial velocity in the GSR frame and its uncertainty at the set of $\phi_1$ points. The full table is available in the supplementary materials.}}
    \label{tab:rv_meas}
\end{table}

\begin{table}
    \centering
    \begin{tabular}{ccc}
\hline 
$\phi_1$ & $\ln {\mathcal S}_{V}$ & $\sigma_{\ln {\mathcal S},V}$ \\
deg & &  \\
\hline 
-80.0 & 1.4 & 0.5 \\
-40.0 & 1.6 & 0.3 \\
40.0 & 1.2 & 0.2 \\
80.0 & 1.7 & 0.2 \\
120.0 & 1.9 & 0.4 \\
\hline 
\end{tabular}

    \caption{\CHANGE{Measurements of the radial velocity dispersion through spline models. We provide measurements of the natural log of the velocity dispersion in \kms and its uncertainty at the set of $\phi_1$ points. The full table is included in the supplementary materials.}}
    \label{tab:rvdisp_meas}
\end{table}

\begin{table}
    \centering
    \begin{tabular}{ccccc}
\hline 
$\phi_1$ & $\mu_{1}$ & $\sigma_{\mu,1}$& $\mu_{2}$  & $\sigma_{\mu,2}$ \\
deg & mas\,yr$^{-1}$ & mas\,yr$^{-1}$ & mas\,yr$^{-1}$ & mas\,yr$^{-1}$\\
\hline 
-80.0 & 1.08 & 0.07 & 0.54 & 0.06\\
-70.0 & 1.65 & 0.03 & 0.92 & 0.03\\
-60.0 & 2.15 & 0.02 & 1.15 & 0.01\\
\hline 
\end{tabular}

    \caption{\CHANGE{Measurement of stream proper motion through spline models. We provide proper motions in $\phi_1$ and $\phi_2$ (not corrected for Solar reflex motion) and their uncertainties at the set of $\phi_1$ points. The full table is included in the supplementary materials.}}
    \label{tab:pm_meas}
\end{table}

\begin{table}
    \centering
    \begin{tabular}{ccccc}
\hline 
$\phi_1$ & $\ln {\mathcal S}_{\mu,1}$ & $\sigma_{\ln {\mathcal S},\mu,1}$& $\ln {\mathcal S}_{\mu,2}$  & $\sigma_{\ln {\mathcal S },\mu,2}$ \\
deg &  & & & \\
\hline 
-80.0 & -3.4 & 1.0 & -4.1 & 1.1\\
-50.0 & -3.2 & 0.3 & -3.4 & 0.5\\
-25.0 & -2.0 & 0.3 & -3.2 & 0.5\\
\hline 
\end{tabular}

    \caption{\CHANGE{Measurement of stream proper motion dispersion. We provide measurements of the natural logarithm of the proper motion dispersion in $\phi_1$ and $\phi_2$ in \masyr together with the uncertainties. The full table is included in the supplementary materials.}}
    \label{tab:pmsig_meas}
\end{table}

\begin{table}
    \centering
    \begin{tabular}{ccc}
\hline 
$\phi_1$ & $\phi_2$ & $\sigma_{\phi,2}$ \\
deg & deg & deg \\
\hline 
-90.0 & -1.79 & 2.80 \\
-80.0 & 0.70 & 0.38 \\
-70.0 & 1.69 & 0.22 \\
\hline 
\end{tabular}

    \caption{\CHANGE{Measurement of stream track on the sky through spline models.  We provide the $\phi_2$ coordinates with the corresponding uncertainty for a set of $\phi_1$ locations. The full table is included in the supplementary materials.}}
    \label{tab:track_meas}
\end{table}

\begin{table}
    \centering
    \begin{tabular}{ccc}
\hline 
$\phi_1$ & $\ln \Sigma_{\phi,2}$ & $\sigma_{\ln \Sigma,\phi,2}$ \\
deg &  & \\
\hline 
-90.0 & -0.13 & 0.28 \\
-60.0 & 0.14 & 0.13 \\
-30.0 & -0.29 & 0.19 \\
\hline 
\end{tabular}

    \caption{\CHANGE{Measurements of stream width on the sky through spline models. We provide the value of the natural logarithm of stream width in degrees and its uncertainty at at the set of $\phi_1$ positions. The full table is included in supplementary materials.}}
    \label{tab:width_meas}
\end{table}

\begin{table}
    \centering
    \begin{tabular}{ccc}
\hline 
$\phi_1$ & $\ln I$ & $\sigma_{\ln I}$ \\
deg &  & \\
\hline 
-90.0 & -11.83 & 2.99 \\
-80.0 & -4.14 & 0.36 \\
-70.0 & -3.12 & 0.20 \\
\hline 
\end{tabular}

    \caption{\CHANGE{Measurements of stream surface brightness from  spline models (see Eq.~\ref{eqn:stream_dens_model}). We provide the logarithm of surface  brightness and its uncertainty at the set of $\phi_1$ positions. The units of surface brightness are stars per bin. The full table is included in the supplementary materials.}}
    \label{tab:density_meas}
\end{table}

\renewcommand{\arraystretch}{1.2}
\begin{table}
\begin{centering}
\begin{tabular}{|c|c|c|c|}
\hline
Parameter & Spherical & Prolate & Oblate \\
\hline $\sigma_{\phi_2}$ (deg) & $0.74^{+0.16}_{-0.17}$ & $0.41^{+0.19}_{-0.17}$ & $0.39^{+0.17}_{-0.15}$ \\
$\sigma_{\rm DM}$ (\masyr) & $0.09^{+0.04}_{-0.03}$ & $0.10^{+0.03}_{-0.03}$ &  $0.07^{+0.03}_{-0.02}$ \\
$\sigma_{\mu_1^*}$ (\masyr) & $0.05^{+0.03}_{-0.02}$ & $0.04^{+0.02}_{-0.02}$ & $0.05^{+0.02}_{-0.02}$ \\
$\sigma_{\mu_2}$ (\masyr) & $0.05^{+0.02}_{-0.01}$ & $0.04^{+0.02}_{-0.02}$ & $0.02^{+0.02}_{-0.01}$ \\
$\sigma_{v_r}$ (\kms) & $4.93^{+3.71}_{-2.70}$ & $2.51^{+2.18}_{-1.35}$ & $3.18^{+4.84}_{-1.76}$\\
\hline
\end{tabular}
\caption{Values of nuisance parameters for the additional scatter in proper motions, radial velocities and stream track determined as part of the model presented in Section~\ref{sec:modelling}. }
\label{tab:nuisance}
\end{centering}
\end{table}
\renewcommand{\arraystretch}{1.}

\section*{APPENDIX C:}
\CHANGE{In supplementary materials we also provide fiducial splines used to select likely members (shown on Figure~\ref{fig:rv_pm}). The files contain sets of points $\phi_1, \phi_2, \mu_{\phi,1},\mu_{\phi,2}, V_{gsr}$ and distance defining fiducial stream tracks. The actual tracks shown on Figure~\ref{fig:rv_pm} can be obtained by linear interpolation of provided points.
}

\section*{APPENDIX D:}
In the supplementary materials, we include {\tt Stan} models used in this paper. 
\begin{itemize}
    \item {\tt rvfit.stan} is the radial velocity model used in Section~\ref{sec:rv_model}
    \item {\tt pmfit.stan} is the proper motion model used in Section~\ref{sec:pm_model}
    \item {\tt distfit.stan} is the distance modulus model for RR Lyrae used in Section~\ref{sec:distance_model}.
    \item {\tt fehmodel.stan} is the model for the [Fe/H] gradients used in Section~\ref{sec:feh_grad}.
    \item{\tt density\_pm\_fast\_reduce.stan} is the proper motion/spatial distribution model described in Section~\ref{sec:density_model}. The model is parallisable across multiple cores.
\end{itemize}

\section*{APPENDIX E:} \label{sec:ELz_snaps}

\begin{figure*}
    \centering
    \includegraphics[width=\textwidth]{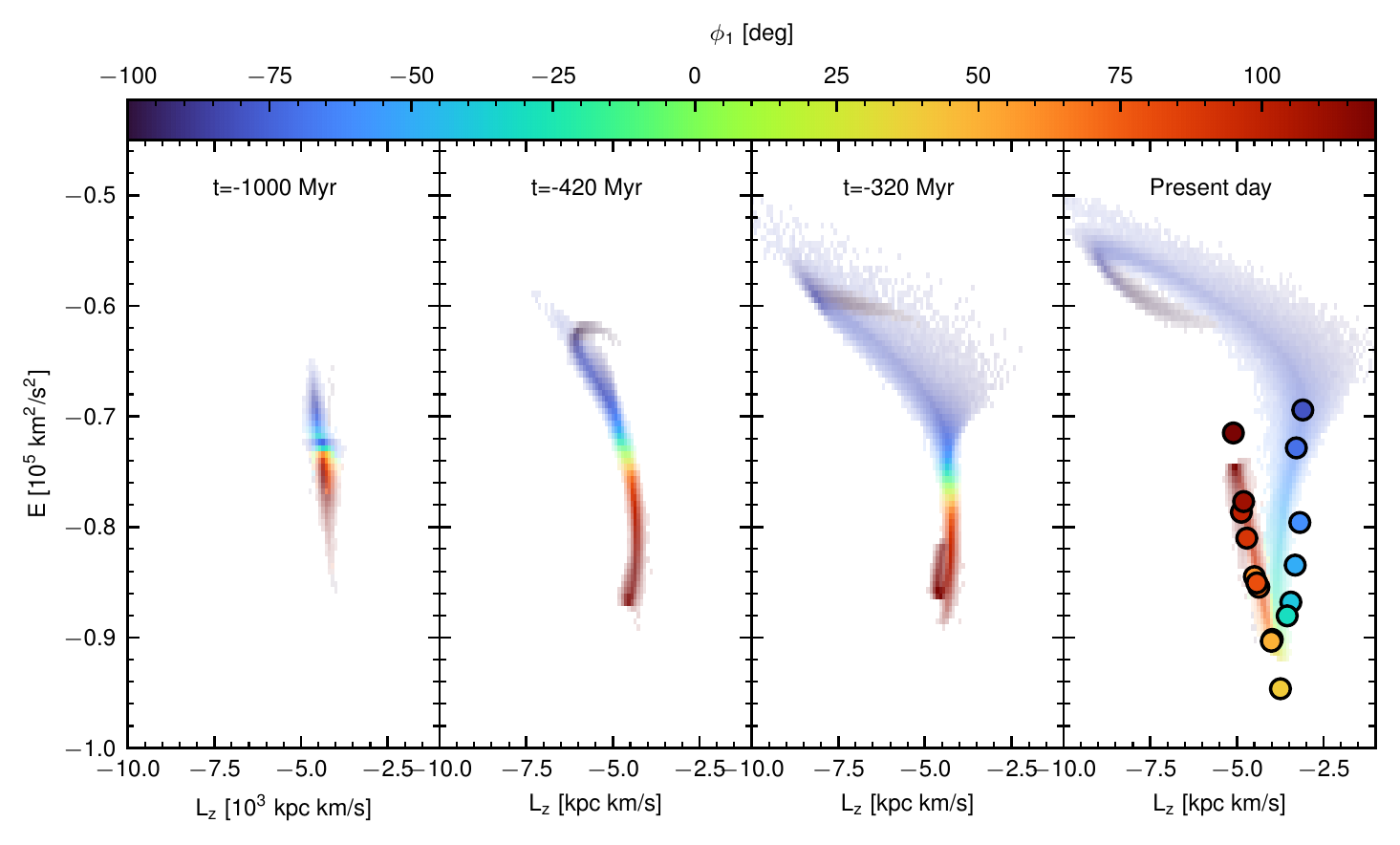}
    \caption{Evolution of the OC stream in energy and angular momentum. From left to right, the panels show the energy versus angular momentum for stream particles at look-back times of $1000$ Myr (i.e. well before the LMC has an effect), $420$ Myr ($\sim 50$ Myr before the LMC's closest approach), $320$ Myr ($\sim 50$ Myr after the LMC's closest approach), and the present day. A movie showing the evolution of energy and angular momentum is available \href{https://youtu.be/TTVaO1nHaGU}{here}.}
    \label{fig:energy_evol}
\end{figure*}

In Figure~\ref{fig:energy_evol} we show the evolution of the OC stream over the past 1 Gyr in energy versus angular momentum. 

\section*{APPENDIX F:}

\begin{figure}
    \centering
    \includegraphics{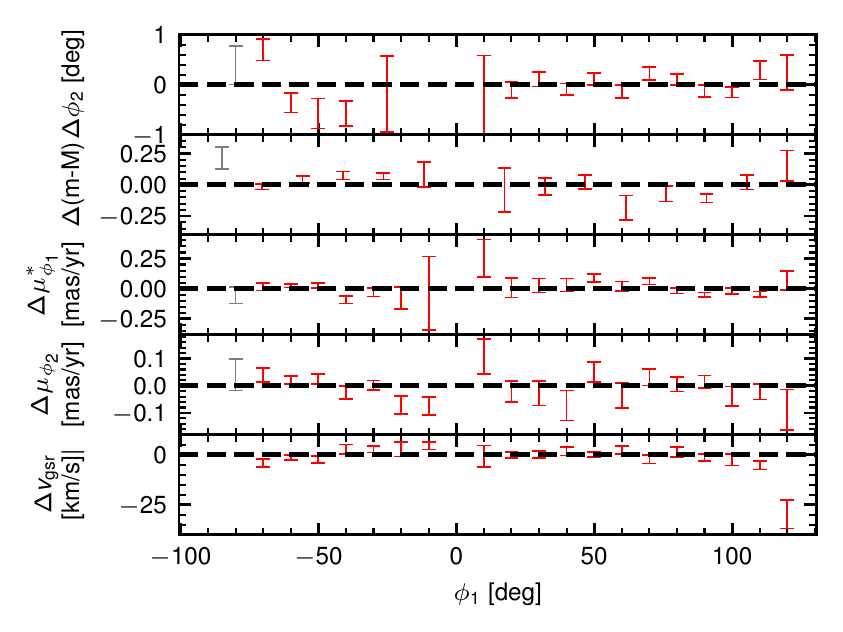}
    \caption{Residuals of the best-fit model shown in Figure~\ref{fig:OC_best}. From top to bottom, the panels show the residuals of the track on the sky, the distance modulus, the two proper motions, and the radial velocity. This fit has a $\chi^2$ per degree of freedom of 2.5, suggesting that our best fit model is not fully describing the data.  }
    \label{fig:oc_residuals}
\end{figure}

The residuals for our best-fit model presented in Figure~\ref{fig:OC_best} are shown in Figure~\ref{fig:oc_residuals}.  The residuals are mostly small with the exception of the rightmost RV measurement and the points at the very left edge of the stream that are the most violently affected by the LMC interaction. Grey points show the points that were not part of the fit.

\bsp	
\label{lastpage}
\end{document}